\documentclass[fleqn,usenatbib]{mnras}

\usepackage{newtxtext,newtxmath}
\usepackage[T1]{fontenc}

\DeclareRobustCommand{\VAN}[3]{#2}
\let\VANthebibliography\thebibliography
\def\thebibliography{\DeclareRobustCommand{\VAN}[3]{##3}\VANthebibliography}

\usepackage{graphicx, subcaption}	% Including figure files
\usepackage{amsmath}	% Advanced maths commands

\usepackage{natbib,color,ctable}
\usepackage{longtable}
\usepackage{threeparttablex}
\usepackage{wasysym}
\usepackage{url}
\usepackage{xcolor}
\usepackage{hyperref}
\usepackage{multirow}

\newcommand{\orcid}[2]{\href{http://orcid.org/#2}{#1}}
\newcommand{\kd}{KMOS$^{\rm 3D}$\,}

\title[Kinematics and mass estimates at $z\sim1$]{Galaxy kinematics and mass estimates at $z\sim1$ from ionised gas and stars}

\author[\"Ubler et al.]{\parbox{\textwidth}{
\orcid{Hannah \"Ubler}{0000-0003-4891-0794}$^{\hyperlink{aff1}{1},\hyperlink{aff2}{2}}$\thanks{E-mail: hu215@cam.ac.uk}, 
\orcid{Natascha M.\ F\"orster Schreiber}{0000-0003-4264-3381}$^{\hyperlink{aff3}{3}}$, 
\orcid{Arjen van der Wel}{0000-0002-5027-0135}$^{\hyperlink{aff4}{4}}$,
\orcid{Rachel Bezanson}{0000-0001-5063-8254}$^{\hyperlink{aff5}{5}}$,
\orcid{Sedona H.~Price}{0000-0002-0108-4176}$^{\hyperlink{aff3}{3},\hyperlink{aff5}{5}}$,
\orcid{Francesco D'Eugenio}{0000-0003-2388-8172}$^{\hyperlink{aff1}{1},\hyperlink{aff2}{2}}$,
\orcid{Emily Wisnioski}{0000-0003-1657-7878}$^{\hyperlink{aff6}{6},\hyperlink{aff7}{7}}$,
\orcid{Reinhard Genzel}{0000-0002-2767-9653}$^{\hyperlink{aff3}{3},\hyperlink{aff8}{8}}$,
\orcid{Linda J. Tacconi}{0000-0002-1485-9401}$^{\hyperlink{aff3}{3}}$,
\orcid{Stijn Wuyts}{0000-0003-3735-1931}$^{\hyperlink{aff9}{9}}$,
\orcid{Thorsten Naab}{0000-0002-7314-2558}$^{\hyperlink{aff10}{10}}$,
\orcid{Dieter Lutz}{0000-0003-0291-9582}$^{\hyperlink{aff3}{3}}$,
\orcid{Caroline M. S. Straatman}{0000-0001-5937-4590}$^{\hyperlink{aff11}{11}}$,
\orcid{T. Taro Shimizu}{0000-0002-2125-4670}$^{\hyperlink{aff3}{3}}$,
\orcid{Ric Davies}{0000-0003-4949-7217}$^{\hyperlink{aff3}{3}}$,
\orcid{Daizhong Liu}{0000-0001-9773-7479}$^{\hyperlink{aff3}{3}}$,
\orcid{J. Trevor Mendel}{0000-0002-6327-9147}$^{\hyperlink{aff6}{6},\hyperlink{aff7}{7}}$
}
\vspace{0.4cm}
\\
\parbox{\textwidth}{
\hypertarget{aff1}{$^{1}$}Kavli Institute for Cosmology, University of Cambridge, Madingley Road, Cambridge, CB3 0HA, UK\\
\hypertarget{aff2}{$^{2}$}Cavendish Laboratory, University of Cambridge, 19 JJ Thomson Avenue, Cambridge, CB3 0HE, UK\\
\hypertarget{aff2}{$^{3}$}Max-Planck-Institut f{\"u}r extraterrestrische Physik, Gie\ss enbachstra\ss e 1, 85748 Garching, Germany\\
\hypertarget{aff4}{$^{4}$}Sterrenkundig Observatorium, Universiteit Gent, Krijgslaan 281 S9, B-9000 Gent, Belgium\\
\hypertarget{aff5}{$^{5}$}Department of Physics and Astronomy and PITT PACC, University of Pittsburgh, Pittsburgh, PA 15260, USA\\
\hypertarget{aff6}{$^{6}$}Research School of Astronomy and Astrophysics, Australian National University, Canberra, ACT 2611, Australia\\
\hypertarget{aff7}{$^{7}$}ARC center of Excellence for All Sky Astrophysics in 3 Dimensions (ASTRO 3D), Australia\\
\hypertarget{aff8}{$^{8}$}Departments of Physics and Astronomy, University of California, Berkeley, CA 94720, USA\\
\hypertarget{aff9}{$^{9}$}Department of Physics, University of Bath, Claverton Down, Bath BA2 7AY, UK\\
\hypertarget{aff10}{$^{10}$}Max-Planck-Institut f{\"u}r Astrophysik, Karl-Schwarzschild-Stra\ss e 1, 85748 Garching, Germany\\
\hypertarget{aff11}{$^{11}$}Department of Physics and Astronomy, Ghent University, Krijgslaan 281 S9, B-9000 Gent, Belgium
}
}

\date{Accepted XXX. Received YYY; in original form ZZZ}

\pubyear{2022}

\begin{document}
\label{firstpage}
\pagerange{\pageref{firstpage}--\pageref{lastpage}}
\maketitle

\begin{abstract}
We compare ionised gas and stellar kinematics of 16 star-forming galaxies ($\log(M_\star/M_\odot)=9.7-11.2$, SFR~$=6-86 M_\odot/yr$) at $z\sim1$ using near-infrared integral field spectroscopy (IFS) of H$\alpha$ emission from the \kd survey and optical slit spectroscopy of stellar absorption and gas emission from the LEGA-C survey. 
H$\alpha$ is dynamically colder than stars, with higher disc rotation velocities (by $\sim45$ per cent) and lower disc velocity dispersions (by a factor $\sim2$). This is similar to trends observed in the local Universe.
We find higher rotational support for H$\alpha$ relative to [OII], potentially explaining systematic offsets in kinematic scaling relations found in the literature.
Regarding dynamical mass measurements, for six galaxies with cumulative mass profiles from Jeans Anisotropic Multi-Gaussian Expansion (JAM) models the H$\alpha$ dynamical mass models agree remarkably well out to $\sim10$~kpc for all but one galaxy (average $\Delta M_{\rm dyn}(R_{e,\rm F814W})<0.1$~dex). 
Simpler dynamical mass estimates based on integrated stellar velocity dispersion are less accurate (standard deviation 0.24~dex). 
Differences in dynamical mass estimates are larger, for example, for galaxies with stronger misalignments of the H$\alpha$ kinematic major axis and the photometric position angle, highlighting the added value of IFS observations for dynamics studies.
The good agreement between the JAM models and the dynamical models based on H$\alpha$ kinematics at $z\sim1$ corroborates the validity of dynamical mass measurements from H$\alpha$ IFS observations, which can be more easily obtained for higher redshift  galaxies.
\end{abstract}


%%%%%%%%%%%%%%%%%%%%%%%%%%%%%%%%%%%%%%%%%%%%%%%%%%

%%%%%%%%%%%%%%%%% BODY OF PAPER %%%%%%%%%%%%%%%%%%

\section{Introduction}\label{s:intro}

The study of galaxy kinematics as a function of cosmic time provides important insights into the evolution of galactic mass budgets and structure \citep[e.g.][]{Sofue01, FS20}. Different kinematic tracers like molecular gas, ionised gas, or stars move in the same galactic potential $\Phi$ and allow for estimates of the galactic dark matter content.
The kinematic signatures of different tracers vary due to their different nature: stars are collision-less while gas is dissipative; different gas phases have different temperatures, turbulent velocities, and might be affected by outflows. The various tracers often have different spatial distributions and probe different regions of the overall potential. Nonetheless, dynamical models based on complementary tracers should give the same mass estimates for systems in equilibrium.

In the local Universe, large interferometric and IFS surveys provide spatially resolved kinematics of stars, and atomic, molecular, and ionised gas. 
Comparative studies of baryonic kinematics at $z=0$, in particular from the EDGE-CALIFA survey \citep{Sanchez12, Bolatto17} and the ATLAS$^{\rm 3D}$ project \citep{Cappellari11}, brought forth the following general results:
(i) rotation velocities are highest and velocity dispersions are lowest for neutral gas, followed by ionised gas, and then stars \citep[e.g.][]{VegaBeltran01, Davis13, Martinsson13a, Bolatto17, Levy18, CrespoGomez21, Girard21}; 
(ii) modelling of stellar kinematics (e.g.\ with axisymmetric Jeans Anisotropic Multi-Gaussian Expansion models; JAM; \citealp{Cappellari08}) produces circular velocity curves that match observed cold molecular gas rotation velocities in regularly rotating galaxies \citep[e.g.][]{Davis13, Leung18}, suggesting that the mass estimates from molecular gas and stars are in agreement;
(iii) the mis-alignment of gas and stellar kinematic major axes with each other and with the morphological major axis is small for the majority of non-interacting systems without strong bars, but generally higher for Early-Type Galaxies (ETGs) \citep[e.g.][]{FalconBarroso06, Sarzi06, Davis11, BarreraBallesteros14, BarreraBallesteros15, Serra14, Bryant19}. 

In contrast, our knowledge of galaxy kinematics at $1<z<3$, when massive galaxies assemble most of their stellar mass \citep[e.g.][]{Moster20}, is dominated by ionised gas observations. Efficient multiplexing near-infrared spectrographs such as KMOS \citep[the K-band Multi-Object Spectrograph;][]{Sharples04, Sharples13} and MOSFIRE \citep[the Multi-Object Spectrometer For Infra-Red Exploration][]{McLean10, McLean12}, trace strong rest-frame optical emission lines in several thousand galaxies up to $z\sim3.5$ \citep[see][for an overview]{FS20}. 
Larger surveys of molecular gas with few hundred $1<z<3$ galaxies focus on integrated quantities \citep[e.g.][and references therein]{Tacconi13, Tacconi18, Freundlich19}, with some exceptions of spatially resolved kinematics for individual galaxies \citep[e.g.][]{Tacconi10, Swinbank11, Genzel13, Uebler18, Girard19, Molina19, Kaasinen20, Lelli23, Liu23, Rizzo23}.
Stellar kinematic observations at $z>1$ were initially obtained almost exclusively for quiescent galaxies, and from slit spectroscopy, focusing on integrated quantities due to signal-to-noise (S/N) considerations \citep[but see][]{Newman15, Newman18, Toft17, Mendel20}. 
The LEGA-C \citep[Large Early Galaxy Astrophysics Census;][]{vdWel16, vdWel21, Straatman18} survey brought a step change in stellar kinematics of distant systems. Thanks to its deep uniform integrations and large sample size, spatially resolved kinematic analyses and modelling have become feasible for few hundred galaxies of all types at $0.6<z<1$ \citep{Bezanson18, vHoudt21, Straatman22, vdWel22}, and integrated kinematic measures for few thousands \citep{vdWel21}.

Multi-tracer observations of galaxy kinematics at $z>0$ are sparse, but overall indicate similar trends as $z=0$ studies. Molecular disc velocity dispersions are lower relative to ionised gas at $z\sim0.2$ (\citealp{Cortese17}, see also \citealp{Molina20}). There are indications that this trend prevails out to $z\sim2$ \citep{Girard19, Uebler19, Liu23, Lelli23}, while some individual galaxies have comparable dispersions \citep{Genzel13, Uebler18, Molina19}. 
Tentative trends of higher stellar disc velocity dispersions compared to ionised gas are seen in the data by \cite{Guerou17} of 17 galaxies at $z\sim0.5$.

A recent study by \cite{Straatman22} compares dynamical mass estimates based on slit observations of ionised gas and stars for 157 galaxies at $0.6<z<1$ from the LEGA-C survey. In that paper, dynamical masses from stellar kinematics are inferred from JAM models \citep{vHoudt21}, and dynamical masses from ionised gas are inferred from pressure-supported disc models, where the latter are found to be systematically lower by 0.15~dex. No correlations of this discrepancy with galaxy properties were found. 
The authors find a similar offset when comparing JAM estimates to mass measurements based on integrated emission line widths.

We revisit the comparison of stellar and ionised gas dynamical mass models with IFS observations of the H${\alpha}$ emission line from the KMOS$^{\rm 3D}$ survey \citep{Wisnioski15, Wisnioski19}. IFS data provide knowledge of the kinematic major axis of a galaxy through velocity and velocity dispersion maps. This allows for the extraction of major axis kinematics, which can be fed directly into dynamical models. However, the kinematic major axis is typically unknown for long-slit observations, and is usually assumed to coincide with the photometric position angle. Further corrections are required if the slit orientation is different from the photometric position angle, as is largely the case for the LEGA-C observations.

In this paper, we present a comparison of kinematics and dynamical mass estimates from IFS observations of the H${\alpha}$ emission line from KMOS$^{\rm 3D}$, and long-slit spectroscopic observations of stars from LEGA-C. 
This paper is organized as follows. Our sample is described in Section~\ref{s:data}. In Section~\ref{s:kinex} we discuss the extraction of kinematic profiles from both surveys, including careful matching of instrumental effects, and the construction of dynamical mass models in Section~\ref{s:modelling}. We discuss results based on stellar and ionised gas kinematics in Section~\ref{s:compasobs}. In Section~\ref{s:mdyn} we compare dynamical mass estimates from the different tracers and methods, and investigate correlations of dynamical mass offset with physical, structural, and kinematic properties of the galaxies. We conclude in Section~\ref{s:conclusions}.

\section{Data}\label{s:data}
\subsection{The Sample}

\begin{figure*}
	\centering
	\includegraphics[width=0.49\textwidth]{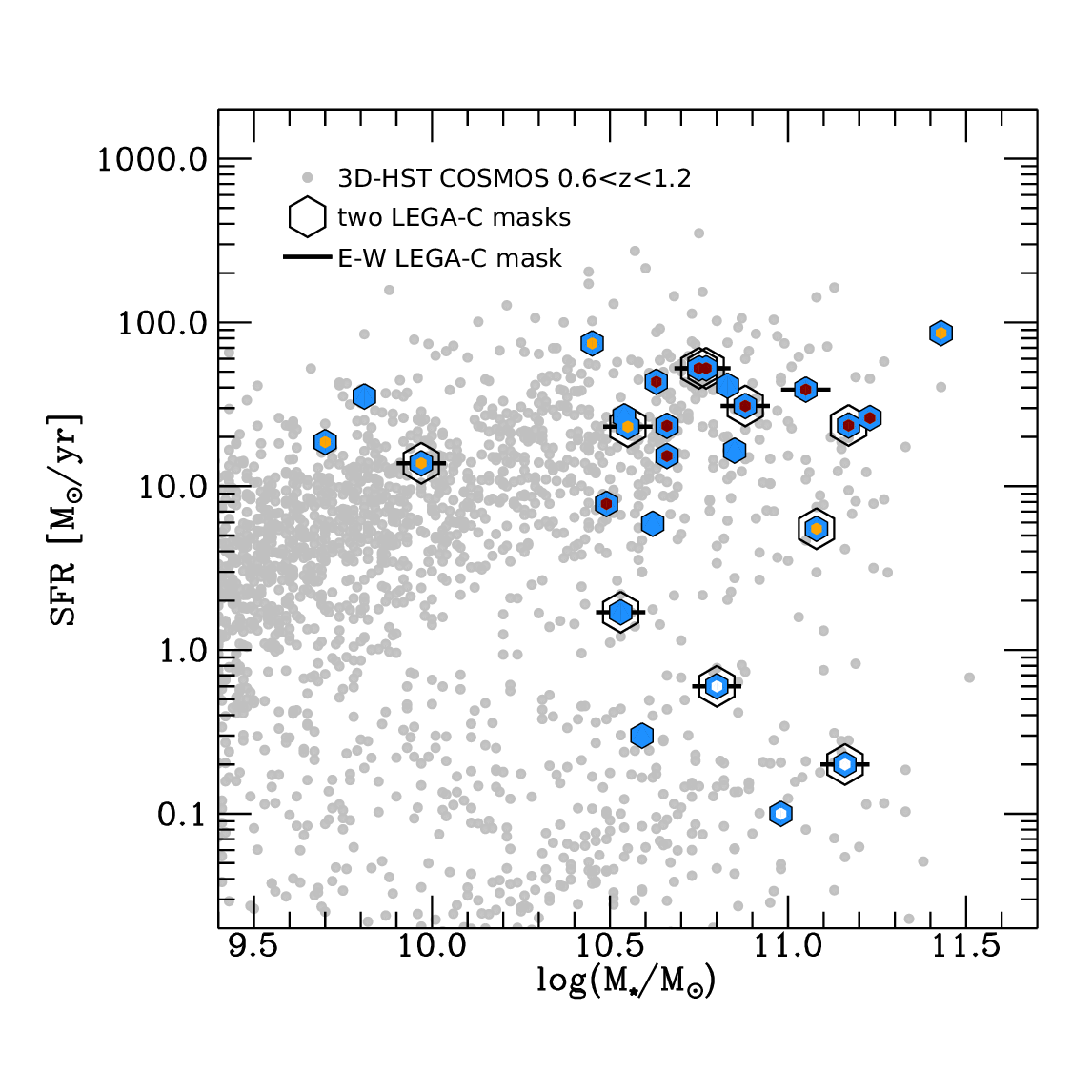}
	\includegraphics[width=0.49\textwidth]{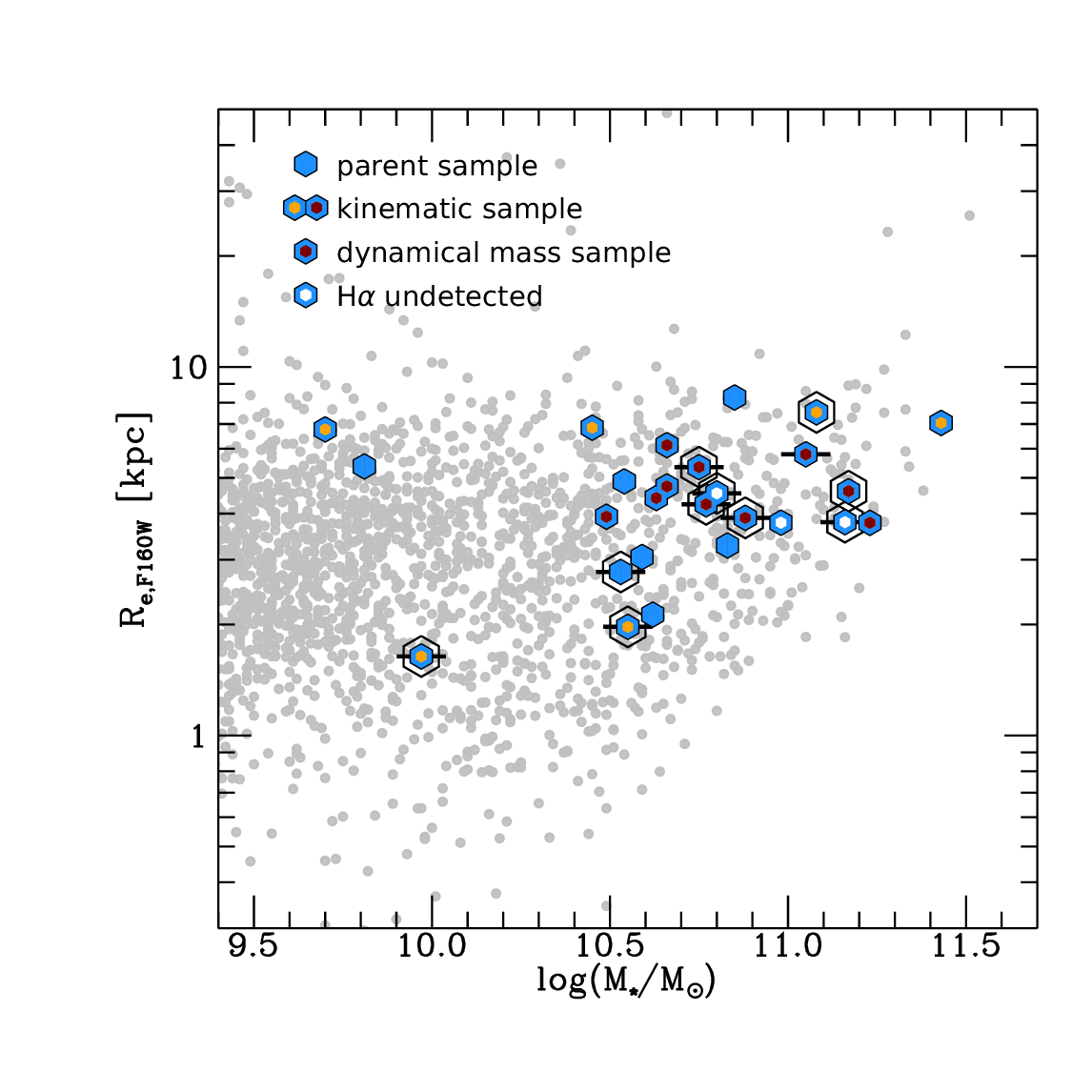}
	\caption{Location of our parent sample (blue symbols) in the $M_{\star}-$ SFR (left) and $M_{\star}-R_e$ (right) planes on top of the underlying galaxy population in the COSMOS field at $0.6<z<1.2$ taken from the 3D-HST catalogue (grey points). Duplicate observations in LEGA-C are indicated by black hexagons, and LEGA-C observations in E-W orientation are indicated by a horizontal bar. Symbols with white centres indicate galaxies formally undetected in H$\alpha$ in the \kd data release. Symbols with orange/red centres indicate galaxies in our kinematics sample, and symbols with red centres indicate galaxies in our dynamical mass sample. Our galaxies span a range in stellar masses, SFRs, and sizes, but most objects are located at the massive end of the $z\sim1$ main sequence and follow the mass-size relation.}
	\label{f:msmr}
\end{figure*}

For our analysis we select galaxies observed within both the \kd and LEGA-C surveys.
There are 26 unique targets common to both surveys, all located in the Cosmic Evolution Survey (COSMOS) field \citep{Scoville07}. Ten of the LEGA-C galaxies have been observed twice with separate mask designs, and nine galaxies have LEGA-C longslit observations oriented in E-W direction in addition to or instead of the default N-S observations. 
The sample spans a range in stellar mass, star-formation rate, and size, as illustrated in Figure~\ref{f:msmr}. Most galaxies are located at the massive end of the $z\sim1$ main sequence and follow the mass-size relation, but a few log$(M_*/M_{\odot})<10$ systems are included, as well as some passive systems. 
The quality of data for both \kd and LEGA-C varies across the sample, primarily due to integration time (for \kd) and observing conditions.

For the comparison of line-of-sight (LOS) kinematics in this work, we focus on 16 galaxies for which we can extract velocities and velocity dispersions across at least 1” along the (pseudo-)slit in both surveys, and for the dynamical mass comparison we utilise 10 galaxies for which dynamical masses can be measured from both surveys (see Sections~\ref{s:kinex} and \ref{s:modelling} for details). We list all galaxies discussed in this work with redshifts and \kd integration times in Table~\ref{t:sample}.

\begin{table*}
	\centering
	\caption{\kd ID with field and 3D-HST v4 catalog object ID \citep{Skelton14}; \kd redshift; \kd on-source integration time in minutes; LEGA-C Mask-ID with mask number and UltraVista catalog object ID \citep{Muzzin13}; second LEGA-C Mask-ID in case of duplicate observations; LEGA-C redshift. The integration times for the LEGA-C observations are all approximately 1200 minutes. LEGA-C Mask-IDs with mask number M101 refer to longslit observations in E-W orientation.}
	\label{t:sample}
		\begin{tabular}{lrrllr}
		\kd ID & \kd $z$ & \kd $t_{\rm int}$ & LEGA-C Mask-ID & second LEGA-C Mask-ID & LEGA-C $z$ \\
		\hline
    COS4\_03493 &      0.679 &       285 & M4\_121150 & - & 0.678\\
    COS4\_13901 &      0.684 &       625 & M1\_131104 & M2\_131104 & 0.683\\
    COS4\_04943 &      0.758 &       230 & M3\_122667 & - & 0.758\\
    COS4\_16227 &      0.796 &       335 & M1\_131985 & - & 0.796\\
    COS4\_09601 &      0.834 &       230 & M5\_126127 & M101\_126127 & 0.834 \\
    COS4\_19648 &      0.889 &       315 & M1\_134839 & - & 0.889\\
    COS4\_25353 &      0.897 &       255 & M1\_139825 & M101\_139825 & 0.897\\
    COS4\_17628 &      0.907 &       100 & M5\_133199 & M101\_133199 & 0.907\\
    COS4\_06487 &      0.907 &       370 & M3\_123575 & M101\_123575 & 0.908\\
    COS4\_05296 &      0.926 &       230 & M4\_122584 & - & 0.925\\
    COS4\_10860 &      0.927 &       230 & M1\_127387 & - & 0.927\\
    COS4\_05238 &      0.936 &       230 & M7\_122836 & - & 0.937\\
    COS4\_09156 &      0.938 &       285 & M3\_125617 & M101\_125617 & 0.939\\
    COS4\_03700 &      0.956 &       230 & M3\_121631 & M4\_121631 & 0.956\\
    COS4\_08096 &      0.979 &       230 & M7\_125257 & - & 0.980\\
    COS4\_12699 &      1.003 &       320 & M101\_128834 & - & 1.005\\
		\hline
	\end{tabular}
\end{table*}

\subsection{The KMOS\texorpdfstring{$^{\rm 3D}$}{3D} Survey}\label{s:k3d}

The \kd survey is a 75-night GTO survey with the multi-IFS KMOS at the VLT, targeting the H$\alpha$+[NII] line emission in 739 $\log(M_*/M_{\odot})>9$ galaxies at $0.6<z<2.7$. The survey is presented by \cite{Wisnioski15, Wisnioski19}, to which we refer the reader for details.

The \kd galaxies were drawn from the {\it Hubble Space Telescope} ({\it HST}) 3D-HST Treasury Survey \citep{Brammer12, Skelton14, Momcheva16}, providing secure spectroscopic or grism redshift for optimal avoidance of skyline contamination at the location of H$\alpha$. A $\log(M_*/M_{\odot})>9$ and $K < 23$~mag selection function was chosen to obtain a population-wide census reducing biases in SFR or colors. Targets are located in COSMOS, GOODS-S (Great Observatories Origins Deep Survey) and UDS (Ultra Deep Survey). High-resolution Wide Field Camera 3 (WFC3) near-IR and Advanced Camera for Surveys (ACS) optical imaging is available from the Cosmic Assembly Near-infrared Deep Extragalactic Legacy Survey \citep[CANDELS;][]{Grogin11, Koekemoer11, vdWel12}, and further multi-wavelength coverage from X-ray through optical, near- to far-infrared, and radio is accessible from e.g.\ \cite{Ueda08, Lutz11, Xue11, Civano12, Magnelli13, Skelton14}.

The publicly released data cubes have a spatial sampling in $x-y-$direction of $0.2\arcsec$ which corresponds to $\sim1.6$ kpc at $z=1$. The wavelength sampling in $z-$direction is 1.7~\AA. The typical near-IR seeing of the \kd data has a FWHM of $0.5\arcsec$, corresponding to $\sim4.0$ kpc at $z=1$. Point-spread function (PSF) images representing the observing conditions for each combined data cube individually are included in the data release, together with both a Gaussian and Moffat parametrization. 
The average spectral resolution for \kd observations in the $YJ$ filter is $R=\lambda/\Delta\lambda=3515$, corresponding to an average instrumental dispersion of $\sigma_{\rm instr}\sim36$~km/s. However, the line-spread function (LSF) of each galaxy, which is close to Gaussian, is determined individually as a function of wavelength, and encoded in the fits header keywords as described by \cite{Wisnioski19}. The average on-source integration time for $z\sim1$ targets in \kd is 5 hours.

Stellar masses $M_*$ and star formation rates (SFRs) for all galaxies are derived from SED fitting following \cite{WuytsS11a}, assuming a \cite{Chabrier03} initial mass function, using \cite{Bruzual03} models with solar metallicity, the reddening law by \cite{Calzetti00}, and constant or exponentially declining star formation histories. Typical uncertainties from SED fitting are $0.15$~dex for stellar masses, and $0.10-0.25$~dex for SFRs.
Structural parameters such as the effective radius $R_e$, the Sérsic index $n_S$, the axis ratio $q=b/a$, the morphological position angle PA$_{\rm morph}$, and for some galaxies the bulge-to-total ratio $B/T$ and corresponding radii and Sérsic indices, are constrained from single-Sérsic or double-Sérsic {\sc{galfit}} \citep{Peng10} models to the CANDELS F160W imaging as presented by \cite{vdWel12, Lang14}.

\subsection{The LEGA-C Survey}\label{s:lgc}

The LEGA-C survey is a 1107-hour public survey with the Visible Multi-Object Spectrograph (VIMOS) at the VLT \citep{LeFevre03}, targeting 3741 galaxies at $0.6<z<1.5$. The survey is presented by \cite{vdWel16, vdWel21, Straatman18}, to which we refer the reader for details.

The LEGA-C targets were drawn from the COSMOS/UltraVISTA $K-$band selected catalogue with photometry \citep{Muzzin13}, and the primary sample consists of $0.6<z<1.0$ galaxies with $K < 20.7 - 7.5\cdot \log[(1 + z)/1.8]$~mag. ACS imaging is available for most targets \citep{Scoville07}.

LEGA-C slits are $1\arcsec$ ($\sim8.0$~kpc at $z\sim1$) wide and typically at least $8\arcsec$ long, oriented in N-S direction, with the exception of one mask having slits oriented in E-W direction. The sampling in the spatial direction is $0.205\arcsec$, and in the wavelength direction 0.6~\AA. The average PSF FWHM measured from Moffat fits is $0.8\arcsec$.
The effective spectral resolution is $R\sim3500$, and therefore comparable to the average KMOS $YJ$ instrumental resolution of 36~km/s. The typical on-source integration time is 20 hours.

Structural parameters such as $R_e$, $n_S$, $q$, and PA$_{\rm morph}$, are constrained from single-Sérsic {\sc{galfit}} models to ACS F814W imaging as presented by \cite{vdWel16, vdWel21}. F814W is chosen because it has the largest overlap with the LEGA-C footprint, and therefore all dynamical mass estimates for LEGA-C discussed below use F814W-based structural parameters (although near-IR WFC3 imaging is also available for the subset studied in this work). 

We note that the main analysis in this work uses the default imaging of the two surveys. We have tested using common imaging information, and we discuss those results where appropriate (see Section~\ref{s:model_diff} and Appendix~\ref{a:f814w_modelling}). Our main conclusions are not affected by this choice.

\begin{figure*}
	\centering
	\includegraphics[width=0.9\textwidth]{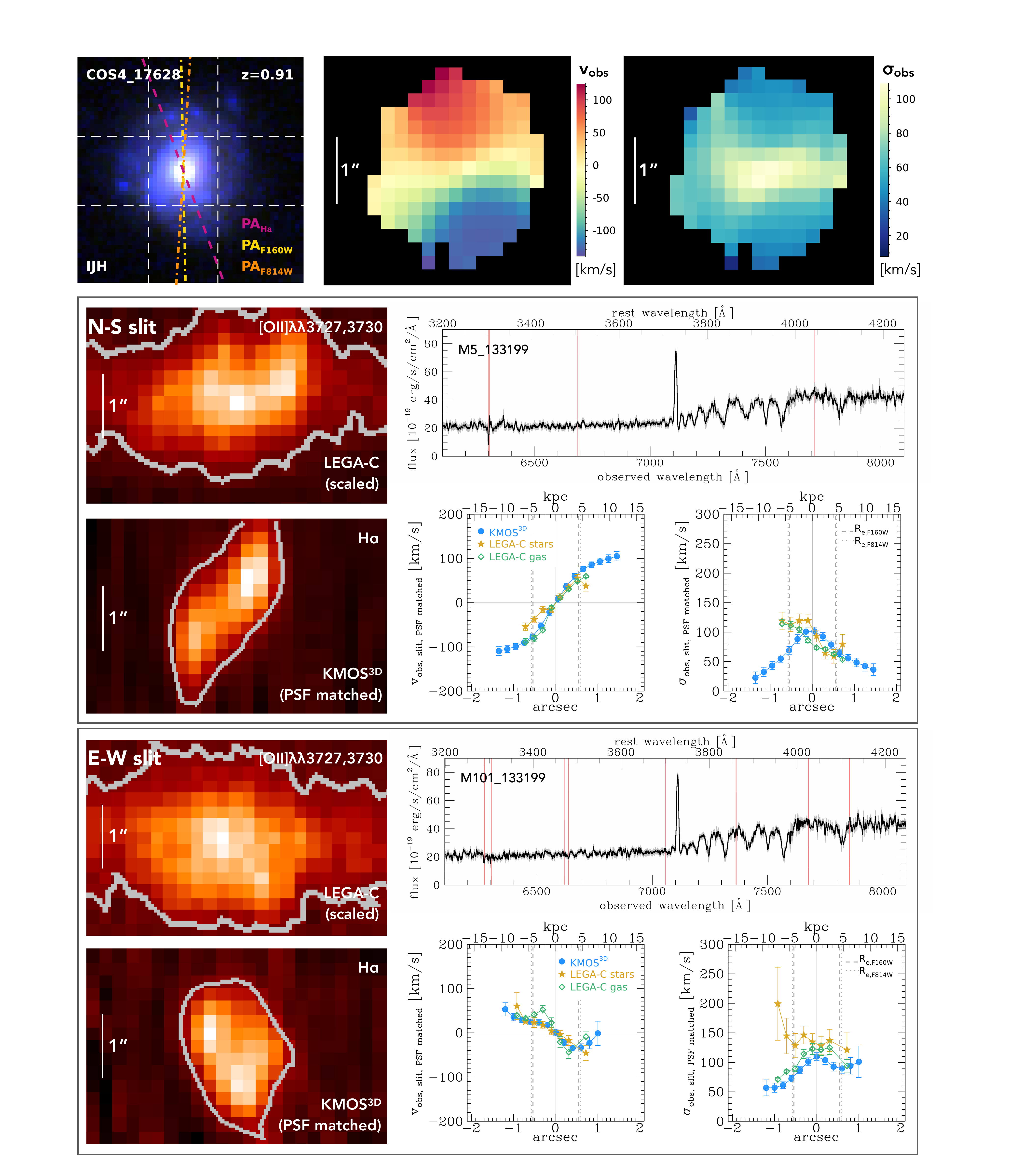}
	\caption{Illustration of the available kinematic and ancillary data for galaxies common to the \kd and LEGA-C surveys.
	Top left: IJH {\it HST} color-composite image, with white dashed lines indicating the $1\arcsec$ wide VIMOS slit positions from LEGA-C, orange and yellow dash-dotted lines indicating the morphological position angles derived from the F814W and F160W passbands, respectively, and the violet dashed line indicating the kinematic position angle determined form the \kd H$\alpha$ IFS data. 
	Top middle, right: H$\alpha$ projected velocity and velocity dispersion maps as derived from the \kd data cubes.
	Middle left (top): cutout around the [OII] emission line of the 2D data from the LEGA-C survey with $S/N=3$ contours. Here, the VIMOS data are rebinned in wavelength direction to match the coarser KMOS sampling.
    Middle left (bottom): 2D pseudo-slit extraction around the H$\alpha$ emission line from the \kd survey with $S/N=3$ contours. Here, the KMOS data cubes are first convolved to match the typically worse VIMOS PSF, and 2D data are subsequently extracted from a pseudo-slit matching the orientation and width of the LEGA-C data (vertical white dashed lines in top left panel).
	Middle right: 1D integrated LEGA-C spectrum collapsed along the N-S slit (top; smoothed for illustrative purposes) with pink vertical lines indicating low quality regions, and 1D velocity and velocity dispersion profiles (bottom) extracted along the N-S (pseudo-)slit from \kd (filled blue circles) and LEGA-C (stars: filled golden stars; gas: open green diamonds); dotted and dashed vertical lines indicate the major axis effective radii derived from the F814W and F160W passbands.
	Bottom panels: same as middle panels, but now for data extracted in E-W direction (horizontal white dashed lines in top left panel).
	White vertical bars indicate $1\arcsec$ in the kinematic maps and PV diagrams.
    The 2D data and 1D collapsed LEGA-C spectra are only shown for illustrative purposes. The 1D LOS kinematic profiles from the fixed (pseudo-)slit extractions (see Sections~\ref{s:kmoskin} and \ref{s:lgckin}) form the basis for the measurements described in Section~\ref{s:measure}.
    This galaxy has a comparable 1D rotation pattern in the \kd and LEGA-C data, however the LEGA-C velocity dispersions appear asymmetric compared to the H$\alpha$ extractions. Differences in the brightness distribution of the [OII] {\it vs.\ }H$\alpha$ emission are apparent in the 2D PV diagrams.}
	\label{f:obsprof}
\end{figure*}

\section{Kinematic extractions and Dynamical models}\label{s:methods}

\subsection{Fixed-slit kinematic extractions}\label{s:kinex}

\subsubsection{Extractions for KMOS$^{3D}$}\label{s:kmoskin}

For an adequate comparison of LOS kinematics, we process the \kd data as follows. First, we match the PSFs of \kd and LEGA-C data for each observation, then we extract position-velocity (PV) diagrams matching the orientation and slit width of the LEGA-C data, and finally we extract from those the 1D kinematic profiles by means of Gaussian fits to the H$\alpha$ line profile in each spatial row.

To match the PSFs, we create for each galaxy a convolution kernel based on the Moffat parametrizations of the PSFs from each pair of observations.\footnote{We use the Photutils \citep{photutils} subpackage for PSF matching {\tt{photutils.psf.matching}} with a top hat window, and verify the convolution based on the model PSFs.} For duplicate observations in LEGA-C, we create separate convolution kernels per observation and galaxy. These kernels are then applied to the \kd data cubes (which have smaller PSF FWHMs for all galaxies in our sample; see also Sections~\ref{s:k3d}, \ref{s:lgc}).

To extract PV diagrams, we place a pseudo-slit with width of $1\arcsec$ on the \kd cubes, oriented in N-S and/or E-W direction, as appropriate. For illustrative purposes, we show examples of such PV diagrams in Figure~\ref{f:obsprof}.
Slit centering is based on the location of the synthesized continuum maps created from the KMOS data cubes. 
Note that the exact VIMOS slit positioning is uncertain, but \cite{vHoudt21} conclude that the typical offset from the galaxy centre should be no more than $1-2$ pixels ($0.2-0.4\arcsec$), based on an analysis of asymmetric light profiles. 
For the purpose of our comparison, we assume that the position of galaxies in the LEGA-C slits aligns well with the \kd continuum centre (but see Section~\ref{s:multiple} for discussion of an outlier).

From the 2D PV diagrams we extract spectra for each row, and determine the H$\alpha$ LOS velocity and LOS velocity dispersion from Gaussian fits to the line profile at the H$\alpha$ position. 
Specifically, we fit the emission line profile in a range $40-80$\AA\ around the systemic line position, and we include or exclude individual fits based on visual inspection. We caution that emission line profiles do not always have a Gaussian shape, as is expected due to beam-smearing and projection effects, potential multi-component structure, but also possibly caused by non-circular motions such as radial flows \citep[see e.g.][]{vdKruit78, Bosma81, Sofue01}. Despite the high data quality from the \kd and LEGA-C surveys for galaxies in this redshift range, such effects cannot be robustly traced for the pixel-based extractions in our sample, and we therefore limit our analysis to the first and second moments.
For the velocity dispersion profiles, we subtract the LSF appropriate for the H$\alpha$ line position in quadrature, to remove instrumental broadening.
An example of the resulting profiles is shown in Figure~\ref{f:obsprof}, and profiles for all galaxies are shown in Appendix~\ref{a:gallery}.

\subsubsection{Extractions for LEGA-C}\label{s:lgckin}

LOS velocity and LOS velocity dispersion profiles for the stellar and ionised gas in LEGA-C are derived as described in detail by \cite{Bezanson18} and \cite{vdWel21}: the 2D PV diagrams are fit in each row (with median S/N~$>2$ per pixel) with pPXF \citep{Cappellari04, Cappellari17} by combining a high-resolution stellar population template and an emission line template. These templates are allowed to shift and broaden independently, delivering independent stellar and ionised gas velocity and velocity dispersion profiles. This procedure takes into account the LSF and removes instrumental broadening from the velocity dispersion profiles. An example for one galaxy is shown in Figure~\ref{f:obsprof}, and profiles for all galaxies are shown in Appendix~\ref{a:gallery}. 
Note that the template for the [OII] doublet consists of two emission lines centred at $\lambda=3727$~\AA\ and $\lambda=3730$~\AA.

For the purpose of our study, we make a few adjustments to the above methodology for individual objects: 
for two observations in our sample (M1\_127387, M5\_126127) we repeat the above fitting procedure without imposing a S/N-cut in order to obtain a radially dependent LOS dispersion profile. 
Another two galaxies (M1/101\_139825, M7\_122836) show strong [NeV]$\lambda3347$, [NeV]$\lambda3427$ and [NeIII]$\lambda3870$ emission. The high-ionisation [NeV]$\lambda3427$ line is a tell-tale signature of a harder ionising radiation field than produced by pure star formation, indicative of AGNs or shocks \citep[e.g.][]{Mignoli13, Feltre16, Vergani18, Kewley19}. Where present in the LEGA-C spectra in our sample, it is kinematically decoupled from other emission lines and centrally concentrated. To extract ionised gas kinematics for these galaxies, we mask the corresponding spectral regions and repeat the above fitting procedure.
The effect on the extracted gas kinematics is substantial, with differences in individual velocity and velocity dispersion measurements of up to 200~km/s (see Appendix~\ref{a:nev} for an example). 
The stellar kinematic measurements are virtually unaffected by this procedure.

For visual comparison of the 2D PV diagrams we further resample the LEGA-C spectra to the (coarser) KMOS wavelength steps. Note that we do not resample in spatial direction due to the very small difference in the KMOS and VIMOS pixel scales of $0.005\arcsec$.

\subsubsection{Measurements}\label{s:measure}

Due to the different radial coverage of the data it is not straight-forward to compare the gas and stellar kinematics in these systems even after matching the observing conditions. To quantify how well the \kd and LEGA-C kinematic data compare to each other, we define the following 1D measurements based on the LOS kinematic profiles (see Figure~\ref{f:obsprof}):

\begin{itemize}
    \item $v_{\rm max}$ is the maximum observed absolute velocity (uncorrected for inclination), and $r_{\rm vmax}$ is the corresponding radius.
    \item $v_{\rm rmax,both}$ is the (mean) velocity at the outermost radius covered by both the \kd and LEGA-C data, and $r_{\rm max,both}$ is the corresponding radius.
    \item $\sigma_{\rm out}$ is the weighted mean observed velocity dispersion of the four outermost measured values (outer two on each side of the profile). Note that this measurement may still be affected by beam smearing, especially for smaller systems.
    \item $\sigma_{\rm rmax,both}$ is the (mean) observed velocity dispersion at $r_{\rm max,both}$.
    \item $v_{\rm rms}$ is an approximation of a classical root mean square velocity, via $v_{\rm rms}^2=v_{\rm max}^2 + \sigma_{\rm out}^2$.
    \item $v_{c,\rm max}$ is an approximation of a circular velocity (here without corrections for inclination and beam-smearing), via $v_{c,\rm max}^2=v_{c}^2(r_{\rm vmax})=v_{\rm max}^2+2\sigma_{\rm out}^2\cdot r_{\rm vmax}/R_d$ \citep[see][for details]{Burkert10}, where we assume that the disc scale length $R_d=R_e/1.68$, with $R_e=R_{e,\rm F160W}$.
\end{itemize}
For the LEGA-C measurements, the above quantities are measured for ionised gas (primarily [OII] and/or H$\beta$) and stellar kinematics individually.

We stress that the above quantities are not derived from modelling, but are based on the LOS kinematics, which for the \kd galaxies have been extracted mimicking the LEGA-C observing conditions and setup.
Therefore, while the LSF is accounted for, the measurements do not include corrections for inclination or beam-smearing. This is to say, intrinsic maximum velocities would be larger, and intrinsic velocity dispersions would be smaller. However, due to our matching of the PSFs, differences due to beam-smearing in the original observations are accounted for, and gas kinematics in \kd and LEGA-C should match if the emission lines trace the same ISM components. 

Furthermore we emphasize that the above observed `maximum' velocities do not necessarily represent the true observed maximum velocities of the galaxies, due to the kinematic major axes generally not being aligned with the slit orientations (see Section~\ref{s:mdynk3d} for \kd kinematic extractions along the kinematic major axis).

The measurements described above cannot be meaningfully performed for all galaxies and observations in the sample. We exclude from the subsequent comparison galaxies for which there are less than five extractions of velocity and velocity dispersion possible along the (pseudo-)slit for either the \kd or LEGA-C data. We further exclude one LEGA-C observation for which the resolved kinematic extractions are contaminated through a secondary object in the slit. 
The final sample includes 16 galaxies, resulting in 20 pairs of observations, including four duplicate observations from LEGA-C with the correspondingly different slit- and PSF-matched extractions from the \kd data cubes.

\subsection{Dynamical modelling}\label{s:modelling}

For our comparison of dynamical mass measurements, we use the data at their native spatial and spectral resolutions, without matching observing conditions between the \kd and LEGA-C surveys. For \kd we build mass models which we fit to the H$\alpha$ major axis kinematics (see Section~\ref{s:mdynk3d}), and for LEGA-C we use published dynamical masses from JAM models (see Section~\ref{s:mdynlgc}) and those computed from integrated stellar velocity dispersions. 
Due to the varying data quality across the sample, robust dynamical models cannot be constructed for all galaxies. Our dynamical mass comparison includes ten galaxies, four of which have two estimates based on integrated stellar velocity dispersion from LEGA-C due to duplicate observations, and six have LEGA-C estimates based on both integrated stellar velocity dispersion and JAM models (see Section~\ref{s:mdynlgc}).

\subsubsection{Modelling for \kd}\label{s:mdynk3d}

For \kd we exploit the 3D information available from the IFS data cubes to build 3D mass models to determine dynamical masses. Specifically, we place a pseudo-slit of width equal to the near-IR PSF FWHM on the continuum-subtracted cube along the kinematic major axis, which is well defined from the 2D projected velocity fields (see Appendix~\ref{a:maps}). From the 2D PV diagrams we then extract 1D profiles of velocity and velocity dispersion by summing rows spanning the PSF FWHM (or half PSF FWHM), and by fitting a Gaussian to the H$\alpha$ line position (see Section~\ref{s:kmoskin}). 

We forward-model the H$\alpha$ major axis kinematics using {\sc{dysmal}} \citep{Cresci09, Davies11, WuytsS16, Uebler18, Price21}, a code that allows for a flexible number of mass components, accounts for finite scale heights and flattened spheroidal potentials \citep{Noordermeer08}, includes effects of pressure support from the turbulent interstellar medium \citep{Burkert10, WuytsS16}, and consistently incorporates the observation-specific PSFs and LSFs.\footnote{
Other 3D forward-modelling tools enabling similar or different functionality (e.g., parametric {\it vs.\ }non-parametric modelling) are, for instance, {\sc{TiRiFiC}} \citep{Jozsa07}, {\sc{KinMS}} \citep{Davis13}, {\sc{GalPaK}}$^{\rm 3D}$ \citep{Bouche15}, or {\sc{3D-Barolo}} \citep{DiTeodoro15}.
} 
Specifically, the mass model created with {\sc{dysmal}} is rotated to match the orientation of the galaxy, projected into observed frame, and convolved with the LSF and PSF measured from the observations. We then extract model 2D PV diagrams and 1D profiles using the same extraction apertures and Gaussian fitting methodology as described above for the data. A detailed description of the functionality of {\sc{dysmal}} is provided by \cite{Price21} in their appendices A.1 and A.2, to which we refer the reader for more details and a mathematical description of the model creation and fitting procedure. 
We emphasise that, by applying the same extraction methodology to the data and the model cube, both data and model are affected in the same way through projection effects.

Due to the heterogeneous data quality in our sample, we consider two basic mass models for the baryonic component: a single Sérsic profile and a bulge-to-disc decomposition. Assuming mass follows light, we fix the structural parameters, specifically $i_{\rm F160W}$, $R_{e,\rm F160W}$ (or $R_{e,\rm F160W,bulge}$ and $R_{e,\rm F160W,disc}$), and $n_{S,\rm F160W}$ (or $n_{S,\rm F160W,bulge}$ and $n_{S,\rm F160W,disc}$), to measurements from {\sc{galfit}} models to the CANDELS F160W imaging as presented by \cite{vdWel12, Lang14, WuytsS16} (see  Section~\ref{s:k3d}). Here, we infer the galaxy inclination $i_{\rm F160W}$ from $q_{\rm F160W}=b/a$ by assuming an intrinsic ratio of scale height to scale length of $q_0=0.2$ \citep[see][]{vdWel14b, WuytsS16, Straatman22}. If including a bulge, we assume an axis ratio of 1 for this component.
We estimate the total baryonic mass $M_{\rm bar}$ by adding the stellar mass $M_{\star}$ from SED modeling and the gas mass $M_{\rm gas}$ based on $M_{\star}$, SFR, and redshift of each galaxy, by utilising the gas mass scaling relations by \cite{Tacconi20}. This estimate is used to centre a Gaussian prior with standard deviation 0.2~dex on the logarithmic total baryonic mass.
The intrinsic velocity dispersion $\sigma_0$ is assumed to be isotropic and constant throughout the disc, supported by deep adaptive optics assisted observations of SFGs at this redshift \citep[see][]{Genzel06, Genzel08, Genzel11, Genzel17, Cresci09, FS18, Uebler19, Liu23}. The value of $\sigma_0$ is a free parameter in our modelling.

All our dynamical models include an NFW \citep{NFW96} dark matter halo. Its total mass $M_{\rm halo}$ is inferred from the dark matter mass fraction within the effective (disc) radius, $f_{\rm DM}(<R_{e,\rm F160W})$, which is a free parameter in our modelling \citep[see appendix A.3 by][for a discussion of prior choices in MCMC fitting, specifically addressing $f_{\rm DM}(<r)$ {\it vs.\ }$M_{\rm halo}$]{Price21}. However we fix the halo concentration parameter $c$ following the fitting functions derived by \cite{Dutton14}, by first assuming a typical dark matter halo mass based on the stellar mass and redshift of each galaxy following \cite{Moster18}. Typical values for our sample are $c\sim6-7$.

In total, we have three free parameters in our basic mass models: the total baryonic mass $M_{\rm bar}$, the intrinsic velocity dispersion $\sigma_0$, and the central dark matter fraction $f_{\rm DM}$. The focus of our study is the enclosed dynamical mass, which we take to be the sum of the best-fit baryonic and dark matter masses as a function of radius: $M_{\rm dyn}(<r)=M_{\rm bar}(<r)+M_{\rm DM}(<r)$. $M_{\rm dyn}(<r)$ is calculated within spherical apertures from the non-spherically symmetric potential of the 3D mass distribution \citep[see][]{Price22}.
We emphasize that the enclosed dynamical mass is relatively insensitive to the detailed partitioning of baryonic and dark matter mass \citep[see also][]{WuytsS16, Price21}.

We fit using Markov Chain Monte Carlo (MCMC) techniques as implemented in {\sc{dysmal}} through the {\sc{emcee}} package \citep{ForemanMackey13}. We use 300 walkers, a burn-in phase of 100 steps, and a run phase of 200 steps (greater than ten times the maximum auto-correlation time of the individual parameters). We adopt the maximum a posteriori values of the fit parameters as our best-fit values, based on a joint analysis of the posteriors for all free parameters \citep[see][]{Price21}. For some galaxies we can only constrain an upper limit on $\sigma_0$ through the upper $2\sigma$ boundary of the marginalised posterior distribution (see Section~\ref{s:corr_kin}).
We show the MCMC posterior distributions for the modelled \kd galaxies in Appendix~\ref{a:corner}.

For two galaxies in our sample with very high H$\alpha$ data quality, COS4\_17628 and COS4\_06487, more sophisticated dynamical models with a higher number of free parameters exist, in particular including fitting of structural parameters \citep{Nestor22}. The dynamical masses at the radii considered in this analysis agree among the various mass models within 0.06~dex.

\subsubsection{Modelling for LEGA-C}\label{s:mdynlgc}

For the LEGA-C dynamical mass estimates, we use the recently published values by \cite{vdWel21} and \cite{vHoudt21} based on the LEGA-C stellar kinematics. We refer the reader to those papers for details, but summarize the most relevant information here.

\cite{vHoudt21} construct axisymmetric Jeans anisotropic models to the LOS stellar rms velocity $v_{\rm rms,\star}$, with $v_{\rm rms,\star}^2(r)=v^2_\star(r)+\sigma^2_\star(r)$, for LEGA-C galaxies for which the morphological position angle is not misaligned with the slit by more than $45^{\circ}$, that are not mergers or irregular galaxies, that are not mid-IR or X-ray-identified AGN\footnote{
Such galaxies are identified by \texttt{FLAG\_SPEC}=1 in the LEGA-C data release \citep[see][]{vdWel21}. Narrow-line or radio-identified AGN are not flagged.}, do not have flux calibration issues, and that have S/N~$>10$ in at least three spatial resolution elements. 
The JAM modelling accounts for beam smearing through comparison of the wavelength-collapsed LEGA-C spectrum with a model light profile, obtained by convolving the F814W imaging with a Moffat kernel and the slit profile. The kinematic position angle is assumed to coincide with the morphological position angle, and any misalignment between the position angle and the slit orientation is taken into account. 
The fitting is performed to the LOS stellar rms velocity $v_{\rm rms,\star}$. In case of asymmetric data points at projected distance $r$, uncertainties on $v_{\rm rms,\star}$ are set to the maximum of the formally measured uncertainties and half the difference between the values at $+r$ and $-r$ \citep[see][]{vHoudt21}.

The JAM models consist of two mass components, a stellar component assuming mass follows light based on F814W imaging, and an NFW dark matter halo. The halo concentration $c$ is tied to the halo mass following \cite{Dutton14}. Free fit parameters are the stellar velocity anisotropy, the stellar mass-to-light ratio $M/L$, the dark matter halo mass as parameterized by the circular velocity, the galaxy inclination, and the slit centering.
Here, the inclination is constrained by a $q-$dependent prior assuming an intrinsic thickness distribution ${\mathcal{N}}(0.41,0.18)$ which is constrained from the full primary LEGA-C sample. \cite{vHoudt21} note that the inclination and the slit centering are typically unconstrained by the data.

Based on the JAM results, dynamical masses are provided out to 20~kpc and/or $2R_{e,\rm F814W}$, if supported by the data, where $R_{e,\rm F814W}$ is the semimajor axis effective radius determined from single-Sérsic {\sc{galfit}} models to the F814W imaging as presented by \cite{vdWel16, vdWel21}. 
Analogous to the \kd modelling, the enclosed dynamical mass $M_{\rm dyn}(<r)$ is calculated within spherical apertures from the non-spherically symmetric potential of the 3D mass distribution.

As described above, Jeans anisotropic models can be built only for a subset of the LEGA-C survey. However, the existing models are used to calibrate more accessible virial mass estimators based on the integrated stellar velocity dispersion \citep[e.g.][]{vdWel06, Hyde09, Taylor10, Cappellari13, Belli17a, Mendel20}. The details of this calibration are described by \cite{vdWel22}. In short, virial masses are computed as
$$M_{\rm vir}=K(n_{S,\rm F814W})\frac{\sigma^2_{\star,\rm vir}R_{e,\rm F814W}}{G},$$
where $n_{S,\rm F814W}$ and $R_{e,\rm F814W}$ are derived from F814W imaging (see Section~\ref{s:lgc}) and $K(n_{S})=8.87-0.831n_{S}+0.0241n_{S}^2$ following \cite{Cappellari06}, $\sigma_{\star,\rm vir}$ is the inclination- and aperture-corrected, integrated stellar velocity dispersion (measured from collapsed 1D spectra), and $G$ is the gravitational constant. 
The correction for $\sigma_{\star,\rm vir}$ is derived by calibration to the JAM dynamical masses, setting $M_{\rm vir}$ as twice the JAM mass within $r=R_{e,\rm F814W}$. Therefore, the virial mass in the above equation is not the enclosed mass within $1~R_e$, and does not correspond to the enclosed mass within a specific radius. This mass should not be confused with the concept of virial mass frequently used in cosmology, that is the mass within a radius that encompasses a fixed overdensity with respect to the critical density of the Universe.

The main focus of our dynamical mass comparison is between the \kd IFS H$\alpha$ models and the LEGA-C JAM and $M_{\rm vir}$ measurements based on stellar kinematics. Two galaxies in our dynamical mass sample are also part of the emission line modelling analysis by \cite{Straatman22}, and we include their results where appropriate. \cite{Straatman22} build a kinematic model where the rotation curve is parametrized by an arctan function, assuming the ionised emission originates from a thick, exponential distribution constrained from the F814W imaging, with a constant and isotropic intrinsic velocity dispersion. The model accounts for beam smearing and misalignment between the slit and PA$_{\rm F814W}$, and the dynamical mass is calculated from the model rotation curve including a pressure support correction. See \cite{Straatman22} for further details.

\subsubsection{Notable differences between the dynamical models}\label{s:model_diff}

Our H$\alpha$ dynamical mass models use structural parameters from F160W imaging, while the stellar dynamical mass measurements use structural parameters from F814W imaging (see Sections~\ref{s:k3d}, \ref{s:lgc}, and Appendix~\ref{a:structure} for further discussion).
To quantify the impact of different structural measurements for our sample, we repeat the dynamical modelling for the \kd galaxies, this time utilizing the $i-$band (F814W) based values for a single-Sérsic baryonic component. We construct two additional sets of dynamical models adopting $R_e$, $n_S$, and $q$ from F814W imaging. 
For the first set we keep the inclination we inferred for our fiducial \kd models. For the second set we re-calculate the inclination based on the observed axis ratio $q_{\rm F814W}$ and using $q_0=0.41$, the prior that is used for the LEGA-C JAM modelling \citep{vHoudt21}.
Overall, the impact on our dynamical mass estimates is minor, and the corresponding results are presented in Appendix~\ref{a:f814w_modelling}: for the first set of alternative models we find an average increase in $M_{\rm dyn}$ of 0.02~dex (standard deviation 0.26~dex); for the second set of alternative models we find an average decrease in $M_{\rm dyn}$ of 0.03~dex (standard deviation 0.25~dex).

Another difference lies in the explicit assumption of mass components. As described in Section~\ref{s:mdynk3d}, for the modelling of the \kd data we estimate total baryonic mass by including a cold gas component derived from the scaling relations by \cite{Tacconi20}. For our sample, such derived gas-to-baryonic-mass fractions are between 2 and 70 per cent, with a mean value of $f_{\rm gas}=0.27$. The LEGA-C JAM modelling assumes only a stellar and a dark matter component (see Section~\ref{s:mdynlgc}). However, although they assume mass follows light, $M/L$ is a free parameter in their fit. The combination of a free $M/L$ and an explicit dark matter halo component therefore allows for an (unconstrained) contribution from gas (following stars) as well. 
The more simplistic $M_{\rm vir}$ calculation does not make any assumptions on the involved mass components, however uses the (corrected) integrated 1D stellar velocity dispersion as a tracer of dynamical mass. As the movement of stars is dictated by the full potential, this includes any contribution from all stars, gas, and dark matter.

\section{Stellar and ionised gas kinematics}\label{s:compasobs}

\begin{figure*}
	\centering
	\includegraphics[width=0.245\textwidth]{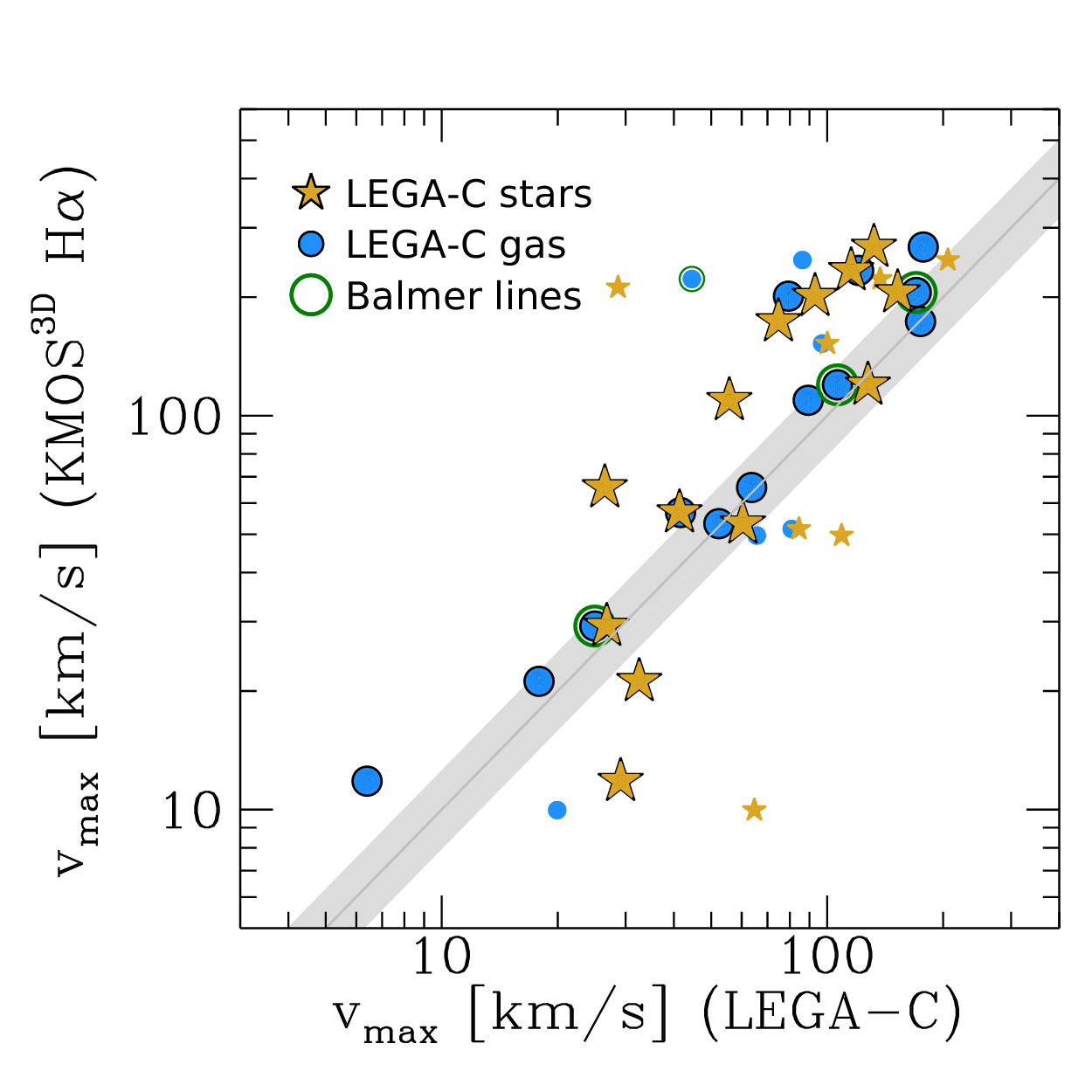}
	\includegraphics[width=0.245\textwidth]{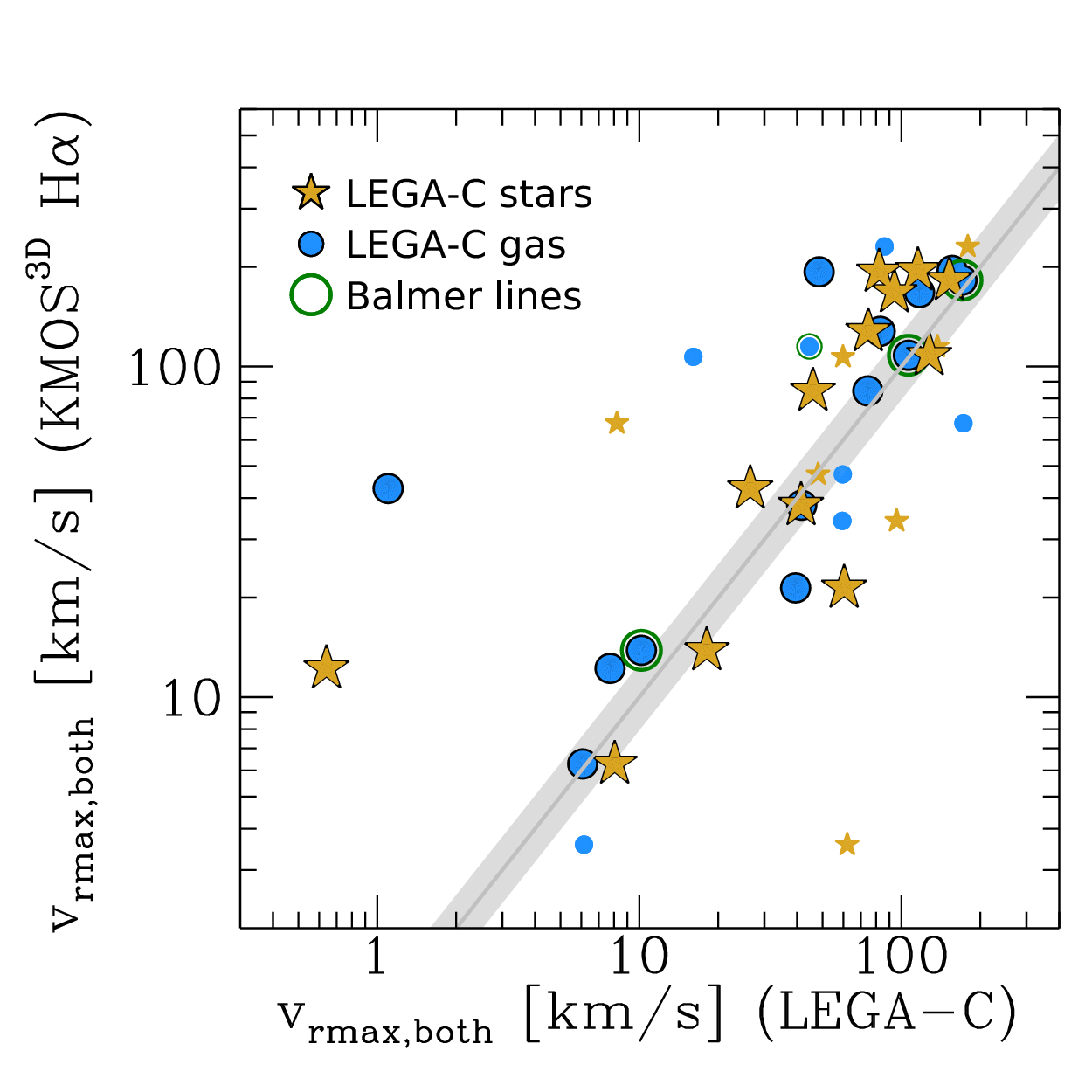}
	\includegraphics[width=0.245\textwidth]{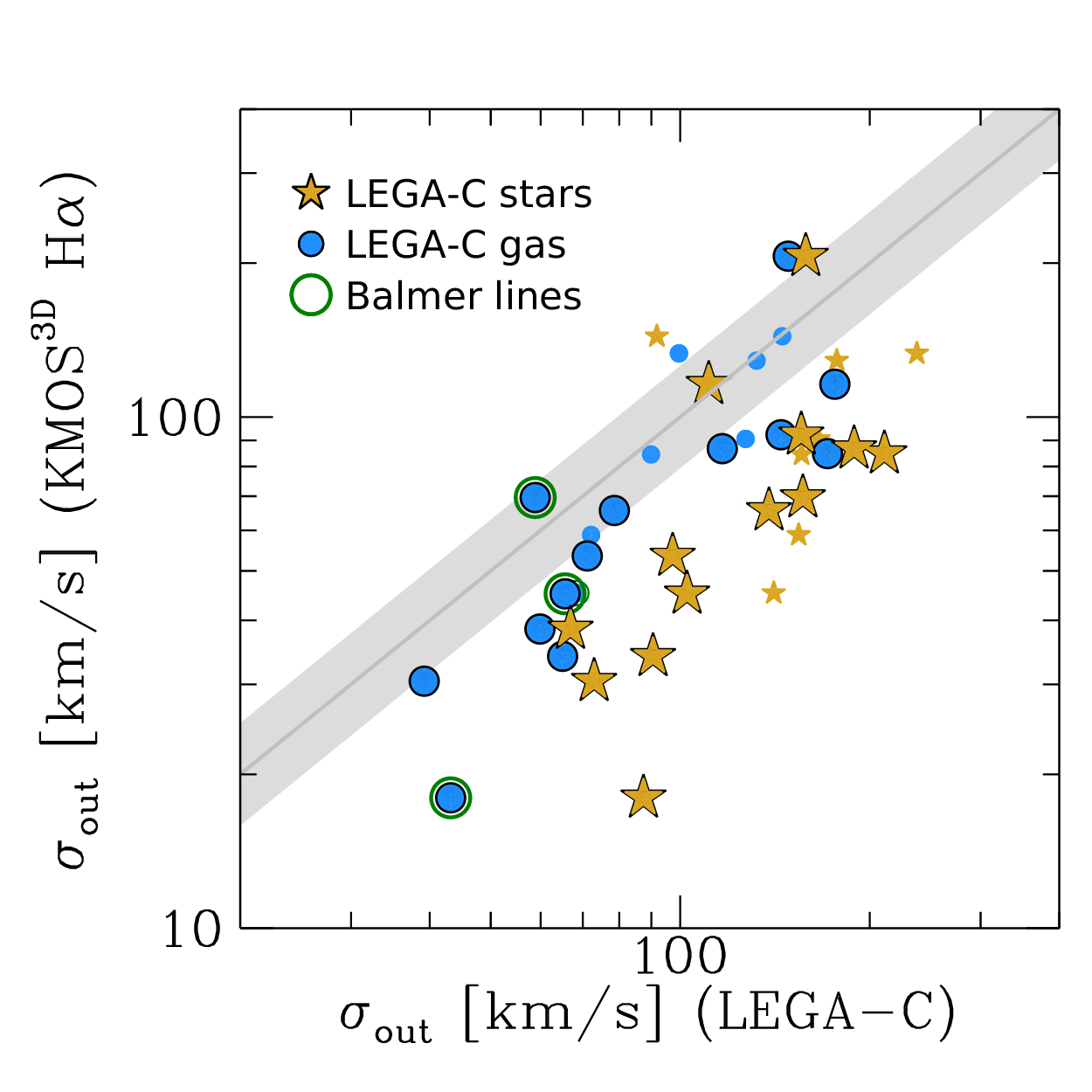}
	\includegraphics[width=0.245\textwidth]{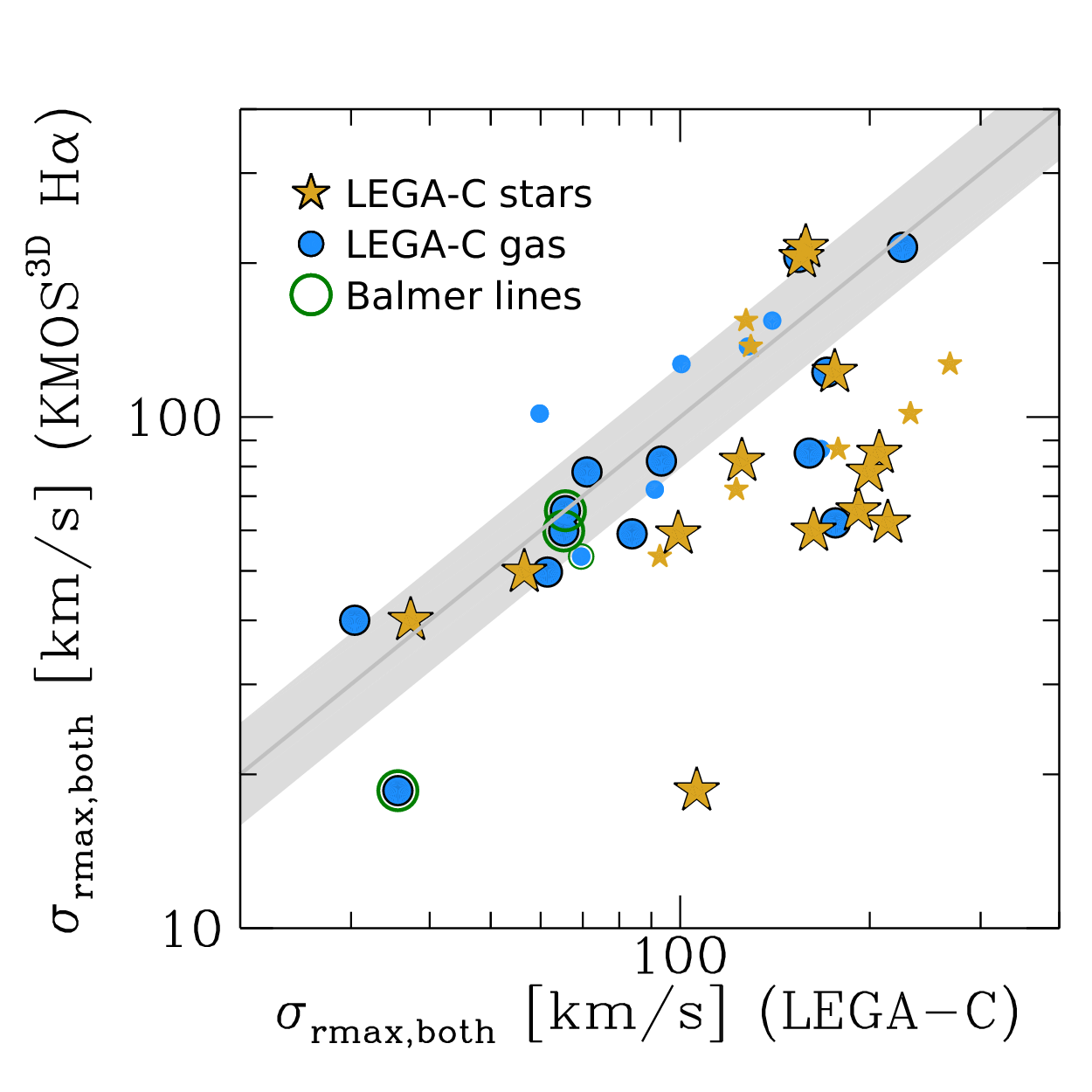}
	\caption{Comparison of LOS kinematic quantities from PSF-matched fixed-slit extractions, as defined in Section~\ref{s:measure}. From left to right: maximum observed absolute rotation velocity, $v_{\rm max}(r_{\rm vmax})$; (average, interpolated) rotation velocity at the outermost radius covered by both \kd and LEGA-C data, $v_{\rm rmax,both}(r_{\rm max,both})$; weighted mean outer velocity dispersion, $\sigma_{\rm out}$; (average, interpolated) velocity dispersion at the outermost radius covered by both \kd and LEGA-C data, $\sigma_{\rm rmax,both}(r_{\rm max,both})$. 
	Golden filled stars compare \kd H$\alpha$ measurements with LEGA-C stellar measurements, and blue filled circles with LEGA-C gas measurements. Green open circles indicate the presence of prominent Balmer lines in the LEGA-C spectra. Larger symbols indicate galaxies for which a dynamical modelling of both the \kd and LEGA-C data is possible (typically higher-quality, more extended data, excluding mergers). 
    The shaded region around the 1:1 line indicates a constant interval of $\pm0.1$~dex in all plots, to highlight differences in scatter between the comparisons.
    Due to duplicate observations in LEGA-C, galaxies can appear multiple times in each panel.
	On average, velocities are larger and velocity dispersions lower for H$\alpha$ compared to stars.
    }
	\label{f:measures1}
\end{figure*}

\begin{figure*}
	\centering
	\includegraphics[width=0.245\textwidth]{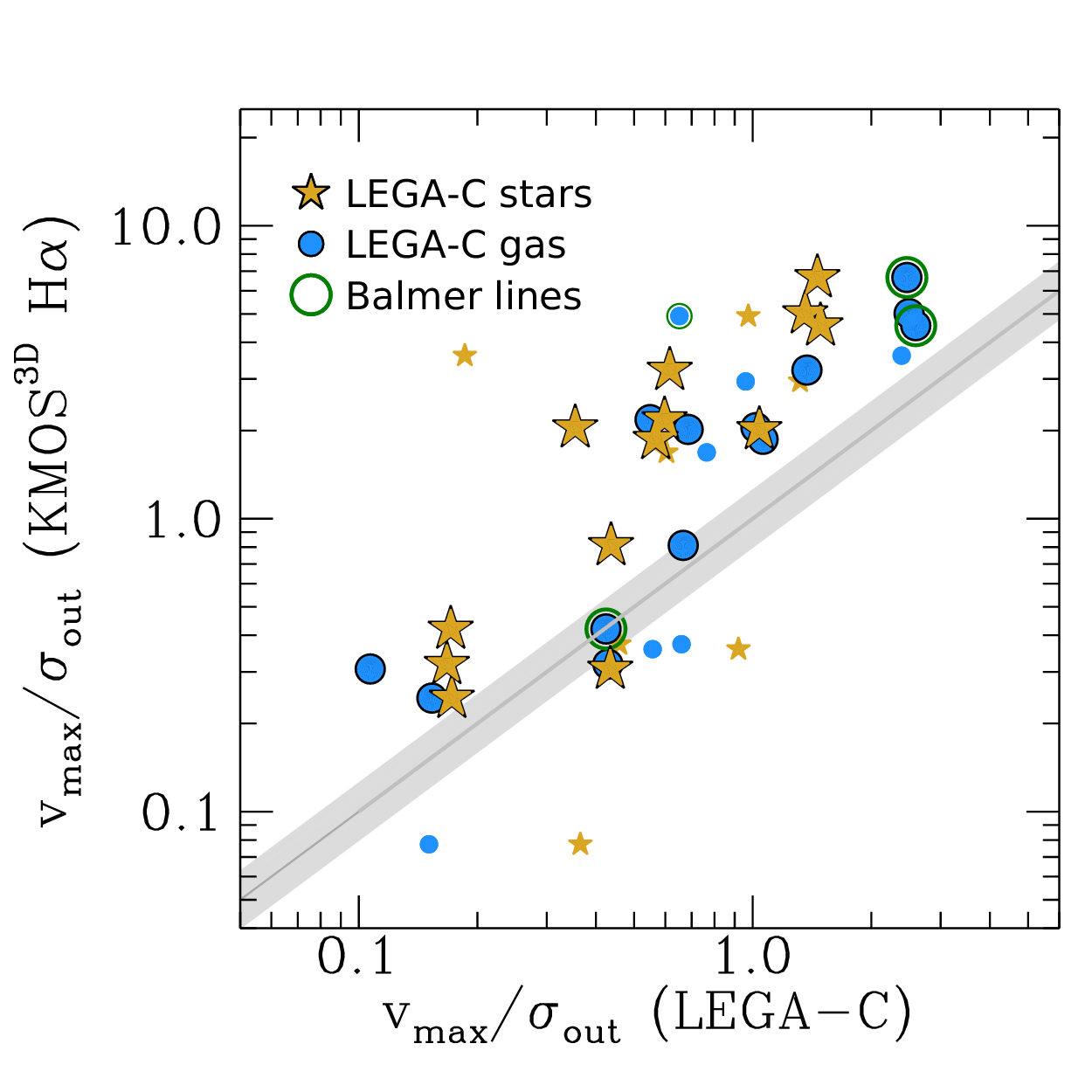}
	\includegraphics[width=0.245\textwidth]{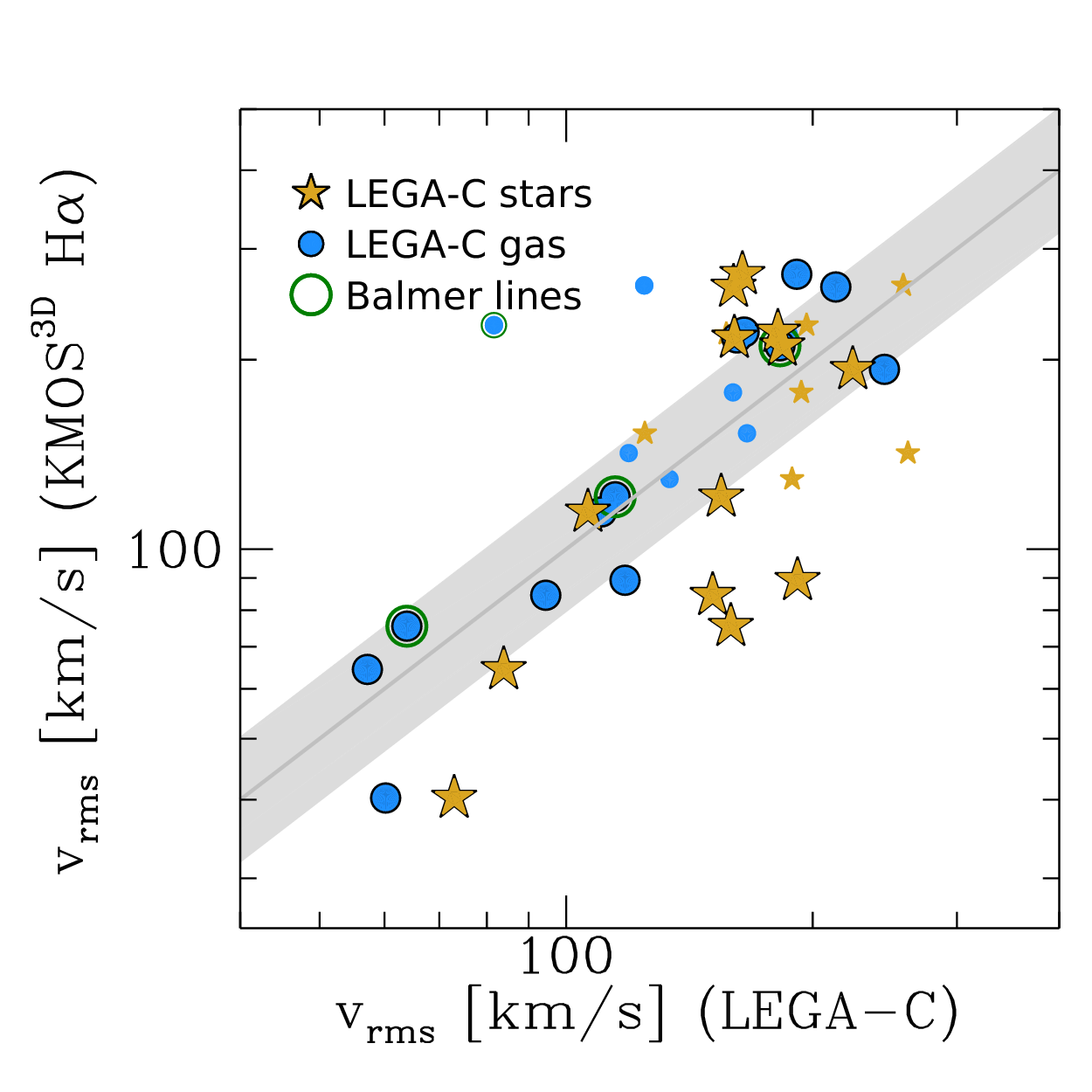}
	\includegraphics[width=0.245\textwidth]{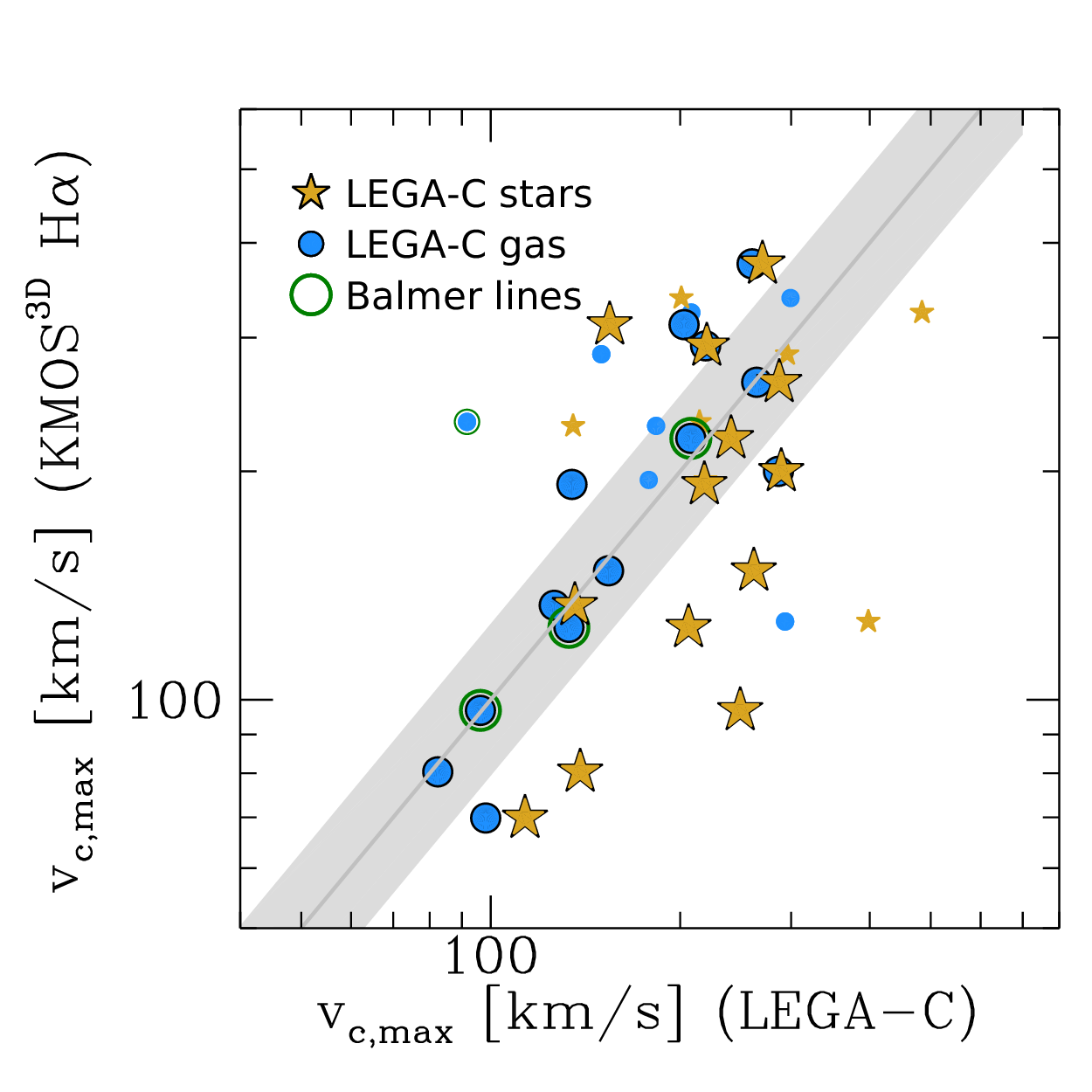}
	\caption{Comparison of LOS kinematic quantities from PSF-matched fixed-slit extractions, defined as in Section~\ref{s:measure}. From left to right: rotational support, $v_{\rm max}/\sigma_{\rm out}$; root mean square velocity, $v_{\rm rms}$; approximation of the circular velocity, $v_{c,\rm max}(r_{\rm vmax})$.
	Symbols are as in Figure~\ref{f:measures1}. 
    The shaded region around the 1:1 line indicates a constant interval of $\pm0.1$~dex in all plots, to highlight differences in scatter between the comparisons. 
	On average, the rotational support measured from H$\alpha$ emission is larger than in stellar and gas measurements from the LEGA-C spectra.
    }
	\label{f:measures2}
\end{figure*}

We now compare the stellar (from LEGA-C) and ionised gas (from KMOS$^{\rm 3D}$) LOS kinematics from matched observing setups (Section~\ref{s:kinex}), that is from fixed slits after matching the individual PSFs for each pair of observations. We also compare different measurements of the ionised gas observed kinematics using the H$\alpha$ line from the \kd observations, and the emission line fits to the full LEGA-C spectrum, typically dominated by [OII] emission.

In Figure~\ref{f:obsprof} we show an example of 2D and 1D kinematic extractions from \kd and LEGA-C data along both a N-S and an E-W (pseudo-)slit of width $1\arcsec$ for one galaxy. For this galaxy, both the ionised gas and stellar velocities from LEGA-C and the H$\alpha$ velocities from KMOS$^{\rm 3D}$ qualitatively agree, with stellar velocities reaching somewhat lower amplitudes. The velocity dispersion profiles are often dissimilar, with asymmetric profiles for the ionised gas and stars from LEGA-C compared to the \kd profile that is centrally peaked as expected for rotating disc kinematics uncorrected for beam-smearing.

It has been shown that [OII] emission does not only trace star formation, but can be related to AGN activity and low-ionisation nuclear emission line regions \citep[LINERS; e.g.][]{Yan06, Yan18, Lemaux10, DaviesRL14, Maseda21}. The differences in shape and intensity between the H$\alpha$ and [OII] emission we see in the 2D position-velocity diagrams and the extracted 1D profiles further suggests that not all [OII] emission is originating from the co-rotating ISM. However, also line blending of the doublet emission complicates the extraction of kinematic information (due to degeneracies between the line amplitudes, widths, and centroids, when no other prominent emission lines are present, as is typical for LEGA-C galaxies at $z>0.9$).

\subsection{Velocities}\label{s:comp_v}
In the left panels of Figure~\ref{f:measures1} we compare maximum velocities ($v_{\rm max}$) and velocities at the outermost common radius ($v_{\rm rmax,both}$) for the \kd and LEGA-C samples, measured from the `observed' kinematics. Corresponding numbers for galaxies for which also a dynamical modelling is possible are listed in Table~\ref{t:statobs}.
On average, the velocities measured from the \kd H$\alpha$ data are larger compared to LEGA-C stars (golden stars in Figure~\ref{f:measures1}) by $\sim40$ per cent. This is similar when comparing to LEGA-C gas (blue circles), but here we also note that the gas velocity measurements agree well for the three non-interacting galaxies where the LEGA-C spectrum includes strong Balmer lines (large symbols with green circles).
Based on a two-sample Kolmogorov-Smirnov statistic, only the maximum velocity of stars is different from the H$\alpha$ $v_{\rm max}$ by more than $1\sigma$.
In general, lower amplitudes in rotation velocity for stars compared to gas are expected based on $z=0$ data (see Section~\ref{s:intro}).

\subsection{Velocity dispersions}\label{s:comp_s}
In the right panels of Figure~\ref{f:measures1} we show the corresponding plots for the outer weighted mean observed velocity dispersion ($\sigma_{\rm out}$), and the (mean) velocity dispersion at the outermost radius common to both data sets ($\sigma_{\rm rmax,both}$). On average, the H$\alpha$ dispersion measurements are lower than the LEGA-C measurements. 
In particular the stellar velocity dispersions are larger by about a factor of two relative to H$\alpha$, and show a significantly different distribution towards higher values by more than $2\sigma$ based on a two-sample Kolmogorov-Smirnov statistic (see also Table~\ref{t:statobs}). 
In addition, there is some indication that, when measured at the same radius, the difference between stellar and H$\alpha$ velocity dispersions is higher for systems with higher stellar velocity dispersion.
In general, higher disc velocity dispersions for stars compared to gas are also expected based on $z=0$ data, as discussed in Section~\ref{s:intro}.

\begin{table}
    \centering
    \caption{Mean difference of the logarithm, log(KMOS$^{\rm 3D}$/LEGA-C), and corresponding standard deviation for various kinematic quantities, comparing \kd H$\alpha$ to LEGA-C stars and gas, respectively, averaged over galaxies for which a dynamical modelling is possible. The quantities are defined in Section~\ref{s:measure} and individual measurements are shown in Figures~\ref{f:measures1} and \ref{f:measures2} (large symbols).}
    \label{t:statobs}
\begin{tabular}{lcccc}
    & \multicolumn{2}{c}{stars} & \multicolumn{2}{c}{gas} \\
    Quantity [dex] & mean & std. dev. & mean  & std. dev. \\
\hline
    $v_{\rm max}$            & 0.13 & 0.24 & 0.13 & 0.12 \\
    $v_{\rm rmax,both}$      & 0.16 & 0.40 & 0.21 & 0.46 \\
    $\sigma_{\rm out}$       & -0.30 & 0.20 & -0.15 & 0.14 \\
    $\sigma_{\rm rmax,both}$ & -0.26 & 0.27 & -0.09 & 0.17 \\
    $v_{\rm max}/\sigma_{\rm out}$  & 0.42 & 0.25 & 0.28 & 0.20 \\
    $v_{\rm rms}$            & -0.05 & 0.20 & 0.02 & 0.10 \\
    $v_{\rm c,max}$          & -0.08 & 0.19 & 0.02 & 0.11 \\
 \hline
\end{tabular}
\end{table}

\subsection{Rotational support}\label{s:comp_vs}

In the left panel of Figure~\ref{f:measures2} we plot the ratio of maximum LOS velocity and outer LOS velocity dispersion ($v_{\rm max}/\sigma_{\rm out}$), as defined in Section~\ref{s:measure}. We stress again that these values are not derived from modelling, but have been measure from fixed slits after PSF matching. I.e., the LSF is accounted for but not any inclination effects or PSF effects, although the latter should be effectively the same for our \kd and LEGA-C extractions, as discussed in Section~\ref{s:kmoskin}. 
Overall the \kd measurements suggest a stronger rotational support in the star-forming ionised gas phase, possibly indicating that the H$\alpha$ line emission is originating from a more disc-like structure, or that it is less affected by non-circular motions (e.g.\ compared to [OII]).
The difference between the H$\alpha$ measurements and the stellar measurements is statistically significant by more than $2\sigma$, and the H$\alpha$ measurements and the LEGA-C gas measurements by more than $1\sigma$.

In addition, we compare several combinations of observed velocity and observed velocity dispersion in the right-hand panels of Figure~\ref{f:measures2}, as described in Section~\ref{s:measure}. The combination of velocity and velocity dispersion into a common probe of the galactic potential results in more similar estimates on average between the \kd and LEGA-C data: both the average offsets and the scatter are reduced (see also Table~\ref{t:statobs}).
We find the best average agreement between stellar and H$\alpha$ data for $v_{\rm rms}$ (second panel in Figure~\ref{f:measures2}). 
We note that \cite{Bezanson18b} find comparable {\it integrated} velocity dispersion for ionised gas and stars within the LEGA-C survey, in qualitative agreement with our result.

\subsection{Implications for compilations of ionised gas kinematics}\label{s:compilations}

The differences in ionised gas velocities and velocity dispersions between the \kd and LEGA-C extractions, where the latter are mostly dominated by [OII] emission, serve as a caution in the combination of samples with different emission lines. The somewhat lower velocities and higher velocity dispersions measured from [OII] compared to H$\alpha$ or other Balmer lines might motivate a revision of literature compilations for the study of galaxy gas kinematics evolution, such as the Tully-Fisher relation \citep{Tully77}, or gas velocity dispersion. 

\begin{figure*}
	\centering
	\includegraphics[width=0.33\textwidth]{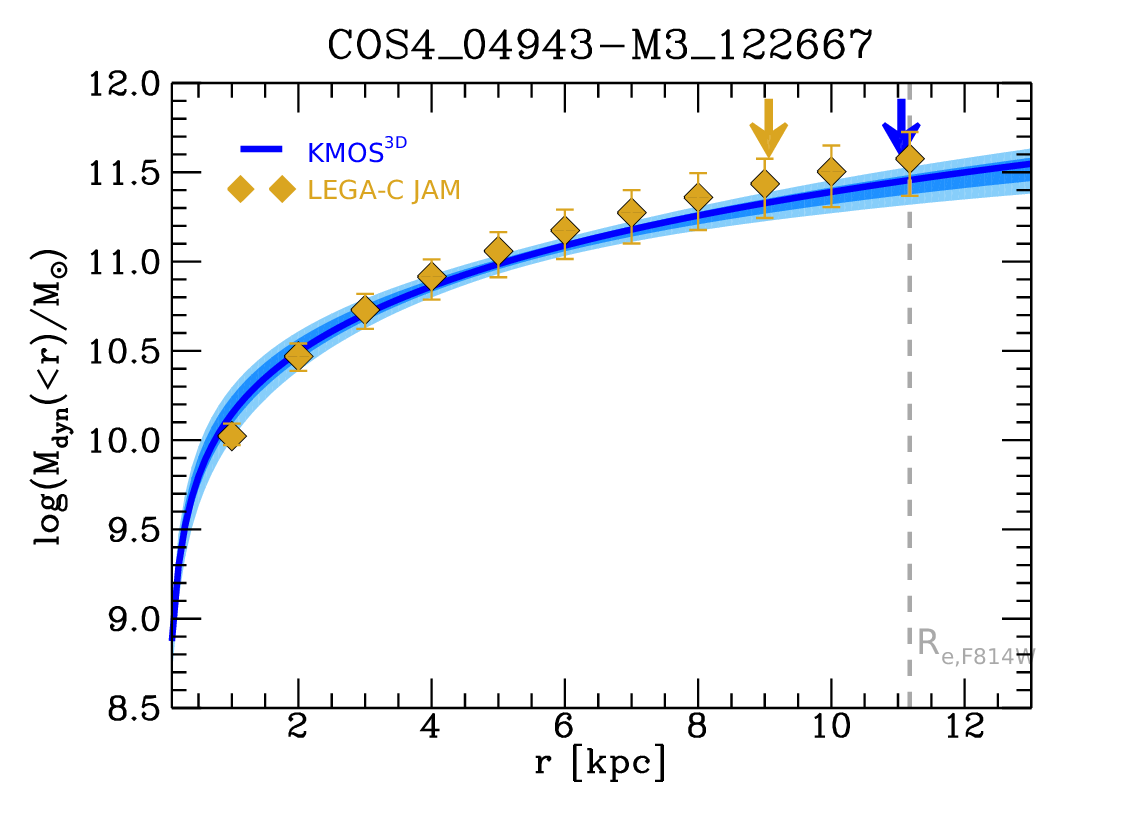}
	\includegraphics[width=0.33\textwidth]{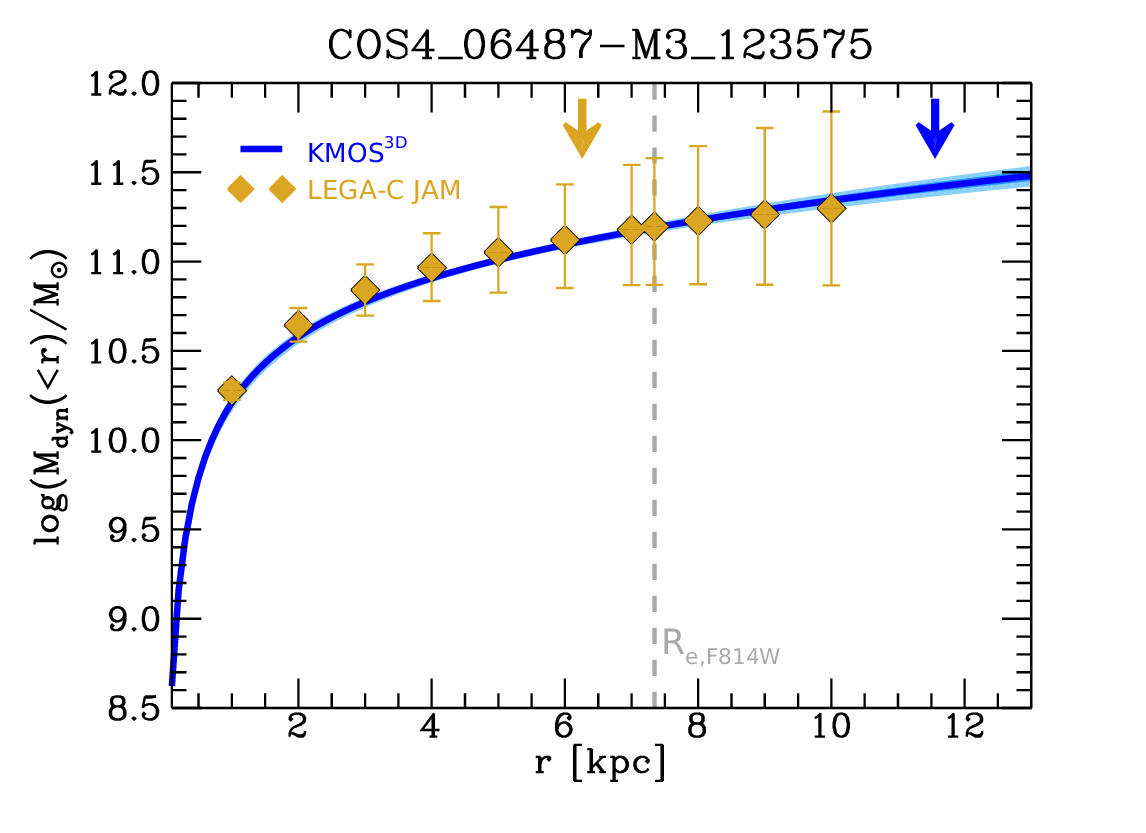}
	\includegraphics[width=0.33\textwidth]{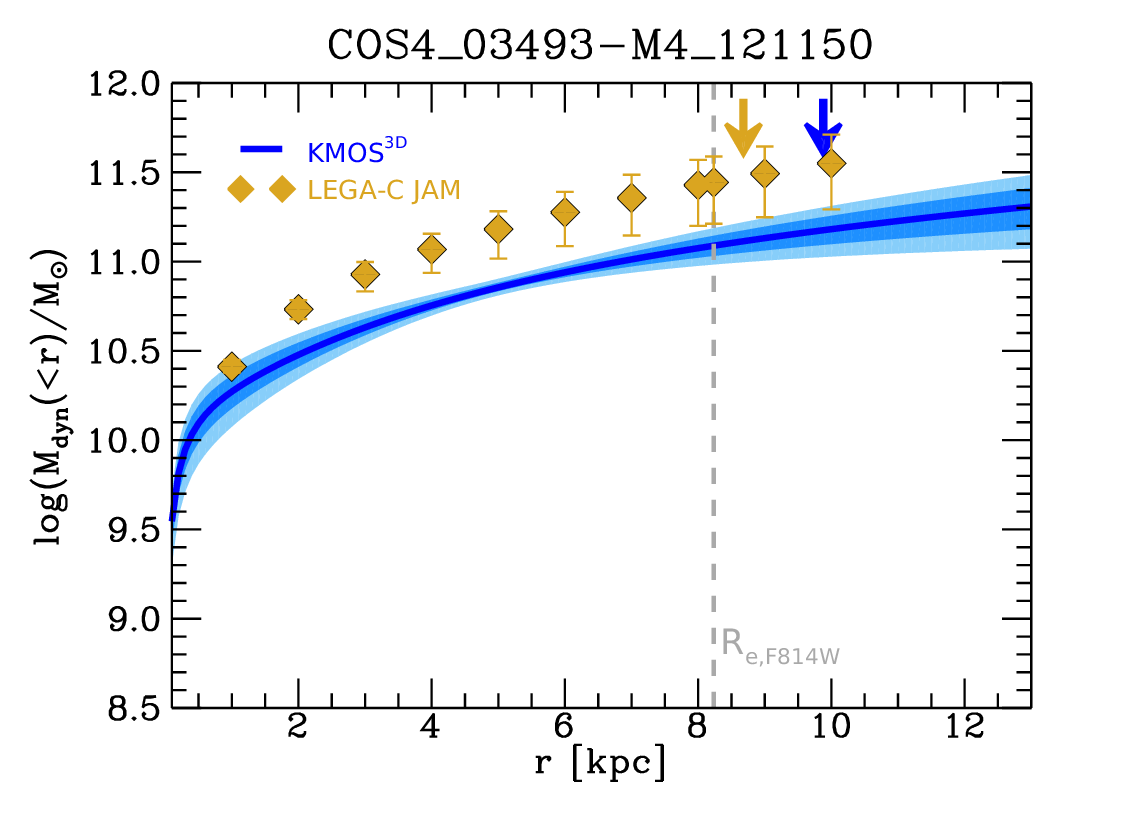}
	\includegraphics[width=0.33\textwidth]{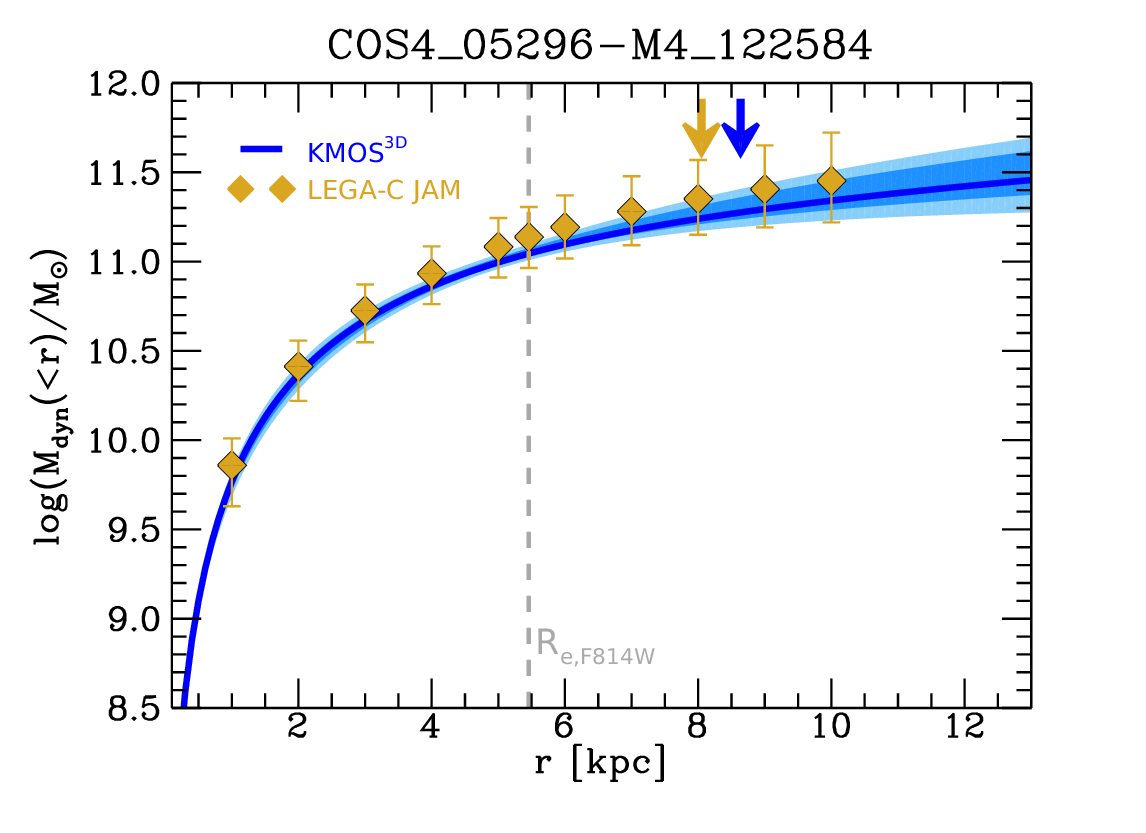}
	\includegraphics[width=0.33\textwidth]{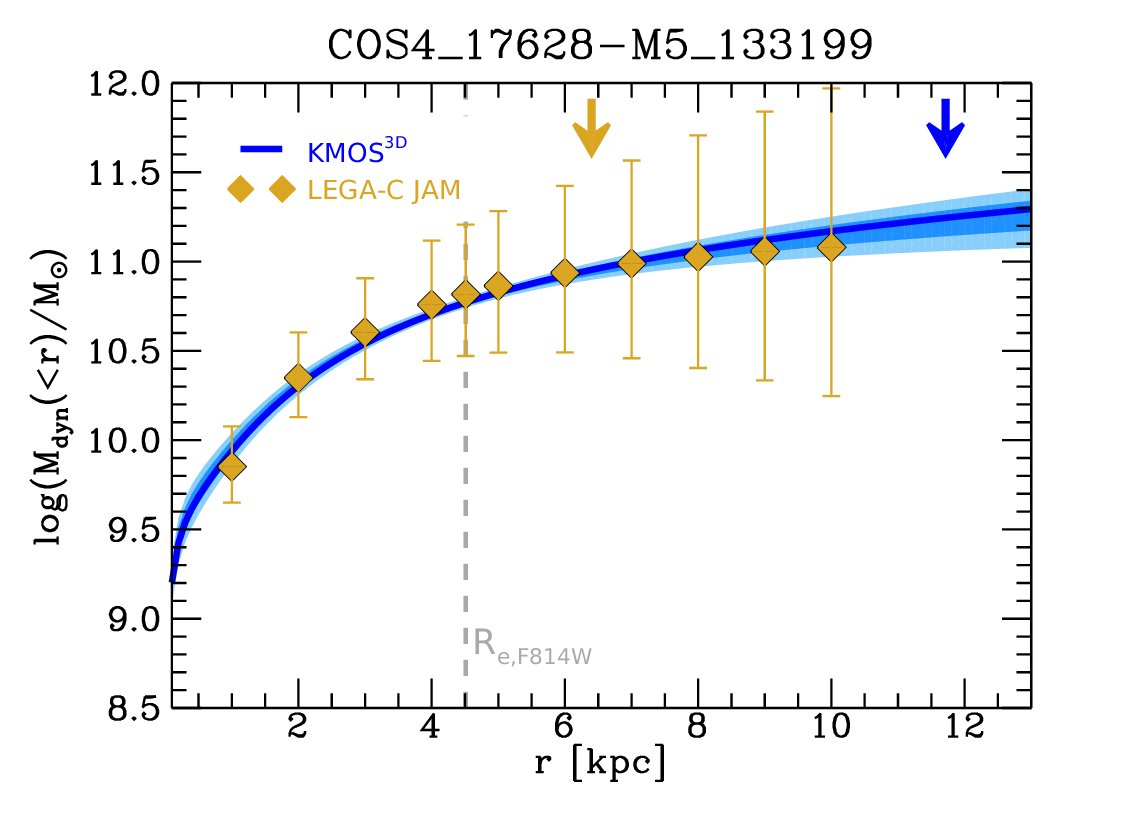}
	\includegraphics[width=0.33\textwidth]{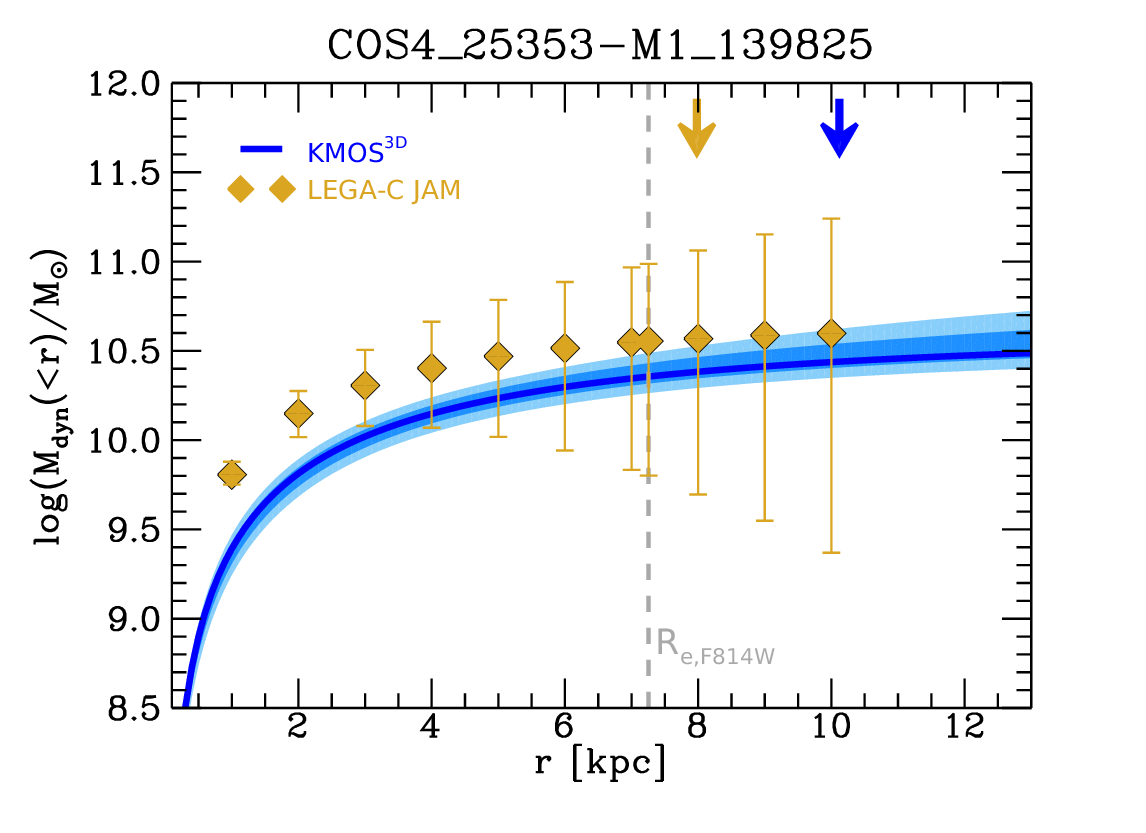}
	\caption{Comparison of cumulative mass profiles from H$\alpha$ dynamical mass models and JAM. The JAM estimates by \citet{vHoudt21} are shown as golden diamonds every kiloparsec out to 10~kpc, and at $R_{e,\rm F814W}$, and the KMOS$^{\rm 3D}$ best-fit enclosed total mass is shown as a blue line, with lighter shading indicating one and two standard deviations as constrained by the full MCMC chains (60000 realisations). The vertical dashed grey line marks $R_{e,\rm F814W}$, and the golden and blue arrows indicate the projected extent of the stellar and H$\alpha$ kinematic data, respectively. Overall, the agreement between the JAM model estimates and the KMOS$^{\rm 3D}$ model estimates is very good, demonstrating that the total mass distribution can be robustly inferred from different modelling techniques and data sets, as long as data quality allows. For galaxy COS4\_03493-M4\_121150 (top right), the JAM model overestimates the dynamical mass (see Section~\ref{s:mcum} for details.)}
	\label{f:mcum}
\end{figure*} 

In fact, the discrepancy in zero-point offset of the stellar mass Tully-Fisher relation at $z\sim1$ among \cite{Miller11, Miller12} and \cite{Tiley16, Uebler17} could be partly due to the use of different tracers.\footnote{
In addition, other important reasons for such differences have been identified \citep[see e.g.\ Appendix~A by][]{Uebler17}: inclusion of galaxies where the peak velocity is not reached by the data, lack of a correction for beam-smearing, or sample selection effects based on e.g.\, $v_{\rm rot}/\sigma_0$ cuts.}
The measurements by \cite{Miller11, Miller12} include or are based on [OII] emission, while \cite{Tiley16, Uebler17} target H$\alpha$. A zero-point difference of about $-0.3$~dex is found between those studies, with a corresponding offset in velocity of $\sim0.1$~dex. Our results suggest that for a velocity of 50~km/s (200~km/s), a systematic velocity offset of up to 0.2~dex (0.05~dex) could be solely due to the use of different gas tracers, potentially resolving the disagreement between those studies.

Considering the recent literature compilation by \cite{Uebler19} of the redshift evolution of intrinsic ionised gas velocity dispersion, our findings indicate that the difference in average disc velocity dispersion at fixed redshift found between some surveys could be due to the use of different emission line tracers. Any systematic difference between ionised gas velocity dispersion measured from H$\alpha$ {\it vs.\,} [OII] or also [OIII], which is known to typically have a higher excitation contribution from narrow-line AGN than H$\alpha$ \citep[e.g.][]{Kauffmann03, DaviesRL14}, could affect evolutionary trends.\footnote{
We note that studies of nearby giant HII regions typically find lower velocity dispersions for [OIII] compared to H$\alpha$, and this has been linked to [OIII] originating from denser regions more deeply embedded in HII regions \citep[e.g.][]{Hippelein86}. \cite{Law22} find a correlation of $\sigma_{\rm [OIII]}/\sigma_{\rm H\alpha}$ with SFR for galaxies in the MaNGA survey \citep{Bundy15}. In their work, velocity dispersions measured from [OIII] are higher relative to H$\alpha$ for SFR$\gtrsim1~M_{\odot}/yr$. Massive main-sequence galaxies at $z\sim1-2$ have typical SFRs of $10-100~M_{\odot}/yr$ \citep[e.g.][]{Whitaker14}. \cite{Law22} also find [OII] velocity dispersions to be systematically higher compared to measurements based on H$\alpha$.}
Indeed, several surveys including [OII] or [OIII] emission lines have average intrinsic velocity dispersion values above the relation derived by \cite{Uebler19}. However for the more challenging measurement of the velocity dispersion the situation is further complicated by different methodologies in accounting for beam-smearing in those studies.\\

In general, considering the full 1D profiles, we find that (i) stellar velocities reach lower amplitudes and average disc stellar velocity dispersions are higher compared to the ionised gas kinematics, reminiscent of local Universe findings; (ii) stellar velocity dispersions and ionised gas velocity dispersion dominated by [OII]$\lambda\lambda3726,3729$ emission are often more asymmetric compared to H$\alpha$; (iii) the correspondence between the \kd H$\alpha$ data and the LEGA-C emission line data is better for LEGA-C spectra including Balmer lines (green circles in Figures~\ref{f:measures1} and \ref{f:measures2}).

Overall, more high-quality data would be beneficial to characterise the differences in ionised gas kinematics provided by different tracers for the same galaxies. 
Upcoming data from {\it James Webb Space Telescope} ({\it JWST}) enabling H$\alpha$ studies up to $z\sim7$ and extensions of $z<3$ ground-based kinematic studies of multiple emission lines with IFUs such as ERIS, MUSE, and KMOS will provide important references.

\section{Dynamical Masses}\label{s:mdyn}

We now proceed with a comparison of dynamical mass measurements from the \kd H$\alpha$ data and the LEGA-C stellar kinematic data. 
In contrast to the previous section, where we have matched the observing conditions between \kd and LEGA-C data, we now use the native spatial and spectral resolution of the data to build the best possible dynamical models based on H$\alpha$ and stars. 

In our comparison of the dynamical modelling results we focus on the inferred enclosed mass close to the range covered by our data. This implies assumptions on the 3D mass distribution, as detailed in Section~\ref{s:modelling}. In general, the movement of gas and stars is governed by the gravitational potential, and one can also directly compare circular velocities, where $v_c^2(r)/r=-\nabla\Phi$ \citep[e.g.][]{Davis13, Leung18}. Other than the enclosed mass, $v_c(r)$ is influenced by the full mass distribution, including at radii beyond $r$. While the mass models for both the \kd and LEGA-C data do make assumptions on the mass distribution beyond radii covered by observations through the explicit modelling of a dark matter halo, the exact halo mass profile is not constrained by our data.

\subsection{Cumulative total mass profiles based on \texorpdfstring{H$\alpha$}{Halpha} and stars} \label{s:mcum}

We begin with a comparison of cumulative mass profiles from JAM models and our best-fit H$\alpha$ dynamical mass models (see Sections~\ref{s:mdynk3d} and \ref{s:mdynlgc}). Figure~\ref{f:mcum} shows the cumulative mass profiles of the six galaxies in our sample for which the data quality is high enough in both stars and H$\alpha$ to construct spatially resolved mass models. The JAM measurements (golden diamonds with error bars indicating one standard deviation) are shown every kiloparsec out to 10~kpc, and at $R_{e,\rm F814W}$, and the \kd models are shown as blue lines, with lighter shading indicating one and two standard deviations, respectively.

It is remarkable to see that for most cases, despite the different techniques, model inputs, and tracers, the constraints on the enclosed mass and its shape are in agreement. This comparison shows that both stars and H$\alpha$ at $z\sim1$ constrain the same total mass distribution over a large range of radii, when high-quality data suitable for dynamical modelling are available. 
We note that the uncertainties for the H$\alpha$ and JAM models are not directly comparable, since the former have fewer free parameters. However, also the extent out to which the model is constrained by the data is larger for H$\alpha$ for all cases discussed here (golden and blue arrows in Figure~\ref{f:mcum} for LEGA-C and KMOS$^{\rm 3D}$, respectively), further reducing uncertainty in the model.

For four of the six galaxies (left panels), the dynamical models agree within their uncertainties from 1~kpc to (at least) 10~kpc, covering a range of $1-2.2~R_{e,\rm F814W}$ ($1.8-2.6~R_{e,\rm F160W}$). 
For one galaxy, COS4\_25353-M1\_139825 (bottom right), the models agree within their uncertainties from 4~kpc to (at least) 10~kpc, while in the central 3~kpc the stellar model yields higher dynamical masses relative to the H$\alpha$ model. This galaxy is seen almost face on, with a difference between the H$\alpha$ kinematic major axis and the F814W position angle of $26.5^{\circ}$. This is also one of the objects with strong [NeV] emission in the central region. We speculate that emission from the AGN could bias the light-weighted estimates of the central density for both models.

There is only one galaxy for which the JAM estimates and the \kd estimates are significantly different over a large range in radius, COS4\_03493-M4\_121150 (top right). At $R_{e,\rm F814W}$, the JAM measurement is higher by $\Delta M_{\rm dyn}=0.35$~dex compared to the H$\alpha$ model. For this highly inclined galaxy ($i\approx68-84^\circ$), the kinematic major axis and the F814W and F160W position angles all align within $1^{\circ}$. However, the F814W structural parameters indicate a high Sérsic index ($n_{S,\rm F814W}=5.1$) and a large disc ($R_{e,\rm F814W}=8.2$~kpc). Yet, adopting structural parameters from F814W imaging for the H$\alpha$ model has negligible effect on the dynamical mass constraints. Instead, it is likely that JAM fits a high $M/L$ to the bright and higher-$S/N$ bulge component, leading to an overestimate of the mass in the extended disc (see also discussion in Section~\ref{s:multiple}).

\subsection{Comparison to measurements at \texorpdfstring{$R_{e,\rm F814W}$}{Re,F814W} based on integrated stellar velocity dispersion}\label{s:compmdyn}

\begin{figure}
	\centering
	\includegraphics[width=\columnwidth]{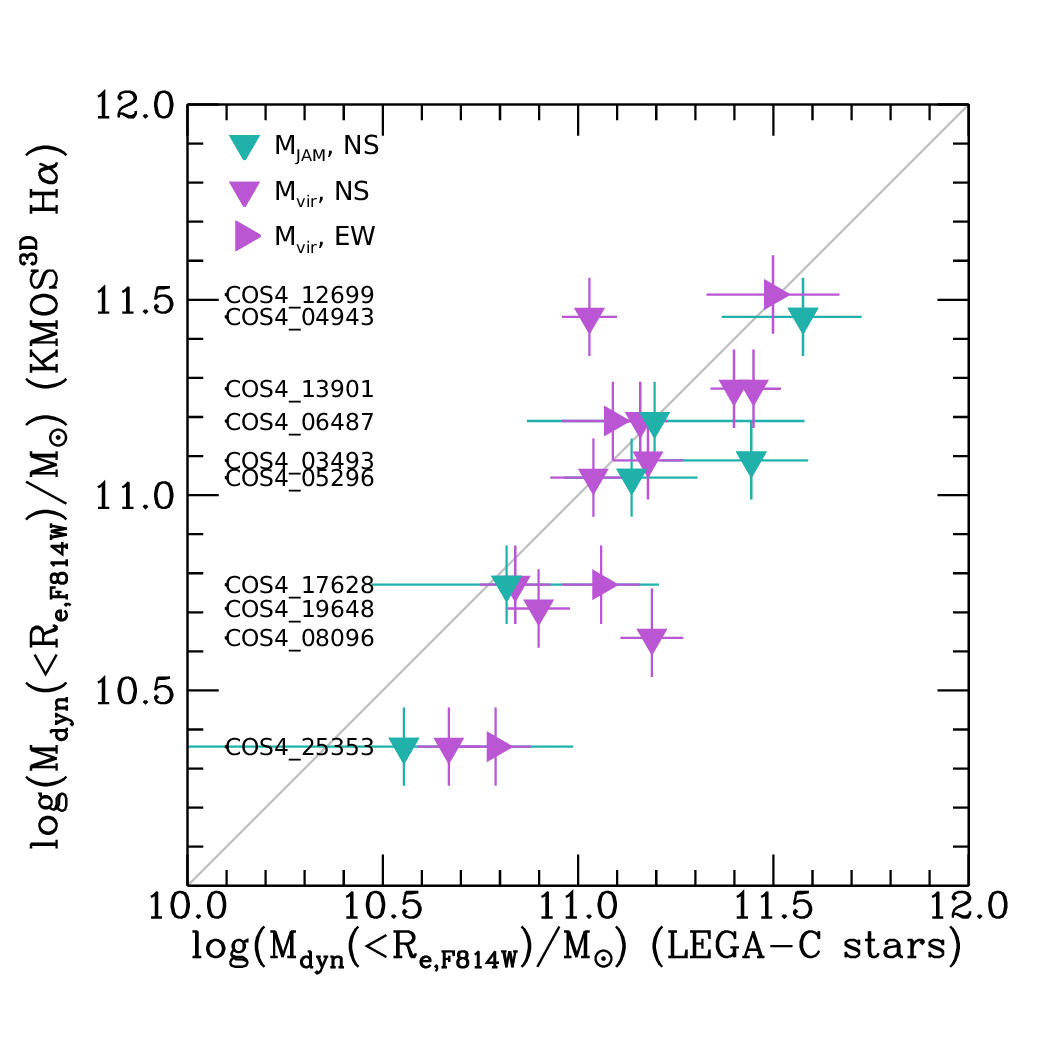}
	\caption{Comparison of dynamical mass estimates within $R_{e,\rm F814W}$ using mass models based on H$\alpha$ for \kd and integrated stellar velocity dispersion \citep[purple;][]{vdWel21} and JAM \citep[teal;][]{vHoudt21} for LEGA-C.
	The direction of triangles indicates the slit orientation of the LEGA-C data from which the estimates were derived. Uncertainties for LEGA-C measurements from integrated velocity dispersion are $1\sigma$, and for JAM models the $16^{\rm th}$ and $84^{\rm th}$ percentiles. For \kd measurements, uncertainties are the $16^{\rm th}$ and $84^{\rm th}$ percentiles, with lower ceiling uncertainties of 0.1~dex.
    Dynamical mass estimates from integrated stellar velocity dispersion are as accurate as the JAM models, with a standard deviation of 0.24~dex when comparing to the \kd H$\alpha$ measurements.}
	\label{f:mdyn}
\end{figure}

The agreement between H$\alpha$ dynamical mass models and measurements based on integrated stellar velocity dispersion is not as good. As described in Section~\ref{s:mdynlgc}, \cite{vdWel21, vdWel22}, have utilised the JAM results for LEGA-C to re-calibrate virial mass measurements based on the integrated stellar velocity dispersion, which is available for a larger number of galaxies. In Figure~\ref{f:mdyn} we compare dynamical masses at $R_{e,\rm 814W}$ from our H$\alpha$ models to LEGA-C estimates based on $\sigma_{\star,\rm vir}$ (purple symbols) for ten galaxies. We choose this radius as it allows for a straight-forward comparison of the \kd models to the LEGA-C $M_{\rm vir}$ values, where $M_{\rm dyn,LEGA-C}(<R_{e,\rm F814W})=0.5\cdot M_{\rm vir}$.\footnote{
We remind the reader that for our \kd models, $M_{\rm dyn}(<R_{e,\rm F814W})$, {\it does not} correspond to the dynamical mass enclosed within the adopted $R_e$ of the best-fit model, as here the fiducial structural parameters are based on the F160W measurements ($R_{e,\rm F814W}$ is typically larger than $R_{e,\rm F160W}$ due to its higher sensitivity for substructure, see discussion in Appendix~\ref{a:structure}).} 

Several objects in Figure~\ref{f:mdyn} have multiple estimates based on the LEGA-C $\sigma_{\star,\rm vir}$ data through duplicate observations, and we indicate the slit orientation of the LEGA-C data through down-wards (N-S) and right-wards (E-W) triangles. We also show JAM measurements within $R_{e,\rm 814W}$ (teal symbols) for the six objects discussed in Section~\ref{s:mcum}.

For the majority of measurement pairs, the dynamical mass estimates based on LEGA-C integrated stellar velocity dispersions are larger than the KMOS$^{\rm 3D}$ estimates, with an average offset of $\Delta M_{\rm dyn}=0.12$~dex. However, in our sample we also find differences in the dynamical mass estimates from the two surveys for individual galaxies of up to 0.55~dex -- amounting to more than a factor of three.
The average offset of $\Delta M_{\rm dyn}=0.12$~dex is somewhat smaller than the average offset between stellar- and emission line-based dynamical mass estimates within the effective radius found within the LEGA-C survey alone. \cite{Straatman22} find an average offset of 0.15~dex when comparing 149 LEGA-C galaxies with log(sSFR/yr)~$>-11$ and $n_S\leq2.5$. 

Due to the small sample size and the duplicate observations and two methods shown for LEGA-C, we concentrate on standard deviations from the mean difference between the various measurement sets to further quantify our results, as listed in Table~\ref{t:stddev}. Considering all observational pairs for which LEGA-C estimates based on integrated velocity dispersion exist, we find a standard deviation of 0.24~dex between the LEGA-C and \kd dynamical mass estimates. The agreement between \kd and LEGA-C JAM is better, with a standard deviation of 0.13~dex (however for a sample of six). 
We note that the reduction in scatter from 0.24 to 0.13 dex is marginally significant ($1.2\sigma$).\footnote{We approximate the significance as follows, where stddev$_i$ is the standard deviation of the mean difference of a set of measurements $i$ as listed in Table~\ref{t:stddev} (second column), and $N_i$ is the sample size of this set (third column):\\
$\sigma=|{\rm stddev}_A-{\rm stddev}_B|/[{\rm stddev}_A^2/(N_A-1)+{\rm stddev}_B^2/(N_B-1)]^{1/2}$
} 
If we further exclude the JAM measurement of galaxy COS4\_03493-M4\_121150 (see Section~\ref{s:mcum}), we find a standard deviation of 0.07~dex when comparing to the \kd estimates. In this case, the reduction in scatter relative to the comparison of \kd models and LEGA-C models based on integrated velocity dispersion has a significance of $2.3\sigma$.

Overall, the discrepancy between \kd and LEGA-C $M_{\rm dyn}$ estimates based on integrated stellar velocity dispersion in our sample is larger than the independent estimate of uncertainties from LEGA-C duplicate observations ($\sigma_{\rm Mvir,dupl}=0.14$), but comparable to the independent estimate of uncertainties from different methods within LEGA-C to determine dynamical mass ($\sigma_{\rm Mvir~vs~JAM}=0.24$). 
For the full LEGA-C survey, the scatter among $M_{\rm vir}$ and JAM measurements for SFGs is lower than our value, with $\sigma_{\rm Mvir~vs~JAM}=0.16$ \citep{vdWel22}. This suggests that our sample includes some outliers in the $M_{\rm vir}$-to-$M_{\rm JAM}$ calibration by \cite{vdWel22}.

\begin{table}
	\centering
	\caption{Standard deviation from the mean difference of the average dynamical mass discrepancy, and average dynamical mass discrepancy $\Delta M_{\rm dyn}$ for various subsets of \kd and LEGA-C data.}
	\label{t:stddev}
	\begin{tabular}{lccl} 
		\hline
		comparison & std. dev.  & $\Delta M_{\rm dyn}$ & sample size\\
		& [dex] & [dex] & \\
		\hline%	
		LEGA-C$_{\rm JAM}$, \kd &  0.13 & 0.14 & 6 \\
		LEGA-C$_{\rm JAM}$, \kd $^{*}$ &  0.07 & 0.09 & 5 \\
		LEGA-C$_{\rm Mvir}$, \kd & 0.24 & 0.12 & 14 (incl.~dupl.)\\
		LEGA-C$_{\rm Mvir}$, duplicates & 0.14 & 0.06 & 4 \\
		LEGA-C$_{\rm Mvir}$, LEGA-C$_{\rm JAM}$ & 0.24 & -0.13 & 6 \\
		\hline
     \multicolumn{4}{p{0.9\columnwidth}}{$^*$ Excluding COS4\_03493-M4\_121150 (see Sections~\ref{s:mcum} and \ref{s:multiple}).} 
	\end{tabular}
\end{table}

\subsection{Notes on LEGA-C duplicate observations and \texorpdfstring{$M_{\rm dyn}$}{Mdyn} estimates from multiple techniques}\label{s:multiple}

Figure~\ref{f:mdyn} shows for several objects multiple LEGA-C measurements of dynamical mass. 
Four galaxies in our dynamical mass sample have been observed with two different masks in LEGA-C, three of which with a different slit orientation. 
The $M_{\rm vir}$ estimates of these duplicate observations agree with each other within the uncertainties for all but one case. In this latter case, the duplicate observations have comparable $S/N$, but the observation which is also in agreement with the \kd measurement is better aligned with the kinematic major axis. For all other cases, the duplicate observation with higher $S/N$ is in better agreement with the \kd measurement. 
This conforms to the expectation that in the absence of asymmetric motions, $S/N$ is more important than alignment for integrated measurements, which are centrally weighted.
This is encouraging not only for existing ground-based surveys, but also for upcoming data from {\it JWST} NIRSpec Multi Shutter Array observations.

For most cases, the JAM measurements are in better agreement with the \kd modelling than the $M_{\rm dyn}$ estimates based on $\sigma_{\star,\rm vir}$. As discussed in Section~\ref{s:mcum}, for galaxy 
COS4\_03493-M4\_121150 JAM predicts a too large dynamical mass within $R_{e,\rm F814W}$ compared to the \kd model ($\Delta M_{\rm dyn}=0.35$~dex), but also compared to the LEGA-C measurement from $\sigma_{\star,\rm vir}$ ($\Delta M_{\rm dyn}=0.26$~dex). 
For this galaxy, spatially resolved modelling of the ionised gas from the LEGA-C survey exists as well \citep{Straatman22}. Prominent emission lines in this LEGA-C slit spectrum are H$\beta$ and [OIII], and the correspondence of the 2D \kd H$\alpha$ pseudo-slit data and the LEGA-C H$\beta$ data is good. The $M_{\rm dyn}(<R_{e,\rm F814W})$ estimate by \cite{Straatman22} agrees with the \kd estimate, further supporting the interpretation that the JAM model overpredicts the dynamical mass in this case.

A second object in our dynamical mass sample has an ionised gas-based dynamical mass estimate from LEGA-C data by \cite{Straatman22}, COS4\_04943-M3\_122667. This is the only object in our sample with a significantly higher dynamical mass measurement from \kd compared to the $M_{\rm dyn}$ measurement from the LEGA-C integrated stellar velocity dispersion ($\Delta M_{\rm dyn}=0.43$~dex). This time, the estimate from the LEGA-C ionised gas data is comparable to the estimate from integrated stellar velocity dispersion, with a difference of only $\Delta M_{\rm dyn}=0.04$~dex. Closer inspection of the LEGA-C data and the \kd pseudo-slit extractions reveals that the kinematic centre is offset by about $0.4\arcsec$ ($\sim2$ pixels) from the central pixel in the LEGA-C 2D data (see Figure~\ref{f:COS3_05062}). While this cannot typically be tested, the shape of the ionised gas velocity and velocity dispersion profiles, here constrained through H$\beta$ emission, can be aligned with the \kd H$\alpha$ kinematic profiles. The resulting spatial shift of the LEGA-C profiles shows that the velocity gradients in the LEGA-C data are underestimated in both ionised gas and stars, providing a plausible explanation for their lower $M_{\rm dyn}$ estimates. The JAM measurement agrees within the uncertainties with the \kd value, likely due to its flexibility of fitting a different centre position (see discussion in Appendix~\ref{a:misalignment}).

\subsection{Correlations with dynamical mass discrepancy}\label{s:corr}

We explore trends in dynamical mass discrepancy with structural parameters, kinematic, and physical properties of our galaxies. Due to the sample size and the comparison of multiple observations or estimates for the same objects, we cannot expect significant correlations. However, we can note a few informative trends that we discuss in this section. 
For this investigation, we exclude the integrated dispersion measurement for galaxy COS4\_04943-M3\_122667, and the JAM measurement for galaxy COS4\_03493-M4\_121150 (see Section~\ref{s:multiple}).

We list Spearman rank correlation coefficients and their significance between the dynamical mass discrepancy and various quantities in Table~\ref{t:corr}.

\begin{table}
    \centering
    \caption{Spearman rank correlation coefficients $\rho_S$ and their significance $\sigma_\rho$ between dynamical mass discrepancy $\Delta M_{\rm dyn}$ as inferred from the \kd and LEGA-C measurements, and various structural, kinematic, and global physical properties of the galaxies. This excludes the integrated dispersion measurement for galaxy COS4\_04943-M3\_122667, and the JAM measurement for galaxy COS4\_03493-M4\_121150 (see Section~\ref{s:multiple}).}
    \label{t:corr}
\begin{tabular}{lcccc}
 \hline
 & \multicolumn{2}{c}{F814W} & \multicolumn{2}{c}{F160W} \\
 Quantity & $\rho_S$ & $\sigma_\rho$ & $\rho_S$ & $\sigma_\rho$ \\
 \hline
    $R_e$ [kpc] & -0.35 & 1.4 & 0.26 & 1.1 \\
    $q$       & 0.54 & 2.2 & 0.81 & 3.3 \\
    $n_S$       & -0.07 & 0.3 & 0.03 & 0.1 \\
    PA$_{\rm morph}$ [deg]    & -0.13 & 0.5 & 0.19 & 0.8 \\
    $R_e$/PSF$_{\rm FWHM}$ $^*$ & -0.24 & 1.0 & -0.49 & 2.0 \\
 \hline
  & \multicolumn{2}{c}{$\rho_S$} & \multicolumn{2}{c}{$\sigma_\rho$} \\
\hline
    $\Psi_{\rm F814W,kin}$ & \multicolumn{2}{c}{0.42} & \multicolumn{2}{c}{1.7} \\
    $\Psi_{\rm kin,slit}$ & \multicolumn{2}{c}{0.24} & \multicolumn{2}{c}{1.0} \\
    $\Psi_{\rm F814W,slit} - \Psi_{\rm kin,slit}$ & \multicolumn{2}{c}{0.22} & \multicolumn{2}{c}{0.9} \\
\hline
    $v_{\rm circ,H\alpha}(r=R_{e,{\rm F814W}})$ [km/s] & \multicolumn{2}{c}{-0.78} & \multicolumn{2}{c}{3.2} \\
    $v_{\rm rot,H\alpha}(r=R_{e,{\rm F814W}})/\sigma_{0,\rm H\alpha}$ & \multicolumn{2}{c}{-0.51} & \multicolumn{2}{c}{2.1} \\
    $v_{\rm rot,H\alpha}(r=R_{e,{\rm F814W}})$ [km/s] & \multicolumn{2}{c}{-0.74} & \multicolumn{2}{c}{3.0} \\
    $\sigma_{0,\rm H\alpha}$ [km/s] & \multicolumn{2}{c}{0.30} & \multicolumn{2}{c}{1.2} \\
 \hline
    SFR [$M_\odot$/yr] & \multicolumn{2}{c}{0.20} & \multicolumn{2}{c}{0.8} \\    
    sSFR [1/Gyr] & \multicolumn{2}{c}{0.24} & \multicolumn{2}{c}{1.0} \\
    $\Sigma_{\rm SFR}$ [$M_\odot$/yr/kpc$^2$] & \multicolumn{2}{c}{0.32} & \multicolumn{2}{c}{1.3} \\
    $\log(M_{\star}/M_\odot)$ & \multicolumn{2}{c}{-0.08} & \multicolumn{2}{c}{0.3} \\
    $\log(M_{\rm bar}/M_\odot)$ & \multicolumn{2}{c}{-0.09} & \multicolumn{2}{c}{0.4} \\
 \hline
      \multicolumn{5}{p{0.9\columnwidth}}{$^*$ For $R_{e,\rm F814W}$ we use the PSF measurements from the LEGA-C data, and for $R_{e,\rm F160W}$ we use the PSF measurements from the \kd data.} 
\end{tabular}
\end{table}

\subsubsection{Structural parameters}\label{s:corr_struct}

\begin{figure*}
	\centering
	\includegraphics[width=0.31\textwidth]{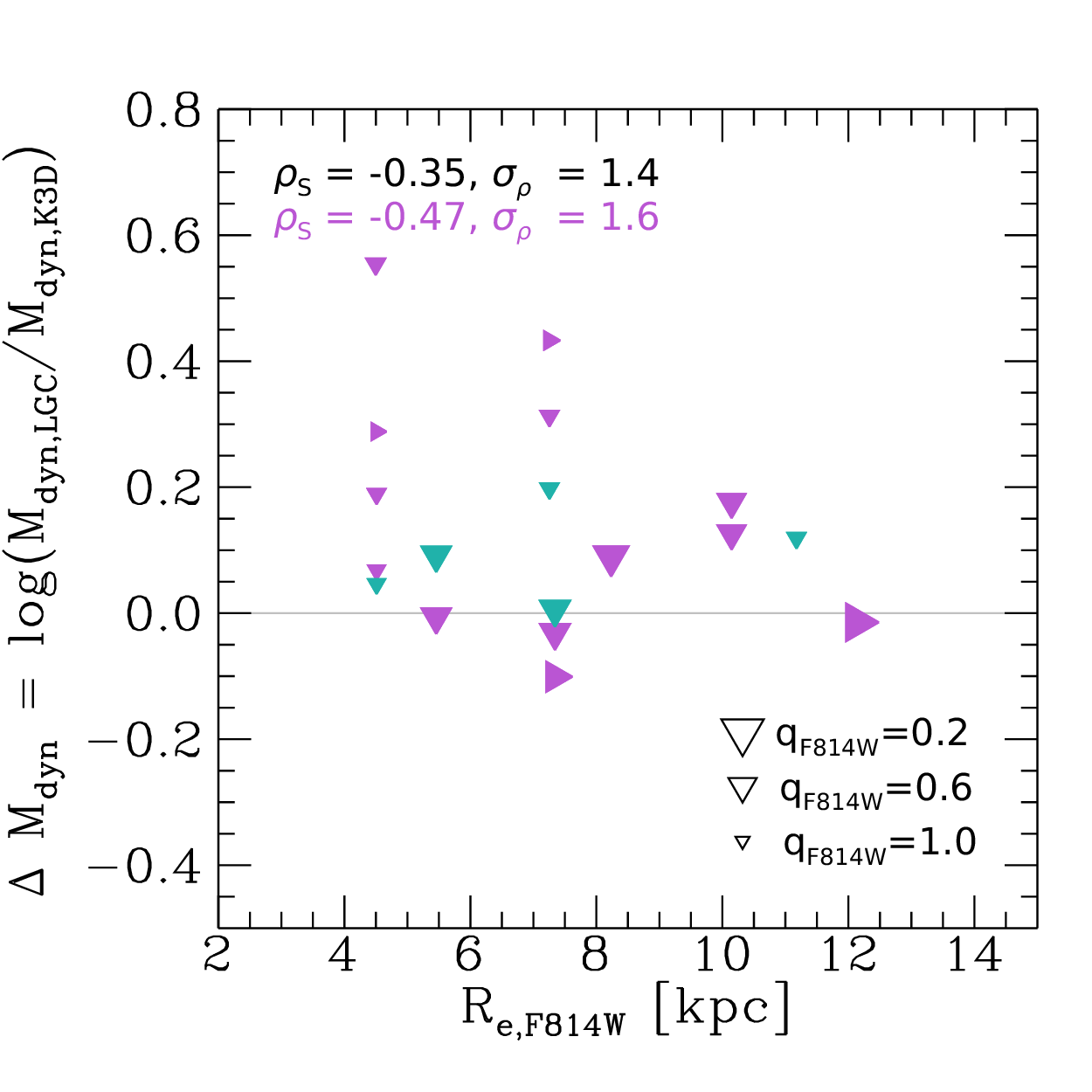}
 \hspace{3mm}
	\includegraphics[width=0.31\textwidth]{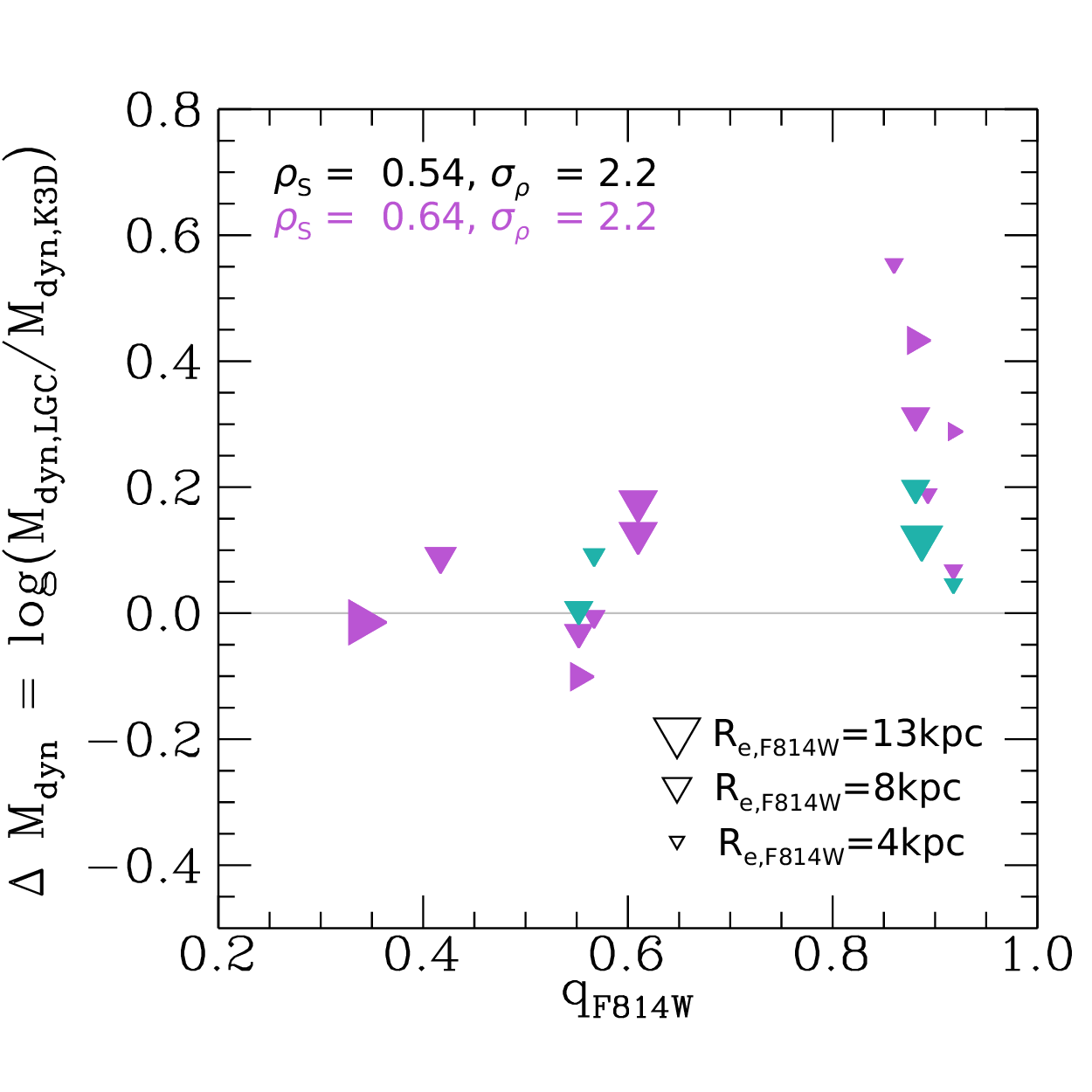}
 \hspace{3mm}
	\includegraphics[width=0.31\textwidth]{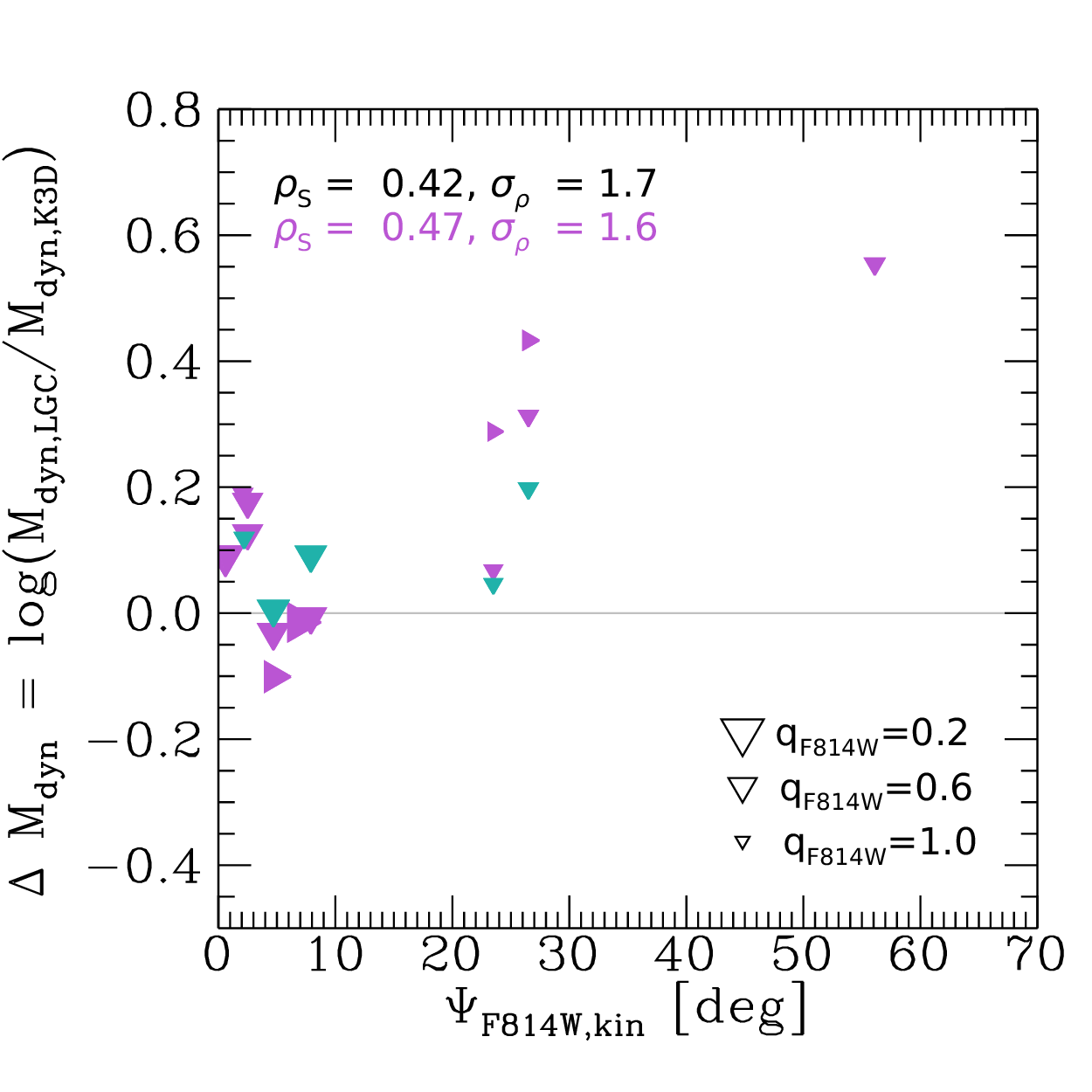}
	\caption{Difference in dynamical mass measurements from LEGA-C stars and \kd H$\alpha$, $\Delta M_{\rm dyn}=\log(M_{\rm dyn,LGC}/M_{\rm dyn,K3D})$, as a function of effective radius $R_{e,\rm F814W}$ (left), projected axis ratio $q_{\rm F814W}$ (middle), and kinematic misalignment $\Psi_{\rm F814W,kin}$ (right). The symbols are the same as in Figure~\ref{f:mdyn}, and symbol size scales with $q_{\rm F814W}$ (left, right) and $R_{e,\rm F814W}$ (middle), as indicated in the panels. 
    We find larger mismatches between dynamical mass estimates for galaxies with stronger kinematic misalignment. These are systems that are seen more face-on.}
	\label{f:dmdyn1}
\end{figure*}

\begin{figure*}
	\centering
	\includegraphics[width=0.31\textwidth]{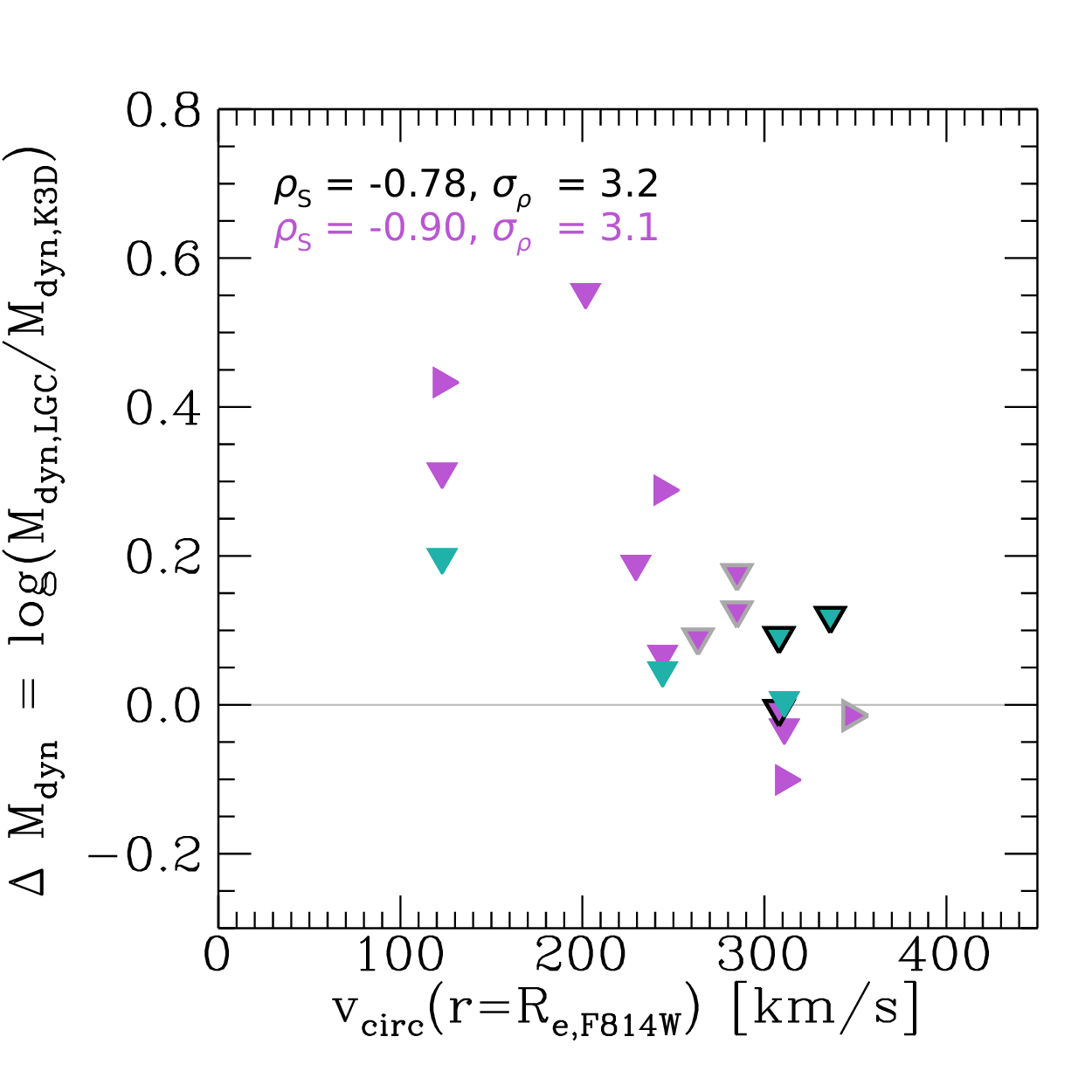}
 \hspace{3mm}
	\includegraphics[width=0.31\textwidth]{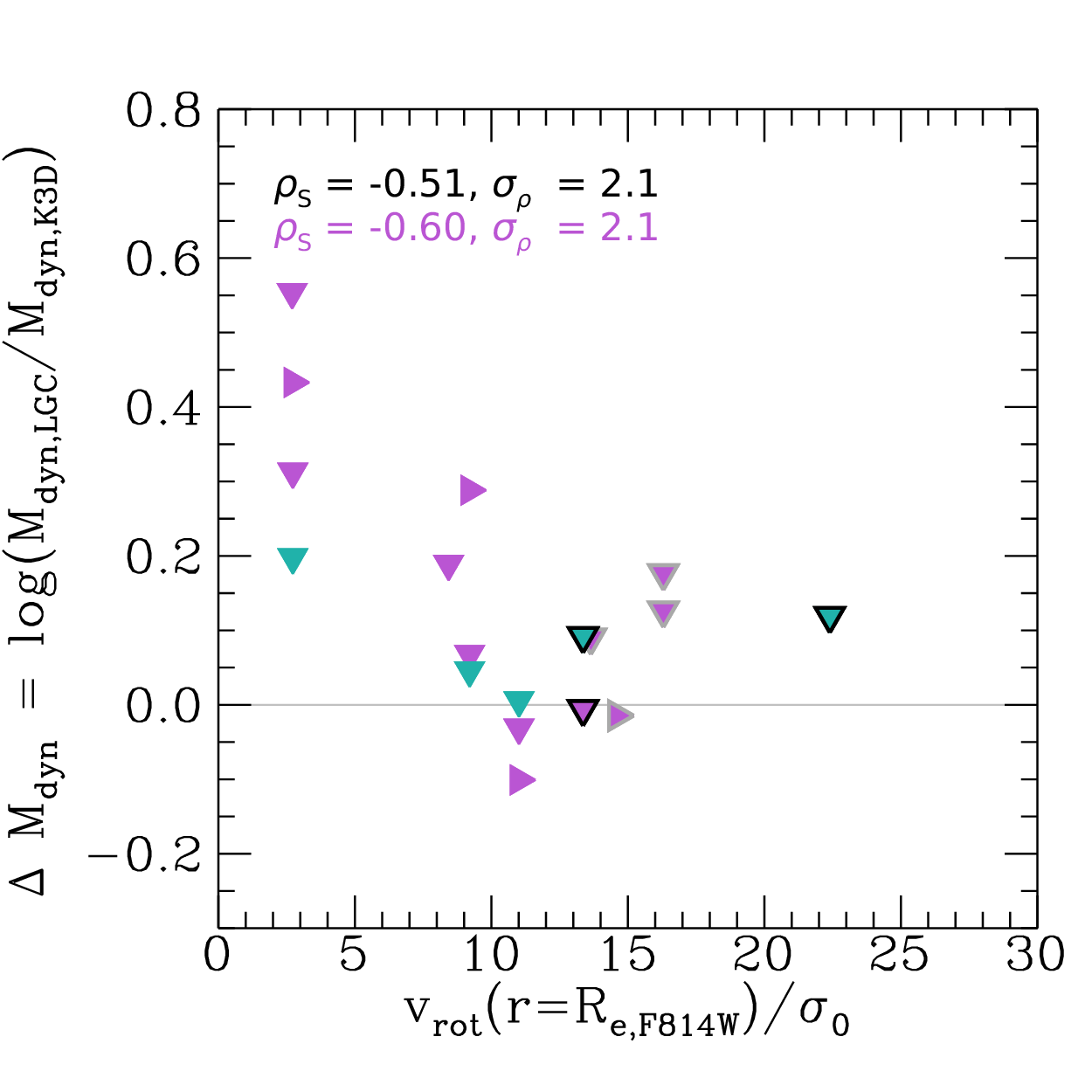}
	 \hspace{3mm}
    \includegraphics[width=0.31\textwidth]{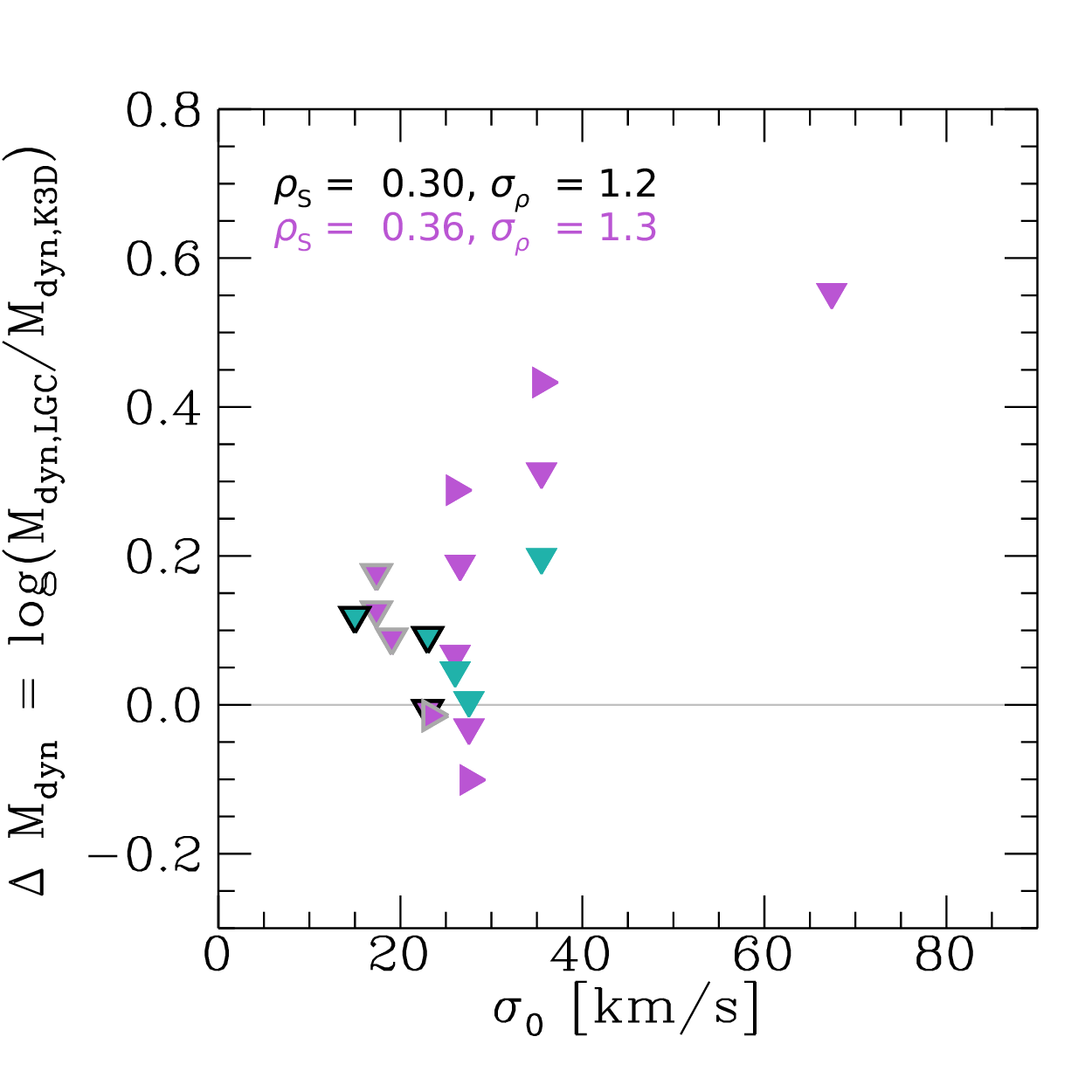}
    \caption{Difference in dynamical mass measurements from LEGA-C stars and \kd H$\alpha$, $\Delta M_{\rm dyn}=\log(M_{\rm dyn,LGC}/M_{\rm dyn,K3D})$, as a function of H$\alpha$ circular velocity $v_{\rm circ}(R_{e,\rm F814W})$ (left), rotational support $v_{\rm rot}(R_{e,\rm F814W})/\sigma_0$ (middle), and intrinsic velocity dispersion $\sigma_0$ (right). The symbols are the same as in Figure~\ref{f:mdyn}. Black outlines indicate two galaxies for which we can only constrain upper limits on $\sigma_0$ from our models to the \kd data, and grey outlines indicate another three galaxies with a non-Gaussian marginalised posterior distribution towards low $\sigma_0$ values (see Section~\ref{s:corr_kin}). We find larger mismatches between dynamical mass estimates primarily for galaxies with lower circular velocities and lower rotational support, as constrained from our best-fit models to the H$\alpha$ data.}
	\label{f:dmdyn3} 
\end{figure*}

Dynamical mass discrepancies are larger for smaller galaxies that are seen more face-on, as illustrated in the left panels of Figure~\ref{f:dmdyn1}. The measured effective radius $R_e$ and the projected axis ratio $q$ are themselves correlated: in the presence of surface brightness limitations, large face-on systems may be missing from a sample, thus imprinting an artificial $R_e$-$q$ correlation \citep[e.g.][]{Driver07, Graham08, Mowla19}. In general, dynamical mass estimates for smaller galaxies are less certain since their kinematics are constrained by fewer independent resolution elements. For more face-on systems it is more difficult to infer robust inclination corrections to the anyhow shallow velocity gradients. Therefore, we interpret the larger discrepancy of H$\alpha$-based and stellar-based $M_{\rm dyn}$ estimates for smaller, more face-on galaxies as being due to less robust measurements.

In the middle panel of Figure~\ref{f:dmdyn1} we also see that those $M_{\rm dyn}$ estimates from LEGA-C that are lower than the \kd estimates have $q<0.6$. This could be a possible effect of overestimated inclinations in the LEGA-C dynamical models due to a larger intrinsic thickness prior (Section~\ref{s:model_diff}), translated to the $M_{\rm dyn}$ measurements from integrated velocity dispersions via the calibration based on the JAM models. However, we do not see evidence for a systematic trend in our data.
We caution though that assumptions on intrinsic thickness and related inclination estimates can introduce systematic effects in dynamical mass measurements (see also discussion in Appendix~\ref{a:structure}). 

For close-to face-on galaxies, it is also harder to determine the position angle from imaging or moment-zero emission line maps. It was shown by \cite{Wisnioski15} that the majority of misalignments between kinematically (H$\alpha$) and morphologically (F160W) determined position angles for $z\sim1-2$ SFGs are found at $q>0.6$. These trends suggest that apparent misalignments between ionised gas kinematics and stellar light are not primarily due to intrinsic physical differences between the warm gas and stellar distributions in galaxies, a possible consequence of e.g.\ misaligned accretion, but are largely due to limitations of photometric measurements for face-on systems. We find a comparable trend in our sample. 

This motivates us to explore in more detail possible correlations between dynamical mass discrepancy and measures of position angle. Here, following \cite{Wisnioski15} and the misalignment diagnostic $\Psi$ by \cite{Franx91}, we define $\sin(\Psi_{\rm F814W,kin}) = |\sin({\rm PA_{F814W}}-{\rm PA_{\rm kin}})|$, where PA$_{\rm kin}$ is measured from the H$\alpha$ IFS data. We find that those galaxies with larger mismatches in their dynamical mass estimates also have stronger kinematic misalignments, with a correlation coefficient $\rho_S=0.42$ and $\sigma_\rho=1.7$.

For $M_{\rm dyn}$ measurements based on JAM models (or the spatially resolved models by \citealt{Straatman22}), we would expect a trend such that kinematic PAs that are more inclined with respect to the slit orientation than the photometric PA would result in underestimated dynamical masses from the stellar data, whereas kinematic PAs that are closer to the slit orientation than the photometric PA would result in overestimated dynamical masses from the stellar data. We find no indication for a corresponding trend based on the JAM measurements only. 
Considering all measurement pairs we find no significant correlation ($\rho_S=0.22$; $\sigma_\rho=0.9$).

Further, we find at most a weak correlation between mismatches of the H$\alpha$ kinematic major axis and the LEGA-C slit position for the $M_{\rm dyn}$ discrepancies based on integrated stellar velocity dispersion measurements ($\rho_S=0.31$; $\sigma_\rho=1.1$).
We also find no (significant) correlation with $M_{\rm dyn}$ discrepancy and the F160W or F814W Sérsic index measurements,  their difference, or estimates of the central 1~kpc stellar surface density.

\subsubsection{Kinematic properties}\label{s:corr_kin}

We consider correlations of dynamical mass discrepancy with kinematic quantities, specifically the H$\alpha$ circular velocity at $R_{e, \rm F814W}$ and the rotational support $v_{\rm rot}/\sigma_0$. These quantities are based on the best-fit dynamical models of the H$\alpha$ data from the \kd survey. We find a correlation with circular velocity ($\rho_S=-0.78$; $\sigma_\rho=3.2$), as illustrated in the left panel of Figure~\ref{f:dmdyn3}. This shows that the difference in $M_{\rm dyn}$ estimates from stars and gas is lower for galaxies with higher circular velocities and stronger dynamical support from rotation in the ionised gas phase (middle panel). Possible explanations could be that systems with higher rotational support are closer to dynamical equilibrium (dynamical equilibrium is the base assumption for all dynamical modelling discussed in this work), or that the pressure support corrections for the H$\alpha$ data are underestimated (cf.\ right panel). The pressure support correction chosen in this work following the self-gravitating disc description by \cite{Burkert10} is stronger than other corrections adopted in the literature, so the latter explanation is unlikely \citep[see e.g.][]{Bouche22, Price22}.

No corresponding significant correlations have been found in the study by \cite{Straatman22} comparing slit-based estimates for both ionised gas and stars, where the kinematic major axis is unknown.

We note that for two galaxies we can only robustly constrain upper limits on the intrinsic velocity dispersion $\sigma_0$ from our models due to the spectral resolution of KMOS (see Figures~\ref{f:c1},\ref{f:c2}, and Section 3.4 by \citealp{Uebler19} for details on the treatment of upper limits). For three more galaxies the MCMC-derived marginalised posterior distribution of $\sigma_0$ is non-Gaussian. If we remove these five galaxies from our calculation of the correlation coefficients, we find $\rho_S=0.47$ and $\sigma_\rho=1.6$ for the correlation between $\Delta M_{\rm dyn}$ and $\sigma_0$, and $\rho_S=-0.87$, $\sigma_\rho=3.0$ and $\rho_S=-0.80$, $\sigma_\rho=2.8$ for the correlations with $v_{\rm circ}(r=R_{e,{\rm F814W}})$ and $v_{\rm rot}(r=R_{e,{\rm F814W}})/\sigma_0$, respectively.

\subsubsection{Global physical properties}

Considering physical properties related to feedback strength such as SFR, sSFR, $\Sigma_{\rm SFR}$, AGN activity, or outflow signatures, we find no (significant) correlations with dynamical mass offset. This suggests that feedback does not play a major role in systematically affecting the dynamical mass estimates differently for H$\alpha$ and stars for galaxies in our sample \citep[see also][]{Straatman22}. However, we remind the reader about the substantial effect of AGN tracer emission line species such as [NeV] on the ionised gas galaxy kinematics extracted from the LEGA-C spectra.

We also caution about the potential impact of the line broadening in integrated line emission spectra induced by the presence of strong outflows (and of important disc velocity dispersion), as discussed by \cite{Wisnioski18} for compact massive SFGs. This underscores the benefit of spatially-resolved emission line kinematic modeling as performed here.

Although (circular) velocity and galaxy mass are connected through the TFR, we find no significant correlation with $\Delta M_{\rm dyn}$ and galaxy mass (see Table~\ref{t:corr}). To some extent, this can be explained by the scatter in the $z\sim1$ TFR \citep[e.g.][]{Uebler17}. However, in our sample the lack of correlation is also driven by the massive, compact galaxy COS4\_08096 having a large dynamical mass discrepancy. If we exclude this galaxy from the calculations, we still find only weak trends, but more along the expected direction for $\log(M_{\star})$ ($\rho_S=-0.28$, $\sigma_\rho=1.1$) and for $\log(M_{\rm bar})$ ($\rho_S=-0.29$, $\sigma_\rho=1.2$).\\

In summary, despite the small sample size our investigation of correlations with dynamical mass discrepancy reveals interesting trends, which should be followed up in future studies.
We find mild correlations in particular with effective radius, projected axis ratio, rotational support in the ionised gas phase, and with kinematic misalignment. 
Among the various quantities explored, we find the strongest correlations between dynamical mass discrepancy and $v_{\rm rot,H\alpha}$, $v_{\rm circ,H\alpha}$, and $q_{\rm F160W}$, with a statistical significance of $\sigma_\rho\geq3$.
This confirms on the one hand the expectation that it is more difficult to constrain robust dynamical masses for galaxies that are smaller, more face-on, and with higher dispersion support \citep[see][for a detailed study]{Wisnioski18}. On the other hand it stresses the importance of spatially-resolved kinematic information to build accurate mass models.

\section{Discussion and Conclusions}\label{s:conclusions}

We have compared kinematics and inferred dynamical masses from ionised gas and stars in 16 star-forming galaxies at $z\sim1$, common to the \kd \citep{Wisnioski15, Wisnioski19} and LEGA-C \citep{vdWel16, vdWel21, Straatman18} surveys. Our main conclusions are as follows:

\begin{itemize}

    \item Comparing stellar and H$\alpha$ kinematic profiles, we find that on average rotation velocities are higher by $\sim45$ per cent and velocity dispersions are lower by a factor of two for H$\alpha$ relative to stars, reminiscent of trends observed in the local Universe (Sections~\ref{s:comp_v} and \ref{s:comp_s}).\\
    
    \item We measure higher rotational support in H$\alpha$ compared to [OII]. This could explain systematic differences found in literature studies of e.g.\ the Tully-Fisher relation when based only on $v_{\rm rot}$ without accounting for pressure support (Sections~\ref{s:comp_vs} and \ref{s:compilations}).\\
    
    \item We find excellent agreement between cumulative total mass profiles constrained from our {\sc{dysmal}} models using H$\alpha$ kinematics and from JAM models to the stellar kinematics, out to at least 10~kpc for five of six galaxies (average $\Delta M_{\rm dyn}(R_{e,\rm F814W})<0.1$~dex, standard deviation 0.07~dex; Section~\ref{s:mcum}). This shows that dynamical masses at $z\sim1$ can be robustly measured from modelling spatially resolved observations, either of stellar or ionised gas kinematics. \\

    \item Simpler dynamical mass estimates based on integrated stellar velocity dispersion are less accurate (standard deviation 0.24~dex; Section~\ref{s:compmdyn}). \\
    
    \item We investigate correlations of dynamical mass offset with galaxy properties and find larger offsets e.g.\ for galaxies with stronger misalignments of photometric and H$\alpha$ kinematic position angles. We find statistically significant correlations with dynamical mass offset and $v_{\rm rot,H\alpha}$, $v_{\rm circ,H\alpha}$, and $q_{\rm F160W}$ (Section~\ref{s:corr}). This highlights the value of 2D spatially resolved kinematic information in inferring dynamical masses. 
            
\end{itemize}

Our comparison of the kinematics of stars and ionised gas reveals differences in their resolved velocities and velocity dispersions that are marginally significant.
Lower rotational support, lower LOS disc velocities, and higher LOS disc velocity dispersions of stars relative to the star-forming gas phase are also seen in modern cosmological simulations (\citealp{Pillepich19}; C. Lagos, priv. comm.).
A possible scenario explaining lower rotational support and higher dispersion in the stellar component is that the observed stars have been born {\it in-situ} from gas with higher velocity dispersions \citep[e.g.][]{Bird21}. 
Unfortunately, the redshift evolution of molecular gas disc velocity dispersion is still poorly constrained through available data \citep[see][]{Uebler19}. 

In general, the collision-less nature of stars allows for a variety of non-circular orbital motions. A higher fraction of low-angular momentum box orbits or $x$-tubes (rotation around the minor axis) can reduce the LOS velocity of stars \citep[e.g.][]{Roettgers14}. The origin of such motions is plausibly connected to assembly history, where more frequent mergers in the past reduce the net angular momentum of the stellar component, in particular if their baryon content is dominated by stars \citep[e.g.][]{Naab14}. 
Such interactions, alongside secular processes like scattering by giant molecular clouds, bars, or spiral arms, could also contribute to disc heating, further increasing the velocity dispersion of the stars \citep[e.g.][]{Jenkins90, Aumer16, Grand16}. 
This is in agreement with our finding of higher $v_{\rm max}/\sigma_{\rm out}$ measured from H$\alpha$ compared to stars. 

Theoretically, the misaligned smooth accretion of gas can also result in different kinematics of gas and stars \citep[e.g.][]{Sales12, Aumer13, Aumer14, Uebler14, vdVoort15, Khim21}. However, such processes typically reduce initially only the net angular momentum of the gas phase. This would correspond to a reduction of $v_{\rm max}/\sigma_{\rm out}$ measured from the star-forming gas relative to the full stellar population, which is not observed in our data set.

Deviating kinematic signatures in gas {\it vs.\ }stars could also be caused by feedback. 
The imprints of stellar- and AGN-driven winds on massive galaxy spectra at $z\sim1-2$ are routinely observed (see e.g.\ \citealp{Rubin10, Shapley03, Newman12b, Carniani15, Harrison16, Zakamska16, Talia17, DaviesRL19, Kakkad20}; and \citealp{Genzel14, FS19, Concas22}, for analyses including \kd data). Such feedback can bias disc kinematic measurements due to the difficulty of disentangling e.g.\ galaxy rotation from outflows in low spectral or spatial resolution observations \citep[see e.g.][]{Wisnioski18, Lelli18}. 
At least three galaxies in our sample show signatures of outflows in their H$\alpha$ spectra, but we could not identify a systematic effect on the spatially-resolved disc kinematic measurements of H$\alpha$ and stars presented in Section~\ref{s:compasobs}. One of these galaxies (COS\_19648-M1\_134839) shows indication of a counter-rotating disc in the stellar {\it vs.} ionised gas components; the difference in $\log(M_{\rm dyn})$ for this object is $\sim0.15$~dex. 
However, we clearly see the impact of feedback processes on the kinematic signatures of specific emission lines (especially [NeV]) deviating from the main disc rotation in the LEGA-C spectra of two objects (see Section~\ref{s:lgckin} and Appendix~\ref{a:nev}). 

Our results on dynamical mass estimates show that data quality and methods play a role for existing differences in dynamical mass estimates of $z\sim1$ galaxies. 
The fact that we find better agreement between the \kd {\sc{dysmal}} and the LEGA-C JAM dynamical mass estimates, as compared to the LEGA-C estimates based on integrated velocity dispersion, demonstrates the advantage of detailed dynamical models leveraging the full structural information available over more approximate estimators. 
The remarkable agreement between spatially resolved dynamical mass estimates from stars and H$\alpha$, and from independent data sets, provides great confidence in our ability to probe the gravitational potential of $z\sim1$ galaxies.
It further suggests that our implementation of the pressure support correction accounting for the turbulent motions in the ionised gas phase is adequate.

At the same time, the residual trends between \kd dynamical mass estimates with {\sc{dysmal}} and LEGA-C dynamical mass estimates from integrated velocity dispersions, particularly with the major axis misalignment of F814W photometry and H$\alpha$ kinematics, could be interpreted as signs of physical processes disturbing global equilibrium for some galaxies. 
A difference in position angle of gas and stars could stem from misaligned smooth accretion, but also from a disruptive merger event in the past \citep[see e.g.][]{Khim21}. If the system has not yet reached a new equilibrium, this could be reflected in deviating dynamical mass estimates from the differently affected baryonic components.
However, galaxies with large misalignment ($\Delta{\rm PA}>20^{\circ}$) in our sample are also seen relatively face-on, indicating that the photometric PA measurements are more uncertain and any intrinsic misalignment is likely smaller. 

Overall, the dynamical mass measurements from LEGA-C stellar kinematics tend to be larger than the measurements from the \kd H$\alpha$ kinematics by 0.12~dex on average \citep[see also][]{Straatman22}. If dynamical mass measurements from stellar kinematics are systematically overestimated, this would reduce mass-to-light ratios inferred from such data and impact conclusions on the initial mass function of galaxies. It could also potentially impact the evolutionary study of the Fundamental Plane \citep{Djorgovski87, Dressler87}. Larger comparison samples at $z>0$ are required to quantify any potential effect.

Larger samples will also be necessary for a statistical assessment of the impact of physical processes on galaxy dynamics at this cosmic epoch. Of further interest would be the extension of our sample towards lower masses, where the shallower potential wells of haloes would allow feedback and accretion processes to have a larger impact on the host galaxy properties. Due to the smaller size of lower-mass galaxies, this would require higher spatial resolution observations than the data presented in this work. This could be achieved with instruments such as ERIS/VLT, and in the future with HARMONI/ELT, or GMTIFS/GMT. Similarly, higher-resolution imaging providing better constrained structural parameters would help in building more accurate dynamical models.
At higher redshifts, higher accretion rates and shallower potential wells may cause larger and more frequent kinematic misalignments. This can be investigated through a combination of {\it JWST}/NIRCam imaging and {\it JWST}/NIRSpec IFS observations.

For a comprehensive assessment of baryonic kinematics and dynamics at $z\sim1$, the high-quality data from the \kd and LEGA-C surveys would ideally be complemented by spatially resolved observations of another independent dynamical tracer, such as CO. With potentially lower disc velocity dispersion than stars and ionised gas, and unaffected by extinction, dynamical masses inferred from molecular gas kinematics could help to determine realistic uncertainties on dynamical masses, and improve our understanding of the role of corrections factors and modelling assumptions required to infer dynamical masses from other baryonic phases.

\section*{Acknowledgements}

We thank the anonymous referee for a careful reading of the manuscript and detailed suggestions. We thank the editor for helpful comments. H{\"U} thanks Sirio Belli, Luca Cortese, Claudia Lagos, and Joop Schaye for insightful discussions on aspects of this work.
We thank Amit Nestor for sharing dynamical mass estimates based on the work by \cite{Nestor22}. We thank Claudia Lagos for sharing a comparison of LOS velocity dispersion from stars and star-forming gas based on calculations described in \cite{Lagos18} utilising the {\sc{eagle}} simulations \citep{Schaye15}. 
H{\"U} gratefully acknowledges support by the Isaac Newton Trust and by the Kavli Foundation through a Newton-Kavli Junior Fellowship.
CMSS acknowledges support from Research Foundation - Flanders (FWO) through Fellowship 12ZC120N.
This research made use of \href{http://www.astropy.org}{Astropy}, a community-developed core Python package for Astronomy \citep{astropy13, astropy18}, and of Photutils, an Astropy package for detection and photometry of astronomical sources \citep{photutils}.

%%%%%%%%%%%%%%%%%%%%%%%%%%%%%%%%%%%%%%%%%%%%%%%%%%
\section*{Data Availability}

The \kd data cubes used in this research are publicly available and accessible at \url{http://www.mpe.mpg.de/ir/KMOS3D} \citep{Wisnioski19}. The 1D spectra and $M_{\rm vir}$ measurements from the LEGA-C survey are published by \cite{vdWel21}. JAM models for the LEGA-C data are published by \cite{vHoudt21}.


\bibliographystyle{mnras}
\bibliography{literature}

\begin{thebibliography}{}
\makeatletter
\relax
\def\mn@urlcharsother{\let\do\@makeother \do\$\do\&\do\#\do\^\do\_\do\%\do\~}
\def\mn@doi{\begingroup\mn@urlcharsother \@ifnextchar [ {\mn@doi@}
  {\mn@doi@[]}}
\def\mn@doi@[#1]#2{\def\@tempa{#1}\ifx\@tempa\@empty \href
  {http://dx.doi.org/#2} {doi:#2}\else \href {http://dx.doi.org/#2} {#1}\fi
  \endgroup}
\def\mn@eprint#1#2{\mn@eprint@#1:#2::\@nil}
\def\mn@eprint@arXiv#1{\href {http://arxiv.org/abs/#1} {{\tt arXiv:#1}}}
\def\mn@eprint@dblp#1{\href {http://dblp.uni-trier.de/rec/bibtex/#1.xml}
  {dblp:#1}}
\def\mn@eprint@#1:#2:#3:#4\@nil{\def\@tempa {#1}\def\@tempb {#2}\def\@tempc
  {#3}\ifx \@tempc \@empty \let \@tempc \@tempb \let \@tempb \@tempa \fi \ifx
  \@tempb \@empty \def\@tempb {arXiv}\fi \@ifundefined
  {mn@eprint@\@tempb}{\@tempb:\@tempc}{\expandafter \expandafter \csname
  mn@eprint@\@tempb\endcsname \expandafter{\@tempc}}}

\bibitem[\protect\citeauthoryear{{Astropy Collaboration} et~al.,}{{Astropy
  Collaboration} et~al.}{2013}]{astropy13}
{Astropy Collaboration} et~al., 2013, \mn@doi [\aap]
  {10.1051/0004-6361/201322068}, \href
  {https://ui.adsabs.harvard.edu/abs/2013A&A...558A..33A} {558, A33}

\bibitem[\protect\citeauthoryear{{Astropy Collaboration} et~al.,}{{Astropy
  Collaboration} et~al.}{2018}]{astropy18}
{Astropy Collaboration} et~al., 2018, \mn@doi [\aj] {10.3847/1538-3881/aabc4f},
  \href {https://ui.adsabs.harvard.edu/abs/2018AJ....156..123A} {156, 123}

\bibitem[\protect\citeauthoryear{{Aumer}, {White}, {Naab}  \&
  {Scannapieco}}{{Aumer} et~al.}{2013}]{Aumer13}
{Aumer} M.,  {White} S.~D.~M.,  {Naab} T.,   {Scannapieco} C.,  2013, \mn@doi
  [\mnras] {10.1093/mnras/stt1230}, \href
  {http://adsabs.harvard.edu/abs/2013MNRAS.434.3142A} {434, 3142}

\bibitem[\protect\citeauthoryear{{Aumer}, {White}  \& {Naab}}{{Aumer}
  et~al.}{2014}]{Aumer14}
{Aumer} M.,  {White} S. D.~M.,   {Naab} T.,  2014, \mn@doi [\mnras]
  {10.1093/mnras/stu818}, \href
  {https://ui.adsabs.harvard.edu/abs/2014MNRAS.441.3679A} {441, 3679}

\bibitem[\protect\citeauthoryear{{Aumer}, {Binney}  \& {Sch{\"o}nrich}}{{Aumer}
  et~al.}{2016}]{Aumer16}
{Aumer} M.,  {Binney} J.,   {Sch{\"o}nrich} R.,  2016, \mn@doi [\mnras]
  {10.1093/mnras/stw1639}, \href
  {https://ui.adsabs.harvard.edu/\#abs/2016MNRAS.462.1697A} {462, 1697}

\bibitem[\protect\citeauthoryear{{Barrera-Ballesteros}
  et~al.,}{{Barrera-Ballesteros} et~al.}{2014}]{BarreraBallesteros14}
{Barrera-Ballesteros} J.~K.,  et~al., 2014, \mn@doi [\aap]
  {10.1051/0004-6361/201423488}, \href
  {https://ui.adsabs.harvard.edu/abs/2014A&A...568A..70B} {568, A70}

\bibitem[\protect\citeauthoryear{{Barrera-Ballesteros}
  et~al.,}{{Barrera-Ballesteros} et~al.}{2015}]{BarreraBallesteros15}
{Barrera-Ballesteros} J.~K.,  et~al., 2015, \mn@doi [\aap]
  {10.1051/0004-6361/201424935}, \href
  {https://ui.adsabs.harvard.edu/abs/2015A&A...582A..21B} {582, A21}

\bibitem[\protect\citeauthoryear{{Belli}, {Newman}  \& {Ellis}}{{Belli}
  et~al.}{2017}]{Belli17a}
{Belli} S.,  {Newman} A.~B.,   {Ellis} R.~S.,  2017, \mn@doi [\apj]
  {10.3847/1538-4357/834/1/18}, \href
  {https://ui.adsabs.harvard.edu/abs/2017ApJ...834...18B} {834, 18}

\bibitem[\protect\citeauthoryear{{Bezanson} et~al.,}{{Bezanson}
  et~al.}{2018a}]{Bezanson18}
{Bezanson} R.,  et~al., 2018a, \mn@doi [\apj] {10.3847/1538-4357/aabc55}, \href
  {https://ui.adsabs.harvard.edu/abs/2018ApJ...858...60B} {858, 60}

\bibitem[\protect\citeauthoryear{{Bezanson} et~al.,}{{Bezanson}
  et~al.}{2018b}]{Bezanson18b}
{Bezanson} R.,  et~al., 2018b, \mn@doi [\apjl] {10.3847/2041-8213/aaf16b},
  \href {https://ui.adsabs.harvard.edu/abs/2018ApJ...868L..36B} {868, L36}

\bibitem[\protect\citeauthoryear{{Bird}, {Loebman}, {Weinberg}, {Brooks},
  {Quinn}  \& {Christensen}}{{Bird} et~al.}{2021}]{Bird21}
{Bird} J.~C.,  {Loebman} S.~R.,  {Weinberg} D.~H.,  {Brooks} A.~M.,  {Quinn}
  T.~R.,   {Christensen} C.~R.,  2021, \mn@doi [\mnras]
  {10.1093/mnras/stab289}, \href
  {https://ui.adsabs.harvard.edu/abs/2021MNRAS.503.1815B} {503, 1815}

\bibitem[\protect\citeauthoryear{{Bolatto} et~al.,}{{Bolatto}
  et~al.}{2017}]{Bolatto17}
{Bolatto} A.~D.,  et~al., 2017, \mn@doi [\apj] {10.3847/1538-4357/aa86aa},
  \href {https://ui.adsabs.harvard.edu/\#abs/2017ApJ...846..159B} {846, 159}

\bibitem[\protect\citeauthoryear{{Bosma}}{{Bosma}}{1981}]{Bosma81}
{Bosma} A.,  1981, \mn@doi [\aj] {10.1086/113063}, \href
  {https://ui.adsabs.harvard.edu/abs/1981AJ.....86.1825B} {86, 1825}

\bibitem[\protect\citeauthoryear{{Bouch{\'e}}, {Carfantan}, {Schroetter},
  {Michel-Dansac}  \& {Contini}}{{Bouch{\'e}} et~al.}{2015}]{Bouche15}
{Bouch{\'e}} N.,  {Carfantan} H.,  {Schroetter} I.,  {Michel-Dansac} L.,
  {Contini} T.,  2015, \mn@doi [\aj] {10.1088/0004-6256/150/3/92}, \href
  {https://ui.adsabs.harvard.edu/abs/2015AJ....150...92B} {150, 92}

\bibitem[\protect\citeauthoryear{{Bouch{\'e}} et~al.,}{{Bouch{\'e}}
  et~al.}{2022}]{Bouche22}
{Bouch{\'e}} N.~F.,  et~al., 2022, \mn@doi [\aap]
  {10.1051/0004-6361/202141762}, \href
  {https://ui.adsabs.harvard.edu/abs/2022A&A...658A..76B} {658, A76}

\bibitem[\protect\citeauthoryear{{Bradley} et~al.,}{{Bradley}
  et~al.}{2022}]{photutils}
{Bradley} L.,  et~al., 2022, {astropy/photutils: 1.5.0}, Zenodo,
  \mn@doi{10.5281/zenodo.6825092}

\bibitem[\protect\citeauthoryear{{Brammer} et~al.,}{{Brammer}
  et~al.}{2012}]{Brammer12}
{Brammer} G.~B.,  et~al., 2012, \mn@doi [\apjs] {10.1088/0067-0049/200/2/13},
  \href {http://adsabs.harvard.edu/abs/2012ApJS..200...13B} {200, 13}

\bibitem[\protect\citeauthoryear{{Bruzual} \& {Charlot}}{{Bruzual} \&
  {Charlot}}{2003}]{Bruzual03}
{Bruzual} G.,  {Charlot} S.,  2003, \mn@doi [\mnras]
  {10.1046/j.1365-8711.2003.06897.x}, \href
  {http://adsabs.harvard.edu/abs/2003MNRAS.344.1000B} {344, 1000}

\bibitem[\protect\citeauthoryear{{Bryant} et~al.,}{{Bryant}
  et~al.}{2019}]{Bryant19}
{Bryant} J.~J.,  et~al., 2019, \mn@doi [\mnras] {10.1093/mnras/sty3122}, \href
  {https://ui.adsabs.harvard.edu/abs/2019MNRAS.483..458B} {483, 458}

\bibitem[\protect\citeauthoryear{{Bundy} et~al.,}{{Bundy}
  et~al.}{2015}]{Bundy15}
{Bundy} K.,  et~al., 2015, \mn@doi [\apj] {10.1088/0004-637X/798/1/7}, \href
  {https://ui.adsabs.harvard.edu/abs/2015ApJ...798....7B} {798, 7}

\bibitem[\protect\citeauthoryear{{Burkert} et~al.,}{{Burkert}
  et~al.}{2010}]{Burkert10}
{Burkert} A.,  et~al., 2010, \mn@doi [\apj] {10.1088/0004-637X/725/2/2324},
  \href {http://adsabs.harvard.edu/abs/2010ApJ...725.2324B} {725, 2324}

\bibitem[\protect\citeauthoryear{{Calzetti}, {Armus}, {Bohlin}, {Kinney},
  {Koornneef}  \& {Storchi-Bergmann}}{{Calzetti} et~al.}{2000}]{Calzetti00}
{Calzetti} D.,  {Armus} L.,  {Bohlin} R.~C.,  {Kinney} A.~L.,  {Koornneef} J.,
   {Storchi-Bergmann} T.,  2000, \mn@doi [\apj] {10.1086/308692}, \href
  {http://adsabs.harvard.edu/abs/2000ApJ...533..682C} {533, 682}

\bibitem[\protect\citeauthoryear{{Cappellari}}{{Cappellari}}{2008}]{Cappellari08}
{Cappellari} M.,  2008, \mn@doi [\mnras] {10.1111/j.1365-2966.2008.13754.x},
  \href {https://ui.adsabs.harvard.edu/abs/2008MNRAS.390...71C} {390, 71}

\bibitem[\protect\citeauthoryear{{Cappellari}}{{Cappellari}}{2017}]{Cappellari17}
{Cappellari} M.,  2017, \mn@doi [\mnras] {10.1093/mnras/stw3020}, \href
  {https://ui.adsabs.harvard.edu/abs/2017MNRAS.466..798C} {466, 798}

\bibitem[\protect\citeauthoryear{{Cappellari} \& {Emsellem}}{{Cappellari} \&
  {Emsellem}}{2004}]{Cappellari04}
{Cappellari} M.,  {Emsellem} E.,  2004, \mn@doi [\pasp] {10.1086/381875}, \href
  {https://ui.adsabs.harvard.edu/abs/2004PASP..116..138C} {116, 138}

\bibitem[\protect\citeauthoryear{{Cappellari} et~al.,}{{Cappellari}
  et~al.}{2006}]{Cappellari06}
{Cappellari} M.,  et~al., 2006, \mn@doi [\mnras]
  {10.1111/j.1365-2966.2005.09981.x}, \href
  {https://ui.adsabs.harvard.edu/abs/2006MNRAS.366.1126C} {366, 1126}

\bibitem[\protect\citeauthoryear{{Cappellari} et~al.,}{{Cappellari}
  et~al.}{2011}]{Cappellari11}
{Cappellari} M.,  et~al., 2011, \mn@doi [\mnras]
  {10.1111/j.1365-2966.2010.18174.x}, \href
  {https://ui.adsabs.harvard.edu/abs/2011MNRAS.413..813C} {413, 813}

\bibitem[\protect\citeauthoryear{{Cappellari} et~al.,}{{Cappellari}
  et~al.}{2013}]{Cappellari13}
{Cappellari} M.,  et~al., 2013, \mn@doi [\mnras] {10.1093/mnras/stt562}, \href
  {http://adsabs.harvard.edu/abs/2013MNRAS.432.1709C} {432, 1709}

\bibitem[\protect\citeauthoryear{{Carniani} et~al.,}{{Carniani}
  et~al.}{2015}]{Carniani15}
{Carniani} S.,  et~al., 2015, \mn@doi [\aap] {10.1051/0004-6361/201526557},
  \href {https://ui.adsabs.harvard.edu/abs/2015A&A...580A.102C} {580, A102}

\bibitem[\protect\citeauthoryear{{Chabrier}}{{Chabrier}}{2003}]{Chabrier03}
{Chabrier} G.,  2003, \mn@doi [Publications of the Astronomical Society of the
  Pacific] {10.1086/376392}, \href
  {https://ui.adsabs.harvard.edu/\#abs/2003PASP..115..763C} {115, 763}

\bibitem[\protect\citeauthoryear{{Civano} et~al.,}{{Civano}
  et~al.}{2012}]{Civano12}
{Civano} F.,  et~al., 2012, \mn@doi [\apjs] {10.1088/0067-0049/201/2/30}, \href
  {https://ui.adsabs.harvard.edu/\#abs/2012ApJS..201...30C} {201, 30}

\bibitem[\protect\citeauthoryear{{Concas} et~al.,}{{Concas}
  et~al.}{2022}]{Concas22}
{Concas} A.,  et~al., 2022, \mn@doi [\mnras] {10.1093/mnras/stac1026}, \href
  {https://ui.adsabs.harvard.edu/abs/2022MNRAS.513.2535C} {513, 2535}

\bibitem[\protect\citeauthoryear{{Cortese}, {Catinella}  \&
  {Janowiecki}}{{Cortese} et~al.}{2017}]{Cortese17}
{Cortese} L.,  {Catinella} B.,   {Janowiecki} S.,  2017, \mn@doi [\apjl]
  {10.3847/2041-8213/aa8cc3}, \href
  {https://ui.adsabs.harvard.edu/abs/2017ApJ...848L...7C} {848, L7}

\bibitem[\protect\citeauthoryear{{Cresci} et~al.,}{{Cresci}
  et~al.}{2009}]{Cresci09}
{Cresci} G.,  et~al., 2009, \mn@doi [\apj] {10.1088/0004-637X/697/1/115}, \href
  {http://adsabs.harvard.edu/abs/2009ApJ...697..115C} {697, 115}

\bibitem[\protect\citeauthoryear{{Crespo G{\'o}mez}, {Piqueras L{\'o}pez},
  {Arribas}, {Pereira-Santaella}, {Colina}  \& {Rodr{\'\i}guez del
  Pino}}{{Crespo G{\'o}mez} et~al.}{2021}]{CrespoGomez21}
{Crespo G{\'o}mez} A.,  {Piqueras L{\'o}pez} J.,  {Arribas} S.,
  {Pereira-Santaella} M.,  {Colina} L.,   {Rodr{\'\i}guez del Pino} B.,  2021,
  \mn@doi [\aap] {10.1051/0004-6361/202039472}, \href
  {https://ui.adsabs.harvard.edu/abs/2021A&A...650A.149C} {650, A149}

\bibitem[\protect\citeauthoryear{{Davies} et~al.,}{{Davies}
  et~al.}{2011}]{Davies11}
{Davies} R.,  et~al., 2011, \mn@doi [\apj] {10.1088/0004-637X/741/2/69}, \href
  {http://adsabs.harvard.edu/abs/2011ApJ...741...69D} {741, 69}

\bibitem[\protect\citeauthoryear{{Davies}, {Kewley}, {Ho}  \&
  {Dopita}}{{Davies} et~al.}{2014}]{DaviesRL14}
{Davies} R.~L.,  {Kewley} L.~J.,  {Ho} I.~T.,   {Dopita} M.~A.,  2014, \mn@doi
  [\mnras] {10.1093/mnras/stu1740}, \href
  {https://ui.adsabs.harvard.edu/abs/2014MNRAS.444.3961D} {444, 3961}

\bibitem[\protect\citeauthoryear{{Davies} et~al.,}{{Davies}
  et~al.}{2019}]{DaviesRL19}
{Davies} R.~L.,  et~al., 2019, \mn@doi [\apj] {10.3847/1538-4357/ab06f1}, \href
  {https://ui.adsabs.harvard.edu/abs/2019ApJ...873..122D} {873, 122}

\bibitem[\protect\citeauthoryear{{Davis} et~al.,}{{Davis}
  et~al.}{2011}]{Davis11}
{Davis} T.~A.,  et~al., 2011, \mn@doi [\mnras]
  {10.1111/j.1365-2966.2011.19355.x}, \href
  {https://ui.adsabs.harvard.edu/abs/2011MNRAS.417..882D} {417, 882}

\bibitem[\protect\citeauthoryear{{Davis} et~al.,}{{Davis}
  et~al.}{2013}]{Davis13}
{Davis} T.~A.,  et~al., 2013, \mn@doi [\mnras] {10.1093/mnras/sts353}, \href
  {https://ui.adsabs.harvard.edu/abs/2013MNRAS.429..534D} {429, 534}

\bibitem[\protect\citeauthoryear{{Di Teodoro} \& {Fraternali}}{{Di Teodoro} \&
  {Fraternali}}{2015}]{DiTeodoro15}
{Di Teodoro} E.~M.,  {Fraternali} F.,  2015, \mn@doi [\mnras]
  {10.1093/mnras/stv1213}, \href
  {https://ui.adsabs.harvard.edu/abs/2015MNRAS.451.3021D} {451, 3021}

\bibitem[\protect\citeauthoryear{{Djorgovski} \& {Davis}}{{Djorgovski} \&
  {Davis}}{1987}]{Djorgovski87}
{Djorgovski} S.,  {Davis} M.,  1987, \mn@doi [\apj] {10.1086/164948}, \href
  {https://ui.adsabs.harvard.edu/abs/1987ApJ...313...59D} {313, 59}

\bibitem[\protect\citeauthoryear{{Dressler}, {Lynden-Bell}, {Burstein},
  {Davies}, {Faber}, {Terlevich}  \& {Wegner}}{{Dressler}
  et~al.}{1987}]{Dressler87}
{Dressler} A.,  {Lynden-Bell} D.,  {Burstein} D.,  {Davies} R.~L.,  {Faber}
  S.~M.,  {Terlevich} R.,   {Wegner} G.,  1987, \mn@doi [\apj]
  {10.1086/164947}, \href
  {https://ui.adsabs.harvard.edu/abs/1987ApJ...313...42D} {313, 42}

\bibitem[\protect\citeauthoryear{{Driver}, {Popescu}, {Tuffs}, {Liske},
  {Graham}, {Allen}  \& {de Propris}}{{Driver} et~al.}{2007}]{Driver07}
{Driver} S.~P.,  {Popescu} C.~C.,  {Tuffs} R.~J.,  {Liske} J.,  {Graham} A.~W.,
   {Allen} P.~D.,   {de Propris} R.,  2007, \mn@doi [\mnras]
  {10.1111/j.1365-2966.2007.11862.x}, \href
  {https://ui.adsabs.harvard.edu/abs/2007MNRAS.379.1022D} {379, 1022}

\bibitem[\protect\citeauthoryear{{Dutton} \& {Macci{\`o}}}{{Dutton} \&
  {Macci{\`o}}}{2014}]{Dutton14}
{Dutton} A.~A.,  {Macci{\`o}} A.~V.,  2014, \mn@doi [\mnras]
  {10.1093/mnras/stu742}, \href
  {http://adsabs.harvard.edu/abs/2014MNRAS.441.3359D} {441, 3359}

\bibitem[\protect\citeauthoryear{{Falc{\'o}n-Barroso}
  et~al.,}{{Falc{\'o}n-Barroso} et~al.}{2006}]{FalconBarroso06}
{Falc{\'o}n-Barroso} J.,  et~al., 2006, \mn@doi [\mnras]
  {10.1111/j.1365-2966.2006.10261.x}, \href
  {https://ui.adsabs.harvard.edu/abs/2006MNRAS.369..529F} {369, 529}

\bibitem[\protect\citeauthoryear{{Feltre}, {Charlot}  \& {Gutkin}}{{Feltre}
  et~al.}{2016}]{Feltre16}
{Feltre} A.,  {Charlot} S.,   {Gutkin} J.,  2016, \mn@doi [\mnras]
  {10.1093/mnras/stv2794}, \href
  {https://ui.adsabs.harvard.edu/abs/2016MNRAS.456.3354F} {456, 3354}

\bibitem[\protect\citeauthoryear{{Foreman-Mackey}, {Hogg}, {Lang}  \&
  {Goodman}}{{Foreman-Mackey} et~al.}{2013}]{ForemanMackey13}
{Foreman-Mackey} D.,  {Hogg} D.~W.,  {Lang} D.,   {Goodman} J.,  2013, \mn@doi
  [\pasp] {10.1086/670067}, \href
  {https://ui.adsabs.harvard.edu/abs/2013PASP..125..306F} {125, 306}

\bibitem[\protect\citeauthoryear{{F{\"o}rster Schreiber} \&
  {Wuyts}}{{F{\"o}rster Schreiber} \& {Wuyts}}{2020}]{FS20}
{F{\"o}rster Schreiber} N.~M.,  {Wuyts} S.,  2020, \mn@doi [\araa]
  {10.1146/annurev-astro-032620-021910}, \href
  {https://ui.adsabs.harvard.edu/abs/2020ARA&A..58..661F} {58, 661}

\bibitem[\protect\citeauthoryear{{F{\"o}rster Schreiber} et~al.,}{{F{\"o}rster
  Schreiber} et~al.}{2018}]{FS18}
{F{\"o}rster Schreiber} N.~M.,  et~al., 2018, \mn@doi [\apjs]
  {10.3847/1538-4365/aadd49}, \href
  {https://ui.adsabs.harvard.edu/#abs/2018ApJS..238...21F} {238, 21}

\bibitem[\protect\citeauthoryear{{F{\"o}rster Schreiber} et~al.,}{{F{\"o}rster
  Schreiber} et~al.}{2019}]{FS19}
{F{\"o}rster Schreiber} N.~M.,  et~al., 2019, \mn@doi [\apj]
  {10.3847/1538-4357/ab0ca2}, \href
  {https://ui.adsabs.harvard.edu/abs/2019ApJ...875...21F} {875, 21}

\bibitem[\protect\citeauthoryear{{Franx}, {Illingworth}  \& {de Zeeuw}}{{Franx}
  et~al.}{1991}]{Franx91}
{Franx} M.,  {Illingworth} G.,   {de Zeeuw} T.,  1991, \mn@doi [\apj]
  {10.1086/170769}, \href
  {https://ui.adsabs.harvard.edu/abs/1991ApJ...383..112F} {383, 112}

\bibitem[\protect\citeauthoryear{{Freundlich} et~al.,}{{Freundlich}
  et~al.}{2019}]{Freundlich19}
{Freundlich} J.,  et~al., 2019, \mn@doi [\aap] {10.1051/0004-6361/201732223},
  \href {https://ui.adsabs.harvard.edu/abs/2019A&A...622A.105F} {622, A105}

\bibitem[\protect\citeauthoryear{{Genzel} et~al.,}{{Genzel}
  et~al.}{2006}]{Genzel06}
{Genzel} R.,  et~al., 2006, \mn@doi [\nat] {10.1038/nature05052}, \href
  {http://adsabs.harvard.edu/abs/2006Natur.442..786G} {442, 786}

\bibitem[\protect\citeauthoryear{{Genzel} et~al.,}{{Genzel}
  et~al.}{2008}]{Genzel08}
{Genzel} R.,  et~al., 2008, \mn@doi [\apj] {10.1086/591840}, \href
  {http://adsabs.harvard.edu/abs/2008ApJ...687...59G} {687, 59}

\bibitem[\protect\citeauthoryear{{Genzel} et~al.,}{{Genzel}
  et~al.}{2011}]{Genzel11}
{Genzel} R.,  et~al., 2011, \mn@doi [\apj] {10.1088/0004-637X/733/2/101}, \href
  {http://adsabs.harvard.edu/abs/2011ApJ...733..101G} {733, 101}

\bibitem[\protect\citeauthoryear{{Genzel} et~al.,}{{Genzel}
  et~al.}{2013}]{Genzel13}
{Genzel} R.,  et~al., 2013, \mn@doi [\apj] {10.1088/0004-637X/773/1/68}, \href
  {http://adsabs.harvard.edu/abs/2013ApJ...773...68G} {773, 68}

\bibitem[\protect\citeauthoryear{{Genzel} et~al.,}{{Genzel}
  et~al.}{2014}]{Genzel14}
{Genzel} R.,  et~al., 2014, \mn@doi [\apj] {10.1088/0004-637X/796/1/7}, \href
  {https://ui.adsabs.harvard.edu/abs/2014ApJ...796....7G} {796, 7}

\bibitem[\protect\citeauthoryear{{Genzel} et~al.,}{{Genzel}
  et~al.}{2017}]{Genzel17}
{Genzel} R.,  et~al., 2017, \mn@doi [\nat] {10.1038/nature21685}, \href
  {http://adsabs.harvard.edu/abs/2017Natur.543..397G} {543, 397}

\bibitem[\protect\citeauthoryear{{Genzel} et~al.,}{{Genzel}
  et~al.}{2020}]{Genzel20}
{Genzel} R.,  et~al., 2020, \mn@doi [\apj] {10.3847/1538-4357/abb0ea}, \href
  {https://ui.adsabs.harvard.edu/abs/2020ApJ...902...98G} {902, 98}

\bibitem[\protect\citeauthoryear{{Girard}, {Dessauges-Zavadsky}, {Combes},
  {Chisholm}, {Patr{\'\i}cio}, {Richard}  \& {Schaerer}}{{Girard}
  et~al.}{2019}]{Girard19}
{Girard} M.,  {Dessauges-Zavadsky} M.,  {Combes} F.,  {Chisholm} J.,
  {Patr{\'\i}cio} V.,  {Richard} J.,   {Schaerer} D.,  2019, \mn@doi [\aap]
  {10.1051/0004-6361/201935896}, \href
  {https://ui.adsabs.harvard.edu/abs/2019A&A...631A..91G} {631, A91}

\bibitem[\protect\citeauthoryear{{Girard} et~al.,}{{Girard}
  et~al.}{2021}]{Girard21}
{Girard} M.,  et~al., 2021, \mn@doi [\apj] {10.3847/1538-4357/abd5b9}, \href
  {https://ui.adsabs.harvard.edu/abs/2021ApJ...909...12G} {909, 12}

\bibitem[\protect\citeauthoryear{{Graham} \& {Worley}}{{Graham} \&
  {Worley}}{2008}]{Graham08}
{Graham} A.~W.,  {Worley} C.~C.,  2008, \mn@doi [\mnras]
  {10.1111/j.1365-2966.2008.13506.x}, \href
  {https://ui.adsabs.harvard.edu/abs/2008MNRAS.388.1708G} {388, 1708}

\bibitem[\protect\citeauthoryear{{Grand}, {Springel}, {G{\'o}mez}, {Marinacci},
  {Pakmor}, {Campbell}  \& {Jenkins}}{{Grand} et~al.}{2016}]{Grand16}
{Grand} R. J.~J.,  {Springel} V.,  {G{\'o}mez} F.~A.,  {Marinacci} F.,
  {Pakmor} R.,  {Campbell} D. J.~R.,   {Jenkins} A.,  2016, \mn@doi [\mnras]
  {10.1093/mnras/stw601}, \href
  {https://ui.adsabs.harvard.edu/\#abs/2016MNRAS.459..199G} {459, 199}

\bibitem[\protect\citeauthoryear{{Grogin} et~al.,}{{Grogin}
  et~al.}{2011}]{Grogin11}
{Grogin} N.~A.,  et~al., 2011, \mn@doi [\apjs] {10.1088/0067-0049/197/2/35},
  \href {http://adsabs.harvard.edu/abs/2011ApJS..197...35G} {197, 35}

\bibitem[\protect\citeauthoryear{{Gu{\'e}rou} et~al.,}{{Gu{\'e}rou}
  et~al.}{2017}]{Guerou17}
{Gu{\'e}rou} A.,  et~al., 2017, \mn@doi [\aap] {10.1051/0004-6361/201730905},
  \href {https://ui.adsabs.harvard.edu/abs/2017A&A...608A...5G} {608, A5}

\bibitem[\protect\citeauthoryear{{Harrison} et~al.,}{{Harrison}
  et~al.}{2016}]{Harrison16}
{Harrison} C.~M.,  et~al., 2016, \mn@doi [\mnras] {10.1093/mnras/stv2727},
  \href {https://ui.adsabs.harvard.edu/abs/2016MNRAS.456.1195H} {456, 1195}

\bibitem[\protect\citeauthoryear{{Hippelein}}{{Hippelein}}{1986}]{Hippelein86}
{Hippelein} H.~H.,  1986, \aap, \href
  {https://ui.adsabs.harvard.edu/abs/1986A&A...160..374H} {160, 374}

\bibitem[\protect\citeauthoryear{{Hyde} \& {Bernardi}}{{Hyde} \&
  {Bernardi}}{2009}]{Hyde09}
{Hyde} J.~B.,  {Bernardi} M.,  2009, \mn@doi [\mnras]
  {10.1111/j.1365-2966.2009.14783.x}, \href
  {https://ui.adsabs.harvard.edu/abs/2009MNRAS.396.1171H} {396, 1171}

\bibitem[\protect\citeauthoryear{{Jenkins} \& {Binney}}{{Jenkins} \&
  {Binney}}{1990}]{Jenkins90}
{Jenkins} A.,  {Binney} J.,  1990, \mnras, \href
  {https://ui.adsabs.harvard.edu/abs/1990MNRAS.245..305J} {245, 305}

\bibitem[\protect\citeauthoryear{{J{\'o}zsa}, {Kenn}, {Klein}  \&
  {Oosterloo}}{{J{\'o}zsa} et~al.}{2007}]{Jozsa07}
{J{\'o}zsa} G.~I.~G.,  {Kenn} F.,  {Klein} U.,   {Oosterloo} T.~A.,  2007,
  \mn@doi [\aap] {10.1051/0004-6361:20066164}, \href
  {https://ui.adsabs.harvard.edu/abs/2007A&A...468..731J} {468, 731}

\bibitem[\protect\citeauthoryear{{Kaasinen} et~al.,}{{Kaasinen}
  et~al.}{2020}]{Kaasinen20}
{Kaasinen} M.,  et~al., 2020, \mn@doi [\apj] {10.3847/1538-4357/aba438}, \href
  {https://ui.adsabs.harvard.edu/abs/2020ApJ...899...37K} {899, 37}

\bibitem[\protect\citeauthoryear{{Kakkad} et~al.,}{{Kakkad}
  et~al.}{2020}]{Kakkad20}
{Kakkad} D.,  et~al., 2020, \mn@doi [\aap] {10.1051/0004-6361/202038551}, \href
  {https://ui.adsabs.harvard.edu/abs/2020A&A...642A.147K} {642, A147}

\bibitem[\protect\citeauthoryear{{Kauffmann} et~al.,}{{Kauffmann}
  et~al.}{2003}]{Kauffmann03}
{Kauffmann} G.,  et~al., 2003, \mn@doi [\mnras]
  {10.1111/j.1365-2966.2003.07154.x}, \href
  {https://ui.adsabs.harvard.edu/abs/2003MNRAS.346.1055K} {346, 1055}

\bibitem[\protect\citeauthoryear{{Kewley}, {Nicholls}  \&
  {Sutherland}}{{Kewley} et~al.}{2019}]{Kewley19}
{Kewley} L.~J.,  {Nicholls} D.~C.,   {Sutherland} R.~S.,  2019, \mn@doi [\araa]
  {10.1146/annurev-astro-081817-051832}, \href
  {https://ui.adsabs.harvard.edu/abs/2019ARA&A..57..511K} {57, 511}

\bibitem[\protect\citeauthoryear{{Khim}, {Yi}, {Pichon}, {Dubois}, {Devriendt},
  {Choi}, {Bryant}  \& {Croom}}{{Khim} et~al.}{2021}]{Khim21}
{Khim} D.~J.,  {Yi} S.~K.,  {Pichon} C.,  {Dubois} Y.,  {Devriendt} J.,  {Choi}
  H.,  {Bryant} J.~J.,   {Croom} S.~M.,  2021, \mn@doi [\apjs]
  {10.3847/1538-4365/abf043}, \href
  {https://ui.adsabs.harvard.edu/abs/2021ApJS..254...27K} {254, 27}

\bibitem[\protect\citeauthoryear{{Koekemoer} et~al.,}{{Koekemoer}
  et~al.}{2011}]{Koekemoer11}
{Koekemoer} A.~M.,  et~al., 2011, \mn@doi [\apjs] {10.1088/0067-0049/197/2/36},
  \href {http://adsabs.harvard.edu/abs/2011ApJS..197...36K} {197, 36}

\bibitem[\protect\citeauthoryear{{Lagos}, {Schaye}, {Bah{\'e}}, {van de Sande},
  {Kay}, {Barnes}, {Davis}  \& {Dalla Vecchia}}{{Lagos} et~al.}{2018}]{Lagos18}
{Lagos} C. d.~P.,  {Schaye} J.,  {Bah{\'e}} Y.,  {van de Sande} J.,  {Kay}
  S.~T.,  {Barnes} D.,  {Davis} T.~A.,   {Dalla Vecchia} C.,  2018, \mn@doi
  [\mnras] {10.1093/mnras/sty489}, \href
  {https://ui.adsabs.harvard.edu/abs/2018MNRAS.476.4327L} {476, 4327}

\bibitem[\protect\citeauthoryear{{Lang} et~al.,}{{Lang} et~al.}{2014}]{Lang14}
{Lang} P.,  et~al., 2014, \mn@doi [\apj] {10.1088/0004-637X/788/1/11}, \href
  {http://adsabs.harvard.edu/abs/2014ApJ...788...11L} {788, 11}

\bibitem[\protect\citeauthoryear{{Lang} et~al.,}{{Lang} et~al.}{2017}]{Lang17}
{Lang} P.,  et~al., 2017, \mn@doi [\apj] {10.3847/1538-4357/aa6d82}, \href
  {http://adsabs.harvard.edu/abs/2017ApJ...840...92L} {840, 92}

\bibitem[\protect\citeauthoryear{{Law} et~al.,}{{Law} et~al.}{2022}]{Law22}
{Law} D.~R.,  et~al., 2022, \mn@doi [\apj] {10.3847/1538-4357/ac5620}, \href
  {https://ui.adsabs.harvard.edu/abs/2022ApJ...928...58L} {928, 58}

\bibitem[\protect\citeauthoryear{{Le F{\`e}vre} et~al.,}{{Le F{\`e}vre}
  et~al.}{2003}]{LeFevre03}
{Le F{\`e}vre} O.,  et~al., 2003, in {Iye} M.,  {Moorwood} A. F.~M.,  eds,
  Society of Photo-Optical Instrumentation Engineers (SPIE) Conference Series
  Vol. 4841, Instrument Design and Performance for Optical/Infrared
  Ground-based Telescopes. pp 1670--1681, \mn@doi{10.1117/12.460959}

\bibitem[\protect\citeauthoryear{{Lelli}, {De Breuck}, {Falkendal},
  {Fraternali}, {Man}, {Nesvadba}  \& {Lehnert}}{{Lelli}
  et~al.}{2018}]{Lelli18}
{Lelli} F.,  {De Breuck} C.,  {Falkendal} T.,  {Fraternali} F.,  {Man} A.
  W.~S.,  {Nesvadba} N. P.~H.,   {Lehnert} M.~D.,  2018, \mn@doi [\mnras]
  {10.1093/mnras/sty1795}, \href
  {https://ui.adsabs.harvard.edu/abs/2018MNRAS.479.5440L} {479, 5440}

\bibitem[\protect\citeauthoryear{{Lelli} et~al.,}{{Lelli}
  et~al.}{2023}]{Lelli23}
{Lelli} F.,  et~al., 2023, \mn@doi [\aap] {10.1051/0004-6361/202245105}, \href
  {https://ui.adsabs.harvard.edu/abs/2023A&A...672A.106L} {672, A106}

\bibitem[\protect\citeauthoryear{{Lemaux}, {Lubin}, {Shapley}, {Kocevski},
  {Gal}  \& {Squires}}{{Lemaux} et~al.}{2010}]{Lemaux10}
{Lemaux} B.~C.,  {Lubin} L.~M.,  {Shapley} A.,  {Kocevski} D.,  {Gal} R.~R.,
  {Squires} G.~K.,  2010, \mn@doi [\apj] {10.1088/0004-637X/716/2/970}, \href
  {https://ui.adsabs.harvard.edu/abs/2010ApJ...716..970L} {716, 970}

\bibitem[\protect\citeauthoryear{{Leung} et~al.,}{{Leung}
  et~al.}{2018}]{Leung18}
{Leung} G. Y.~C.,  et~al., 2018, \mn@doi [\mnras] {10.1093/mnras/sty288}, \href
  {https://ui.adsabs.harvard.edu/abs/2018MNRAS.477..254L} {477, 254}

\bibitem[\protect\citeauthoryear{{Levy} et~al.,}{{Levy} et~al.}{2018}]{Levy18}
{Levy} R.~C.,  et~al., 2018, \mn@doi [\apj] {10.3847/1538-4357/aac2e5}, \href
  {https://ui.adsabs.harvard.edu/\#abs/2018ApJ...860...92L} {860, 92}

\bibitem[\protect\citeauthoryear{{Liu} et~al.,}{{Liu} et~al.}{2023}]{Liu23}
{Liu} D.,  et~al., 2023, \mn@doi [\apj] {10.3847/1538-4357/aca46b}, \href
  {https://ui.adsabs.harvard.edu/abs/2023ApJ...942...98L} {942, 98}

\bibitem[\protect\citeauthoryear{{Lutz} et~al.,}{{Lutz} et~al.}{2011}]{Lutz11}
{Lutz} D.,  et~al., 2011, \mn@doi [\aap] {10.1051/0004-6361/201117107}, \href
  {http://adsabs.harvard.edu/abs/2011A%26A...532A..90L} {532, A90}

\bibitem[\protect\citeauthoryear{{Magnelli} et~al.,}{{Magnelli}
  et~al.}{2013}]{Magnelli13}
{Magnelli} B.,  et~al., 2013, \mn@doi [\aap] {10.1051/0004-6361/201321371},
  \href {http://adsabs.harvard.edu/abs/2013A%26A...553A.132M} {553, A132}

\bibitem[\protect\citeauthoryear{{Martinsson}, {Verheijen}, {Westfall},
  {Bershady}, {Schechtman-Rook}, {Andersen}  \& {Swaters}}{{Martinsson}
  et~al.}{2013}]{Martinsson13a}
{Martinsson} T.~P.~K.,  {Verheijen} M.~A.~W.,  {Westfall} K.~B.,  {Bershady}
  M.~A.,  {Schechtman-Rook} A.,  {Andersen} D.~R.,   {Swaters} R.~A.,  2013,
  \mn@doi [\aap] {10.1051/0004-6361/201220515}, \href
  {http://adsabs.harvard.edu/abs/2013A%26A...557A.130M} {557, A130}

\bibitem[\protect\citeauthoryear{{Maseda} et~al.,}{{Maseda}
  et~al.}{2021}]{Maseda21}
{Maseda} M.~V.,  et~al., 2021, \mn@doi [\apj] {10.3847/1538-4357/ac2bfe}, \href
  {https://ui.adsabs.harvard.edu/abs/2021ApJ...923...18M} {923, 18}

\bibitem[\protect\citeauthoryear{{McLean} et~al.,}{{McLean}
  et~al.}{2010}]{McLean10}
{McLean} I.~S.,  et~al., 2010, in {McLean} I.~S.,  {Ramsay} S.~K.,   {Takami}
  H.,  eds,  Society of Photo-Optical Instrumentation Engineers (SPIE)
  Conference Series Vol. 7735, Ground-based and Airborne Instrumentation for
  Astronomy III. p. 77351E, \mn@doi{10.1117/12.856715}

\bibitem[\protect\citeauthoryear{{McLean} et~al.,}{{McLean}
  et~al.}{2012}]{McLean12}
{McLean} I.~S.,  et~al., 2012, in {McLean} I.~S.,  {Ramsay} S.~K.,   {Takami}
  H.,  eds,  Society of Photo-Optical Instrumentation Engineers (SPIE)
  Conference Series Vol. 8446, Ground-based and Airborne Instrumentation for
  Astronomy IV. p. 84460J, \mn@doi{10.1117/12.924794}

\bibitem[\protect\citeauthoryear{{Mendel} et~al.,}{{Mendel}
  et~al.}{2020}]{Mendel20}
{Mendel} J.~T.,  et~al., 2020, \mn@doi [\apj] {10.3847/1538-4357/ab9ffc}, \href
  {https://ui.adsabs.harvard.edu/abs/2020ApJ...899...87M} {899, 87}

\bibitem[\protect\citeauthoryear{{Mignoli} et~al.,}{{Mignoli}
  et~al.}{2013}]{Mignoli13}
{Mignoli} M.,  et~al., 2013, \mn@doi [\aap] {10.1051/0004-6361/201220846},
  \href {https://ui.adsabs.harvard.edu/abs/2013A&A...556A..29M} {556, A29}

\bibitem[\protect\citeauthoryear{{Miller}, {Bundy}, {Sullivan}, {Ellis}  \&
  {Treu}}{{Miller} et~al.}{2011}]{Miller11}
{Miller} S.~H.,  {Bundy} K.,  {Sullivan} M.,  {Ellis} R.~S.,   {Treu} T.,
  2011, \mn@doi [\apj] {10.1088/0004-637X/741/2/115}, \href
  {http://adsabs.harvard.edu/abs/2011ApJ...741..115M} {741, 115}

\bibitem[\protect\citeauthoryear{{Miller}, {Ellis}, {Sullivan}, {Bundy},
  {Newman}  \& {Treu}}{{Miller} et~al.}{2012}]{Miller12}
{Miller} S.~H.,  {Ellis} R.~S.,  {Sullivan} M.,  {Bundy} K.,  {Newman} A.~B.,
  {Treu} T.,  2012, \mn@doi [\apj] {10.1088/0004-637X/753/1/74}, \href
  {http://adsabs.harvard.edu/abs/2012ApJ...753...74M} {753, 74}

\bibitem[\protect\citeauthoryear{{Molina}, {Ibar}, {Smail}, {Swinbank},
  {Villard}, {Escala}, {Sobral}  \& {Hughes}}{{Molina} et~al.}{2019}]{Molina19}
{Molina} J.,  {Ibar} E.,  {Smail} I.,  {Swinbank} A.~M.,  {Villard} E.,
  {Escala} A.,  {Sobral} D.,   {Hughes} T.~M.,  2019, \mn@doi [\mnras]
  {10.1093/mnras/stz1643}, \href
  {https://ui.adsabs.harvard.edu/abs/2019MNRAS.487.4856M} {487, 4856}

\bibitem[\protect\citeauthoryear{{Molina} et~al.,}{{Molina}
  et~al.}{2020}]{Molina20}
{Molina} J.,  et~al., 2020, \mn@doi [\aap] {10.1051/0004-6361/202039008}, \href
  {https://ui.adsabs.harvard.edu/abs/2020A&A...643A..78M} {643, A78}

\bibitem[\protect\citeauthoryear{{Momcheva} et~al.,}{{Momcheva}
  et~al.}{2016}]{Momcheva16}
{Momcheva} I.~G.,  et~al., 2016, \mn@doi [\apjs] {10.3847/0067-0049/225/2/27},
  \href {http://adsabs.harvard.edu/abs/2016ApJS..225...27M} {225, 27}

\bibitem[\protect\citeauthoryear{{Moster}, {Naab}  \& {White}}{{Moster}
  et~al.}{2018}]{Moster18}
{Moster} B.~P.,  {Naab} T.,   {White} S. D.~M.,  2018, \mn@doi [\mnras]
  {10.1093/mnras/sty655}, \href
  {https://ui.adsabs.harvard.edu/\#abs/2018MNRAS.477.1822M} {477, 1822}

\bibitem[\protect\citeauthoryear{{Moster}, {Naab}  \& {White}}{{Moster}
  et~al.}{2020}]{Moster20}
{Moster} B.~P.,  {Naab} T.,   {White} S. D.~M.,  2020, \mn@doi [\mnras]
  {10.1093/mnras/staa3019}, \href
  {https://ui.adsabs.harvard.edu/abs/2020MNRAS.499.4748M} {499, 4748}

\bibitem[\protect\citeauthoryear{{Mowla}, {Nelson}, {van Dokkum}  \&
  {Tadaki}}{{Mowla} et~al.}{2019}]{Mowla19}
{Mowla} L.~A.,  {Nelson} E.~J.,  {van Dokkum} P.,   {Tadaki} K.-i.,  2019,
  \mn@doi [\apjl] {10.3847/2041-8213/ab54d1}, \href
  {https://ui.adsabs.harvard.edu/abs/2019ApJ...886L..28M} {886, L28}

\bibitem[\protect\citeauthoryear{{Muzzin} et~al.,}{{Muzzin}
  et~al.}{2013}]{Muzzin13}
{Muzzin} A.,  et~al., 2013, \mn@doi [\apjs] {10.1088/0067-0049/206/1/8}, \href
  {https://ui.adsabs.harvard.edu/abs/2013ApJS..206....8M} {206, 8}

\bibitem[\protect\citeauthoryear{{Naab} et~al.,}{{Naab} et~al.}{2014}]{Naab14}
{Naab} T.,  et~al., 2014, \mn@doi [\mnras] {10.1093/mnras/stt1919}, \href
  {https://ui.adsabs.harvard.edu/abs/2014MNRAS.444.3357N} {444, 3357}

\bibitem[\protect\citeauthoryear{{Navarro}, {Frenk}  \& {White}}{{Navarro}
  et~al.}{1996}]{NFW96}
{Navarro} J.~F.,  {Frenk} C.~S.,   {White} S.~D.~M.,  1996, \mn@doi [\apj]
  {10.1086/177173}, \href {http://adsabs.harvard.edu/abs/1996ApJ...462..563N}
  {462, 563}

\bibitem[\protect\citeauthoryear{{Nelson} et~al.,}{{Nelson}
  et~al.}{2016}]{NelsonE16a}
{Nelson} E.~J.,  et~al., 2016, \mn@doi [\apjl] {10.3847/2041-8205/817/1/L9},
  \href {http://adsabs.harvard.edu/abs/2016ApJ...817L...9N} {817, L9}

\bibitem[\protect\citeauthoryear{{Nelson} et~al.,}{{Nelson}
  et~al.}{2021}]{NelsonE21}
{Nelson} E.~J.,  et~al., 2021, \mn@doi [\mnras] {10.1093/mnras/stab2131}, \href
  {https://ui.adsabs.harvard.edu/abs/2021MNRAS.508..219N} {508, 219}

\bibitem[\protect\citeauthoryear{{Nestor Shachar} et~al.,}{{Nestor Shachar}
  et~al.}{2022}]{Nestor22}
{Nestor Shachar} A.,  et~al., 2022, arXiv e-prints, \href
  {https://ui.adsabs.harvard.edu/abs/2022arXiv220912199N} {p. arXiv:2209.12199}

\bibitem[\protect\citeauthoryear{{Newman} et~al.,}{{Newman}
  et~al.}{2012}]{Newman12b}
{Newman} S.~F.,  et~al., 2012, \mn@doi [\apj] {10.1088/0004-637X/761/1/43},
  \href {https://ui.adsabs.harvard.edu/abs/2012ApJ...761...43N} {761, 43}

\bibitem[\protect\citeauthoryear{{Newman}, {Belli}  \& {Ellis}}{{Newman}
  et~al.}{2015}]{Newman15}
{Newman} A.~B.,  {Belli} S.,   {Ellis} R.~S.,  2015, \mn@doi [\apjl]
  {10.1088/2041-8205/813/1/L7}, \href
  {https://ui.adsabs.harvard.edu/abs/2015ApJ...813L...7N} {813, L7}

\bibitem[\protect\citeauthoryear{{Newman}, {Belli}, {Ellis}  \&
  {Patel}}{{Newman} et~al.}{2018}]{Newman18}
{Newman} A.~B.,  {Belli} S.,  {Ellis} R.~S.,   {Patel} S.~G.,  2018, \mn@doi
  [\apj] {10.3847/1538-4357/aacd4f}, \href
  {https://ui.adsabs.harvard.edu/abs/2018ApJ...862..126N} {862, 126}

\bibitem[\protect\citeauthoryear{{Noordermeer}}{{Noordermeer}}{2008}]{Noordermeer08}
{Noordermeer} E.,  2008, \mn@doi [\mnras] {10.1111/j.1365-2966.2008.12837.x},
  \href {http://adsabs.harvard.edu/abs/2008MNRAS.385.1359N} {385, 1359}

\bibitem[\protect\citeauthoryear{{Peng}, {Ho}, {Impey}  \& {Rix}}{{Peng}
  et~al.}{2010}]{Peng10}
{Peng} C.~Y.,  {Ho} L.~C.,  {Impey} C.~D.,   {Rix} H.-W.,  2010, \mn@doi [\aj]
  {10.1088/0004-6256/139/6/2097}, \href
  {http://adsabs.harvard.edu/abs/2010AJ....139.2097P} {139, 2097}

\bibitem[\protect\citeauthoryear{{Pillepich} et~al.,}{{Pillepich}
  et~al.}{2019}]{Pillepich19}
{Pillepich} A.,  et~al., 2019, \mn@doi [\mnras] {10.1093/mnras/stz2338}, \href
  {https://ui.adsabs.harvard.edu/abs/2019MNRAS.490.3196P} {490, 3196}

\bibitem[\protect\citeauthoryear{{Price} et~al.,}{{Price}
  et~al.}{2021}]{Price21}
{Price} S.~H.,  et~al., 2021, \mn@doi [\apj] {10.3847/1538-4357/ac22ad}, \href
  {https://ui.adsabs.harvard.edu/abs/2021ApJ...922..143P} {922, 143}

\bibitem[\protect\citeauthoryear{{Price} et~al.,}{{Price}
  et~al.}{2022}]{Price22}
{Price} S.~H.,  et~al., 2022, arXiv e-prints, \href
  {https://ui.adsabs.harvard.edu/abs/2022arXiv220706442P} {p. arXiv:2207.06442}

\bibitem[\protect\citeauthoryear{{Rizzo} et~al.,}{{Rizzo}
  et~al.}{2023}]{Rizzo23}
{Rizzo} F.,  et~al., 2023, \mn@doi [arXiv e-prints]
  {10.48550/arXiv.2303.16227}, \href
  {https://ui.adsabs.harvard.edu/abs/2023arXiv230316227R} {p. arXiv:2303.16227}

\bibitem[\protect\citeauthoryear{{R{\"o}ttgers}, {Naab}  \&
  {Oser}}{{R{\"o}ttgers} et~al.}{2014}]{Roettgers14}
{R{\"o}ttgers} B.,  {Naab} T.,   {Oser} L.,  2014, \mn@doi [\mnras]
  {10.1093/mnras/stu1762}, \href
  {https://ui.adsabs.harvard.edu/abs/2014MNRAS.445.1065R} {445, 1065}

\bibitem[\protect\citeauthoryear{{Rubin}, {Weiner}, {Koo}, {Martin},
  {Prochaska}, {Coil}  \& {Newman}}{{Rubin} et~al.}{2010}]{Rubin10}
{Rubin} K. H.~R.,  {Weiner} B.~J.,  {Koo} D.~C.,  {Martin} C.~L.,  {Prochaska}
  J.~X.,  {Coil} A.~L.,   {Newman} J.~A.,  2010, \mn@doi [\apj]
  {10.1088/0004-637X/719/2/1503}, \href
  {https://ui.adsabs.harvard.edu/abs/2010ApJ...719.1503R} {719, 1503}

\bibitem[\protect\citeauthoryear{{Sales}, {Navarro}, {Theuns}, {Schaye},
  {White}, {Frenk}, {Crain}  \& {Dalla Vecchia}}{{Sales}
  et~al.}{2012}]{Sales12}
{Sales} L.~V.,  {Navarro} J.~F.,  {Theuns} T.,  {Schaye} J.,  {White} S. D.~M.,
   {Frenk} C.~S.,  {Crain} R.~A.,   {Dalla Vecchia} C.,  2012, \mn@doi [\mnras]
  {10.1111/j.1365-2966.2012.20975.x}, \href
  {https://ui.adsabs.harvard.edu/abs/2012MNRAS.423.1544S} {423, 1544}

\bibitem[\protect\citeauthoryear{{S{\'a}nchez} et~al.,}{{S{\'a}nchez}
  et~al.}{2012}]{Sanchez12}
{S{\'a}nchez} S.~F.,  et~al., 2012, \mn@doi [\aap]
  {10.1051/0004-6361/201117353}, \href
  {https://ui.adsabs.harvard.edu/abs/2012A&A...538A...8S} {538, A8}

\bibitem[\protect\citeauthoryear{{Sarzi} et~al.,}{{Sarzi}
  et~al.}{2006}]{Sarzi06}
{Sarzi} M.,  et~al., 2006, \mn@doi [\mnras] {10.1111/j.1365-2966.2005.09839.x},
  \href {https://ui.adsabs.harvard.edu/abs/2006MNRAS.366.1151S} {366, 1151}

\bibitem[\protect\citeauthoryear{{Schaye} et~al.,}{{Schaye}
  et~al.}{2015}]{Schaye15}
{Schaye} J.,  et~al., 2015, \mn@doi [\mnras] {10.1093/mnras/stu2058}, \href
  {http://adsabs.harvard.edu/abs/2015MNRAS.446..521S} {446, 521}

\bibitem[\protect\citeauthoryear{{Scoville} et~al.,}{{Scoville}
  et~al.}{2007}]{Scoville07}
{Scoville} N.,  et~al., 2007, \mn@doi [\apjs] {10.1086/516585}, \href
  {https://ui.adsabs.harvard.edu/abs/2007ApJS..172....1S} {172, 1}

\bibitem[\protect\citeauthoryear{{Serra} et~al.,}{{Serra}
  et~al.}{2014}]{Serra14}
{Serra} P.,  et~al., 2014, \mn@doi [\mnras] {10.1093/mnras/stt2496}, \href
  {https://ui.adsabs.harvard.edu/abs/2014MNRAS.444.3388S} {444, 3388}

\bibitem[\protect\citeauthoryear{{Shapley}, {Steidel}, {Pettini}  \&
  {Adelberger}}{{Shapley} et~al.}{2003}]{Shapley03}
{Shapley} A.~E.,  {Steidel} C.~C.,  {Pettini} M.,   {Adelberger} K.~L.,  2003,
  \mn@doi [\apj] {10.1086/373922}, \href
  {https://ui.adsabs.harvard.edu/abs/2003ApJ...588...65S} {588, 65}

\bibitem[\protect\citeauthoryear{{Sharples} et~al.,}{{Sharples}
  et~al.}{2004}]{Sharples04}
{Sharples} R.~M.,  et~al., 2004, in {Moorwood} A. F.~M.,  {Iye} M.,  eds,
  Society of Photo-Optical Instrumentation Engineers (SPIE) Conference Series
  Vol. 5492, Ground-based Instrumentation for Astronomy. pp 1179--1186,
  \mn@doi{10.1117/12.550495}

\bibitem[\protect\citeauthoryear{{Sharples} et~al.,}{{Sharples}
  et~al.}{2013}]{Sharples13}
{Sharples} R.,  et~al., 2013, The Messenger, \href
  {http://adsabs.harvard.edu/abs/2013Msngr.151...21S} {151, 21}

\bibitem[\protect\citeauthoryear{{Skelton} et~al.,}{{Skelton}
  et~al.}{2014}]{Skelton14}
{Skelton} R.~E.,  et~al., 2014, \mn@doi [\apjs] {10.1088/0067-0049/214/2/24},
  \href {http://adsabs.harvard.edu/abs/2014ApJS..214...24S} {214, 24}

\bibitem[\protect\citeauthoryear{{Sofue} \& {Rubin}}{{Sofue} \&
  {Rubin}}{2001}]{Sofue01}
{Sofue} Y.,  {Rubin} V.,  2001, \mn@doi [\araa]
  {10.1146/annurev.astro.39.1.137}, \href
  {http://adsabs.harvard.edu/abs/2001ARA%26A..39..137S} {39, 137}

\bibitem[\protect\citeauthoryear{{Straatman} et~al.,}{{Straatman}
  et~al.}{2018}]{Straatman18}
{Straatman} C. M.~S.,  et~al., 2018, \mn@doi [\apjs]
  {10.3847/1538-4365/aae37a}, \href
  {https://ui.adsabs.harvard.edu/abs/2018ApJS..239...27S} {239, 27}

\bibitem[\protect\citeauthoryear{{Straatman} et~al.,}{{Straatman}
  et~al.}{2022}]{Straatman22}
{Straatman} C. M.~S.,  et~al., 2022, \mn@doi [\apj] {10.3847/1538-4357/ac4e18},
  \href {https://ui.adsabs.harvard.edu/abs/2022ApJ...928..126S} {928, 126}

\bibitem[\protect\citeauthoryear{{Swinbank} et~al.,}{{Swinbank}
  et~al.}{2011}]{Swinbank11}
{Swinbank} A.~M.,  et~al., 2011, \mn@doi [\apj] {10.1088/0004-637X/742/1/11},
  \href {http://adsabs.harvard.edu/abs/2011ApJ...742...11S} {742, 11}

\bibitem[\protect\citeauthoryear{{Tacchella} et~al.,}{{Tacchella}
  et~al.}{2018}]{Tacchella18}
{Tacchella} S.,  et~al., 2018, \mn@doi [\apj] {10.3847/1538-4357/aabf8b}, \href
  {https://ui.adsabs.harvard.edu/abs/2018ApJ...859...56T} {859, 56}

\bibitem[\protect\citeauthoryear{{Tacconi} et~al.,}{{Tacconi}
  et~al.}{2010}]{Tacconi10}
{Tacconi} L.~J.,  et~al., 2010, \mn@doi [\nat] {10.1038/nature08773}, \href
  {http://adsabs.harvard.edu/abs/2010Natur.463..781T} {463, 781}

\bibitem[\protect\citeauthoryear{{Tacconi} et~al.,}{{Tacconi}
  et~al.}{2013}]{Tacconi13}
{Tacconi} L.~J.,  et~al., 2013, \mn@doi [\apj] {10.1088/0004-637X/768/1/74},
  \href {https://ui.adsabs.harvard.edu/#abs/2013ApJ...768...74T} {768, 74}

\bibitem[\protect\citeauthoryear{{Tacconi} et~al.,}{{Tacconi}
  et~al.}{2018}]{Tacconi18}
{Tacconi} L.~J.,  et~al., 2018, \mn@doi [\apj] {10.3847/1538-4357/aaa4b4},
  \href {http://adsabs.harvard.edu/abs/2018ApJ...853..179T} {853, 179}

\bibitem[\protect\citeauthoryear{{Tacconi}, {Genzel}  \& {Sternberg}}{{Tacconi}
  et~al.}{2020}]{Tacconi20}
{Tacconi} L.~J.,  {Genzel} R.,   {Sternberg} A.,  2020, \mn@doi [\araa]
  {10.1146/annurev-astro-082812-141034}, \href
  {https://ui.adsabs.harvard.edu/abs/2020ARA&A..58..157T} {58, 157}

\bibitem[\protect\citeauthoryear{{Talia} et~al.,}{{Talia}
  et~al.}{2017}]{Talia17}
{Talia} M.,  et~al., 2017, \mn@doi [\mnras] {10.1093/mnras/stx1788}, \href
  {https://ui.adsabs.harvard.edu/abs/2017MNRAS.471.4527T} {471, 4527}

\bibitem[\protect\citeauthoryear{{Taylor}, {Franx}, {Brinchmann}, {van der Wel}
   \& {van Dokkum}}{{Taylor} et~al.}{2010}]{Taylor10}
{Taylor} E.~N.,  {Franx} M.,  {Brinchmann} J.,  {van der Wel} A.,   {van
  Dokkum} P.~G.,  2010, \mn@doi [\apj] {10.1088/0004-637X/722/1/1}, \href
  {https://ui.adsabs.harvard.edu/abs/2010ApJ...722....1T} {722, 1}

\bibitem[\protect\citeauthoryear{{Tiley} et~al.,}{{Tiley}
  et~al.}{2016}]{Tiley16}
{Tiley} A.~L.,  et~al., 2016, \mn@doi [\mnras] {10.1093/mnras/stw936}, \href
  {http://adsabs.harvard.edu/abs/2016MNRAS.460..103T} {460, 103}

\bibitem[\protect\citeauthoryear{{Toft} et~al.,}{{Toft} et~al.}{2017}]{Toft17}
{Toft} S.,  et~al., 2017, \mn@doi [\nat] {10.1038/nature22388}, \href
  {https://ui.adsabs.harvard.edu/abs/2017Natur.546..510T} {546, 510}

\bibitem[\protect\citeauthoryear{{Tully} \& {Fisher}}{{Tully} \&
  {Fisher}}{1977}]{Tully77}
{Tully} R.~B.,  {Fisher} J.~R.,  1977, \aap, \href
  {http://adsabs.harvard.edu/abs/1977A%26A....54..661T} {54, 661}

\bibitem[\protect\citeauthoryear{{{\"U}bler}, {Naab}, {Oser}, {Aumer}, {Sales}
  \& {White}}{{{\"U}bler} et~al.}{2014}]{Uebler14}
{{\"U}bler} H.,  {Naab} T.,  {Oser} L.,  {Aumer} M.,  {Sales} L.~V.,   {White}
  S.~D.~M.,  2014, \mn@doi [\mnras] {10.1093/mnras/stu1275}, \href
  {http://adsabs.harvard.edu/abs/2014MNRAS.443.2092U} {443, 2092}

\bibitem[\protect\citeauthoryear{{{\"U}bler} et~al.,}{{{\"U}bler}
  et~al.}{2017}]{Uebler17}
{{\"U}bler} H.,  et~al., 2017, \mn@doi [\apj] {10.3847/1538-4357/aa7558}, \href
  {http://adsabs.harvard.edu/abs/2017ApJ...842..121U} {842, 121}

\bibitem[\protect\citeauthoryear{{{\"U}bler} et~al.,}{{{\"U}bler}
  et~al.}{2018}]{Uebler18}
{{\"U}bler} H.,  et~al., 2018, \mn@doi [\apjl] {10.3847/2041-8213/aaacfa},
  \href {http://adsabs.harvard.edu/abs/2018ApJ...854L..24U} {854, L24}

\bibitem[\protect\citeauthoryear{{{\"U}bler} et~al.,}{{{\"U}bler}
  et~al.}{2019}]{Uebler19}
{{\"U}bler} H.,  et~al., 2019, \mn@doi [\apj] {10.3847/1538-4357/ab27cc}, \href
  {https://ui.adsabs.harvard.edu/abs/2019ApJ...880...48U} {880, 48}

\bibitem[\protect\citeauthoryear{{Ueda} et~al.,}{{Ueda} et~al.}{2008}]{Ueda08}
{Ueda} Y.,  et~al., 2008, \mn@doi [\apjs] {10.1086/591083}, \href
  {https://ui.adsabs.harvard.edu/\#abs/2008ApJS..179..124U} {179, 124}

\bibitem[\protect\citeauthoryear{{Vega Beltr{\'a}n}, {Pizzella}, {Corsini},
  {Funes}, {Zeilinger}, {Beckman}  \& {Bertola}}{{Vega Beltr{\'a}n}
  et~al.}{2001}]{VegaBeltran01}
{Vega Beltr{\'a}n} J.~C.,  {Pizzella} A.,  {Corsini} E.~M.,  {Funes} J.~G.,
  {Zeilinger} W.~W.,  {Beckman} J.~E.,   {Bertola} F.,  2001, \mn@doi [\aap]
  {10.1051/0004-6361:20010625}, \href
  {https://ui.adsabs.harvard.edu/abs/2001A&A...374..394V} {374, 394}

\bibitem[\protect\citeauthoryear{{Vergani} et~al.,}{{Vergani}
  et~al.}{2018}]{Vergani18}
{Vergani} D.,  et~al., 2018, \mn@doi [\aap] {10.1051/0004-6361/201732495},
  \href {https://ui.adsabs.harvard.edu/abs/2018A&A...620A.193V} {620, A193}

\bibitem[\protect\citeauthoryear{{Whitaker} et~al.,}{{Whitaker}
  et~al.}{2014}]{Whitaker14}
{Whitaker} K.~E.,  et~al., 2014, \mn@doi [\apj] {10.1088/0004-637X/795/2/104},
  \href {http://adsabs.harvard.edu/abs/2014ApJ...795..104W} {795, 104}

\bibitem[\protect\citeauthoryear{{Wilman} et~al.,}{{Wilman}
  et~al.}{2020}]{Wilman20}
{Wilman} D.~J.,  et~al., 2020, \mn@doi [\apj] {10.3847/1538-4357/ab7914}, \href
  {https://ui.adsabs.harvard.edu/abs/2020ApJ...892....1W} {892, 1}

\bibitem[\protect\citeauthoryear{{Wisnioski} et~al.,}{{Wisnioski}
  et~al.}{2015}]{Wisnioski15}
{Wisnioski} E.,  et~al., 2015, \mn@doi [\apj] {10.1088/0004-637X/799/2/209},
  \href {http://adsabs.harvard.edu/abs/2015ApJ...799..209W} {799, 209}

\bibitem[\protect\citeauthoryear{{Wisnioski} et~al.,}{{Wisnioski}
  et~al.}{2018}]{Wisnioski18}
{Wisnioski} E.,  et~al., 2018, \mn@doi [\apj] {10.3847/1538-4357/aab097}, \href
  {https://ui.adsabs.harvard.edu/\#abs/2018ApJ...855...97W} {855, 97}

\bibitem[\protect\citeauthoryear{{Wisnioski} et~al.,}{{Wisnioski}
  et~al.}{2019}]{Wisnioski19}
{Wisnioski} E.,  et~al., 2019, \mn@doi [\apj] {10.3847/1538-4357/ab4db8}, \href
  {https://ui.adsabs.harvard.edu/abs/2019ApJ...886..124W} {886, 124}

\bibitem[\protect\citeauthoryear{{Wuyts} et~al.,}{{Wuyts}
  et~al.}{2011}]{WuytsS11a}
{Wuyts} S.,  et~al., 2011, \mn@doi [\apj] {10.1088/0004-637X/738/1/106}, \href
  {http://adsabs.harvard.edu/abs/2011ApJ...738..106W} {738, 106}

\bibitem[\protect\citeauthoryear{{Wuyts} et~al.,}{{Wuyts}
  et~al.}{2012}]{WuytsS12}
{Wuyts} S.,  et~al., 2012, \mn@doi [\apj] {10.1088/0004-637X/753/2/114}, \href
  {https://ui.adsabs.harvard.edu/\#abs/2012ApJ...753..114W} {753, 114}

\bibitem[\protect\citeauthoryear{{Wuyts} et~al.,}{{Wuyts}
  et~al.}{2016}]{WuytsS16}
{Wuyts} S.,  et~al., 2016, \mn@doi [\apj] {10.3847/0004-637X/831/2/149}, \href
  {http://adsabs.harvard.edu/abs/2016ApJ...831..149W} {831, 149}

\bibitem[\protect\citeauthoryear{{Xue} et~al.,}{{Xue} et~al.}{2011}]{Xue11}
{Xue} Y.~Q.,  et~al., 2011, \mn@doi [\apjs] {10.1088/0067-0049/195/1/10}, \href
  {https://ui.adsabs.harvard.edu/\#abs/2011ApJS..195...10X} {195, 10}

\bibitem[\protect\citeauthoryear{{Yan}}{{Yan}}{2018}]{Yan18}
{Yan} R.,  2018, \mn@doi [\mnras] {10.1093/mnras/sty502}, \href
  {https://ui.adsabs.harvard.edu/abs/2018MNRAS.481..467Y} {481, 467}

\bibitem[\protect\citeauthoryear{{Yan}, {Newman}, {Faber}, {Konidaris}, {Koo}
  \& {Davis}}{{Yan} et~al.}{2006}]{Yan06}
{Yan} R.,  {Newman} J.~A.,  {Faber} S.~M.,  {Konidaris} N.,  {Koo} D.,
  {Davis} M.,  2006, \mn@doi [\apj] {10.1086/505629}, \href
  {https://ui.adsabs.harvard.edu/abs/2006ApJ...648..281Y} {648, 281}

\bibitem[\protect\citeauthoryear{{Zakamska} et~al.,}{{Zakamska}
  et~al.}{2016}]{Zakamska16}
{Zakamska} N.~L.,  et~al., 2016, \mn@doi [\mnras] {10.1093/mnras/stw718}, \href
  {https://ui.adsabs.harvard.edu/abs/2016MNRAS.459.3144Z} {459, 3144}

\bibitem[\protect\citeauthoryear{{van Houdt} et~al.,}{{van Houdt}
  et~al.}{2021}]{vHoudt21}
{van Houdt} J.,  et~al., 2021, \mn@doi [\apj] {10.3847/1538-4357/ac1f29}, \href
  {https://ui.adsabs.harvard.edu/abs/2021ApJ...923...11V} {923, 11}

\bibitem[\protect\citeauthoryear{{van de Voort}, {Davis}, {Kere{\v{s}}},
  {Quataert}, {Faucher-Gigu{\`e}re}  \& {Hopkins}}{{van de Voort}
  et~al.}{2015}]{vdVoort15}
{van de Voort} F.,  {Davis} T.~A.,  {Kere{\v{s}}} D.,  {Quataert} E.,
  {Faucher-Gigu{\`e}re} C.-A.,   {Hopkins} P.~F.,  2015, \mn@doi [\mnras]
  {10.1093/mnras/stv1217}, \href
  {https://ui.adsabs.harvard.edu/abs/2015MNRAS.451.3269V} {451, 3269}

\bibitem[\protect\citeauthoryear{{van der Kruit} \& {Allen}}{{van der Kruit} \&
  {Allen}}{1978}]{vdKruit78}
{van der Kruit} P.~C.,  {Allen} R.~J.,  1978, \mn@doi [Annual Review of
  Astronomy and Astrophysics] {10.1146/annurev.aa.16.090178.000535}, \href
  {https://ui.adsabs.harvard.edu/\#abs/1978ARA&A..16..103V} {16, 103}

\bibitem[\protect\citeauthoryear{{van der Wel}, {Franx}, {Wuyts}, {van Dokkum},
  {Huang}, {Rix}  \& {Illingworth}}{{van der Wel} et~al.}{2006}]{vdWel06}
{van der Wel} A.,  {Franx} M.,  {Wuyts} S.,  {van Dokkum} P.~G.,  {Huang} J.,
  {Rix} H.~W.,   {Illingworth} G.~D.,  2006, \mn@doi [\apj] {10.1086/508128},
  \href {https://ui.adsabs.harvard.edu/abs/2006ApJ...652...97V} {652, 97}

\bibitem[\protect\citeauthoryear{{van der Wel} et~al.,}{{van der Wel}
  et~al.}{2012}]{vdWel12}
{van der Wel} A.,  et~al., 2012, \mn@doi [\apjs] {10.1088/0067-0049/203/2/24},
  \href {http://adsabs.harvard.edu/abs/2012ApJS..203...24V} {203, 24}

\bibitem[\protect\citeauthoryear{{van der Wel} et~al.,}{{van der Wel}
  et~al.}{2014}]{vdWel14b}
{van der Wel} A.,  et~al., 2014, \mn@doi [\apj] {10.1088/2041-8205/792/1/L6},
  \href {https://ui.adsabs.harvard.edu/\#abs/2014ApJ...792L...6V} {792, L6}

\bibitem[\protect\citeauthoryear{{van der Wel} et~al.,}{{van der Wel}
  et~al.}{2016}]{vdWel16}
{van der Wel} A.,  et~al., 2016, \mn@doi [\apjs] {10.3847/0067-0049/223/2/29},
  \href {https://ui.adsabs.harvard.edu/abs/2016ApJS..223...29V} {223, 29}

\bibitem[\protect\citeauthoryear{{van der Wel} et~al.,}{{van der Wel}
  et~al.}{2021}]{vdWel21}
{van der Wel} A.,  et~al., 2021, \mn@doi [\apjs] {10.3847/1538-4365/ac1356},
  \href {https://ui.adsabs.harvard.edu/abs/2021ApJS..256...44V} {256, 44}

\bibitem[\protect\citeauthoryear{{van der Wel} et~al.,}{{van der Wel}
  et~al.}{2022}]{vdWel22}
{van der Wel} A.,  et~al., 2022, \mn@doi [\apj] {10.3847/1538-4357/ac83c5},
  \href {https://ui.adsabs.harvard.edu/abs/2022ApJ...936....9W} {936, 9}

\makeatother
\end{thebibliography}

%%%%%%%%%%%%%%%%%%%%%%%%%%%%%%%%%%%%%%%%%%%%%%%%%%

%%%%%%%%%%%%%%%%% APPENDICES %%%%%%%%%%%%%%%%%%%%%

\appendix

\section{Dynamical mass estimates for KMOS\texorpdfstring{$^{\rm 3D}$}{3D} assuming structural properties from F814W imaging}\label{a:f814w_modelling} %A

\begin{figure*}
	\centering
	\includegraphics[width=0.45\textwidth]{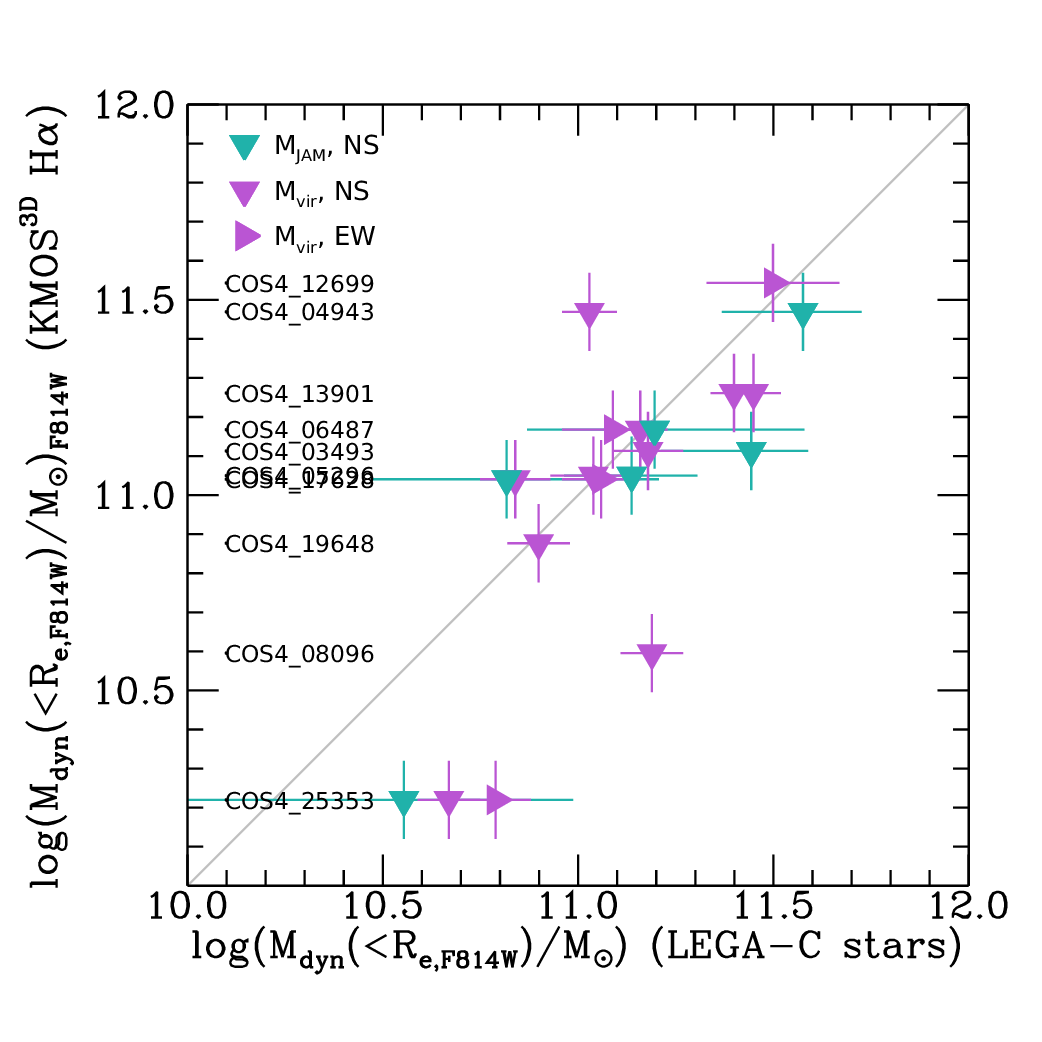}
	\includegraphics[width=0.45\textwidth]{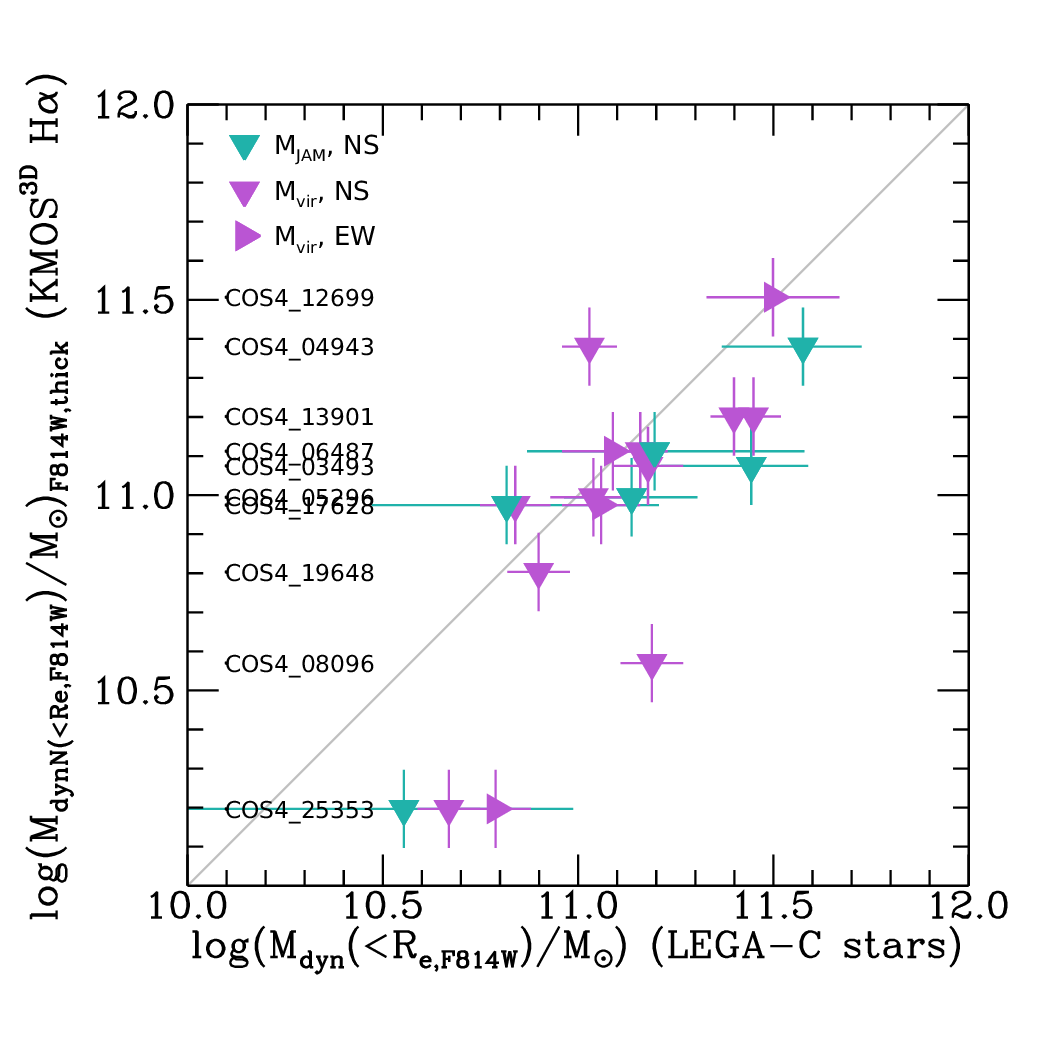}
	\caption{Comparison of dynamical mass estimates within the F814W effective radius using mass models for \kd and integrated stellar velocity dispersion \citep[purple;][]{vdWel21} and JAM models \citep[teal;][]{vHoudt21} for LEGA-C, as in Figure~\ref{f:mdyn}. Here, we use also for the \kd models structural parameters as constrained from F814 imaging (instead of F160W), specifically $R_e$, $n_S$, and $q=b/a$ (left panel). For the comparison in the right panel we additionally infer the (fixed) inclination from $q_{\rm F814W}$ by assuming an intrinsic thickness of $q_0=0.41$, mimicking more closely the assumptions of the LEGA-C JAM modelling.
    The direction of triangles indicates the slit orientation of the LEGA-C data from which the estimates were derived. Uncertainties for LEGA-C measurements from integrated velocity dispersion are $1\sigma$, and for JAM models the $16^{\rm th}$ and $84^{\rm th}$ percentiles. For \kd measurements, uncertainties are the $16^{\rm th}$ and $84^{\rm th}$ percentiles, with lower ceiling uncertainties of 0.1~dex.
	Overall, we find similar average offsets and standard deviations between the dynamical mass estimates from \kd and LEGA-C as for our fiducial models.}
	\label{f:compmdyn_structure}
\end{figure*}

In Figure~\ref{f:compmdyn_structure}, we show comparisons of $M_{\rm dyn}$ estimates from LEGA-C and from KMOS$^{\rm 3D}$, analogous to Figure~\ref{f:mdyn}, but now adopting structural parameters constrained from F814W imaging for two additional sets of \kd models , specifically $R_e$, $n_S$, and $q$.
For the first set (left panel of Figure~\ref{f:compmdyn_structure}) we keep the inclination we inferred for our fiducial \kd models. For the second set (right panel) we re-calculate the inclination based on the observed axis ratio $q_{\rm F814W}$ and using $q_0=0.41$, the prior that is used for the LEGA-C JAM modelling (see Section~\ref{s:model_diff} and \citealp{vHoudt21}).

Overall, adopting the F814W structural parameters has a minor effect on the inferred dynamical masses. For the first set of alternative models we find an average increase in $M_{\rm dyn}$ of 0.02~dex compared to our fiducial model. Correspondingly, the average offset in dynamical mass compared to the LEGA-C estimates decreases from $\Delta M_{\rm dyn}=0.12$ to $\Delta M_{\rm dyn}=0.10$, but the standard deviation increases slightly to 0.26~dex, considering all multiple observations and both the $M_{\rm vir}$ and JAM estimates (0.24~dex for our fiducial setup).
For the second set of alternative models, on the other hand, we find an average decrease in $M_{\rm dyn}$ of 0.03~dex compared to our fiducial model. Again correspondingly, the average offset in dynamical mass compared to the LEGA-C estimates increases to $\Delta M_{\rm dyn}=0.15$, and the standard deviation increases slightly to 0.25~dex.

However, comparing the two alternative models with {\it each other}, we do see a systematic effect, as expected: keeping all other assumptions on structural parameters fixed, assuming an intrinsically thicker structure leads to lower inferred dynamical masses. In fact, all dynamical masses inferred from the second set of alternative models (right panel in Figure~\ref{f:compmdyn_structure}) are lower than those inferred from the first set of alternative models (left panel in Figure~\ref{f:compmdyn_structure}).

\section{2D and 1D profiles from fixed (pseudo-)slit extractions after PSF-matching}\label{a:gallery}

In Figures~\ref{f:gallery1}-\ref{f:gallery3} we show the {\it HST} images, the 2D PV diagrams from fixed (pseudo-)slit extractions after PSF-matching, the corresponding 1D LOS velocities and velocity dispersion profiles, and the integrated 1D spectra for the LEGA-C data for all galaxies, except COS4\_17628-M5/101\_133199 for which these panels are shown in Figure~\ref{f:obsprof}. The 2D PV diagrams and 1D LOS profiles are not extracted along the kinematic major axes. The 1D LOS kinematic profiles from the fixed (pseudo-)slit extractions (see Sections~\ref{s:kmoskin} and \ref{s:lgckin}) form the basis for the measurements described in Section~\ref{s:measure}, while the other panels are added only for illustrative purposes. We add here a few notes on individual objects.

COS4\_03493: The LEGA-C spectrum of this galaxy contains H$\beta$ and higher-order hydrogen lines, for which we expect similar kinematics compared to the \kd H$\alpha$ emission. The high central LOS dispersion for the PSF-matched \kd data suggest that the PSF of the LEGA-C observations could be overestimated in this case.

COS4\_04943: The LEGA-C spectrum of this galaxy contains H$\beta$ and higher-order hydrogen lines, for which we expect similar kinematics compared to the \kd H$\alpha$ emission. This allows us to align the \kd and LEGA-C extracted kinematics with high accuracy. For this object, we shift the LEGA-C kinematic centre by $\sim0.4\arcsec$ to align the kinematics with the \kd data (see Appendix~\ref{a:misalignment} and Figure~\ref{f:COS3_05062} for details).

COS4\_16227: This object had a wrong segmentation map, combining the bright compact source shown in the centre of the HST image together with the extended blue object to the centre left. The H$\alpha$ line emission comes from the blue object in the centre left, and is connected to H$\alpha$ emission from the blue object on the bottom left. These galaxies are likely undergoing a merger. Emission from both these objects falls into the pseudo-slit after PSF-matching. The central compact object is an AGN at $z\sim1.33$ based on bright and broad [O{\sc{iii}}] emission in the \kd and LEGA-C data. The LEGA-C spectrum of the centre left galaxy contains H$\beta$ and higher-order hydrogen lines, for which we expect similar kinematics compared to the \kd H$\alpha$ emission, and indeed the alignment of the \kd and LEGA-C gas kinematic extractions shows that the pPXF fitting is performed on the blue object in the centre left. The mismatch of morphological PA and kinematic PA for this source is likely driven by the bright central object and the wrong segmentation map.

COS4\_19648: The LEGA-C spectrum of this galaxy contains bright H$\beta$ and higher-order hydrogen lines, for which we expect similar kinematics compared to the \kd H$\alpha$ emission. This allows us to align the \kd and LEGA-C extracted kinematics with high accuracy. For this object, we shift the LEGA-C kinematic centre by $\sim0.3\arcsec$ to align the kinematics with the \kd data. There is indication of counter-rotating gas and stellar discs from the extracted LOS velocities. Unfortunately, the LEGA-C N-S slit observations are almost perpendicular to the H$\alpha$ kinematic major axis, and we observed only a shallow velocity gradient in the (pseudo-)slit extractions.

COS4\_25353: The LEGA-C spectrum of this galaxy shows bright and kinematically decoupled [NeV]$\lambda3347$, [NeV]$\lambda3427$ and [NeIII]$\lambda3870$ emission which influences the pPXF fit for the gas kinematics, if not masked. We show the 1D extractions after masking the Neon emission. See Appendix~\ref{a:nev} and Figures~\ref{f:nev_pv} and \ref{f:nev_prof} for more details.

COS4\_05238: The LEGA-C spectrum of this galaxy shows bright and kinematically decoupled [NeV]$\lambda3427$ and [NeIII]$\lambda3870$ emission which influences the pPXF fit for the gas kinematics, if not masked. We show the 1D extractions after masking the Neon emission.

\begin{figure*}
	\centering
	\includegraphics[width=0.85\textwidth]{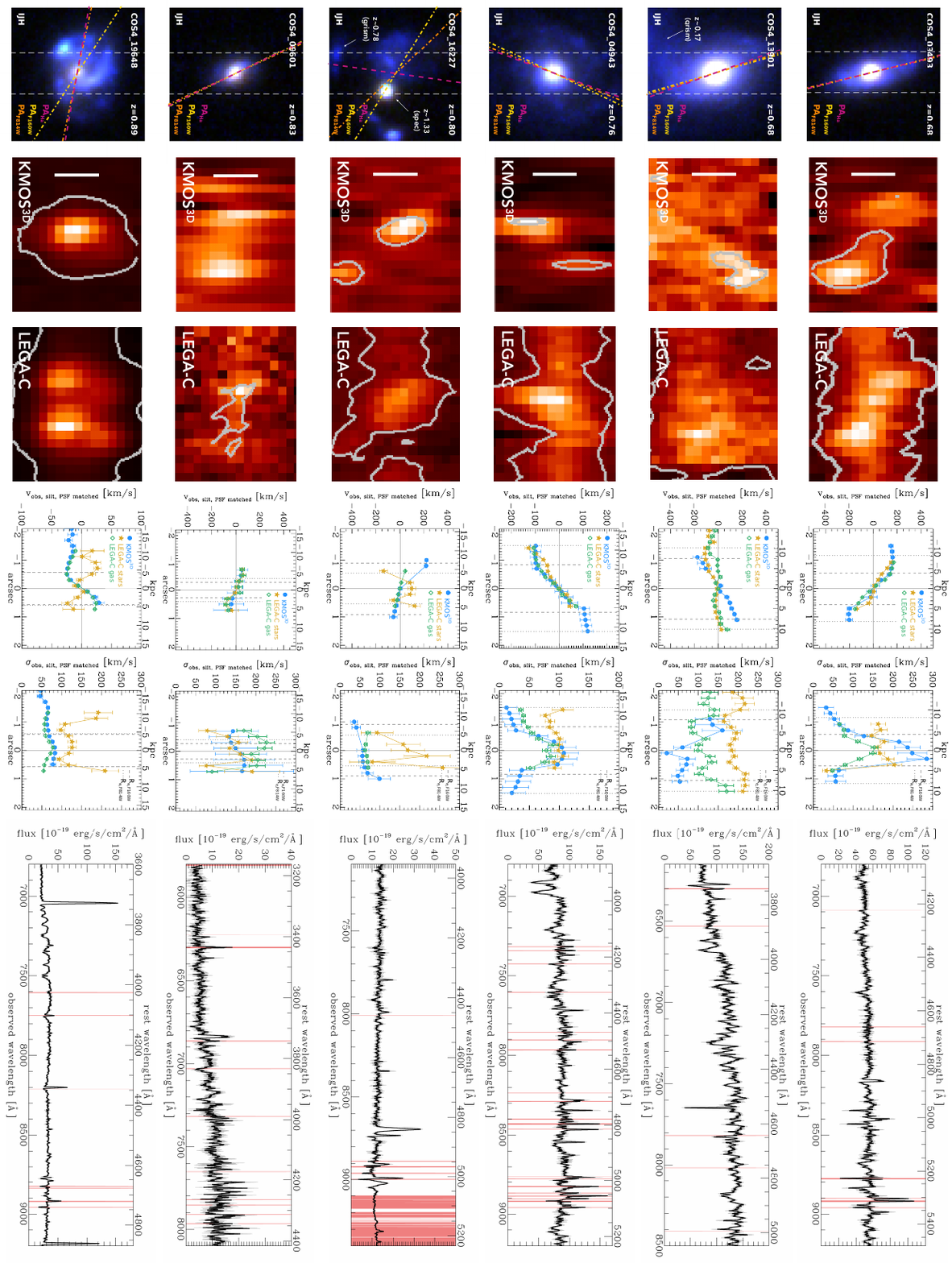}
	\caption{IJH {\it HST} color-composite image, 2D PV diagrams from fixed (pseudo-)slit extractions after PSF-matching with $S/N=3$ contours, corresponding 1D LOS velocities and velocity dispersion profiles, and integrated 1D spectra for the LEGA-C data. See Appendix~\ref{a:gallery} and Fig.~\ref{f:obsprof} for details. The 1D LOS kinematic profiles from the fixed (pseudo-)slit extractions (see Sections~\ref{s:kmoskin} and \ref{s:lgckin}) form the basis for the measurements described in Section~\ref{s:measure}.}
	\label{f:gallery1}
\end{figure*} 
\begin{figure*}
	\centering
	\includegraphics[width=0.85\textwidth]{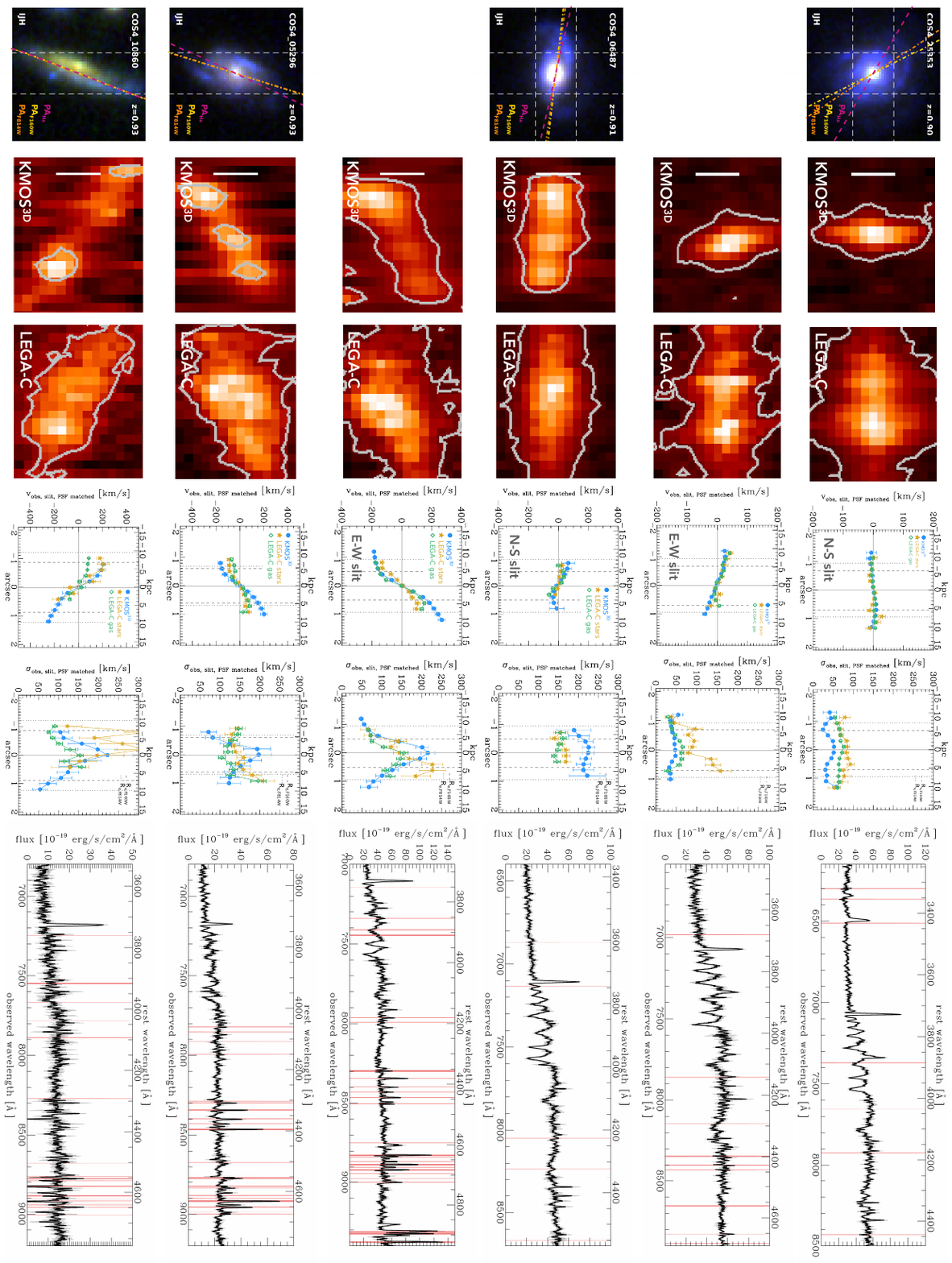}
	\caption{Continuation of Figure~\ref{f:gallery1}. IJH {\it HST} color-composite image, 2D PV diagrams from fixed (pseudo-)slit extractions after PSF-matching with $S/N=3$ contours, corresponding 1D LOS velocities and velocity dispersion profiles, and integrated 1D spectra for the LEGA-C data. See Appendix~\ref{a:gallery} and Fig.~\ref{f:obsprof} for details. The 1D LOS kinematic profiles from the fixed (pseudo-)slit extractions (see Sections~\ref{s:kmoskin} and \ref{s:lgckin}) form the basis for the measurements described in Section~\ref{s:measure}.}
	\label{f:gallery2}
\end{figure*} 
\begin{figure*}
	\centering
	\includegraphics[width=0.85\textwidth]{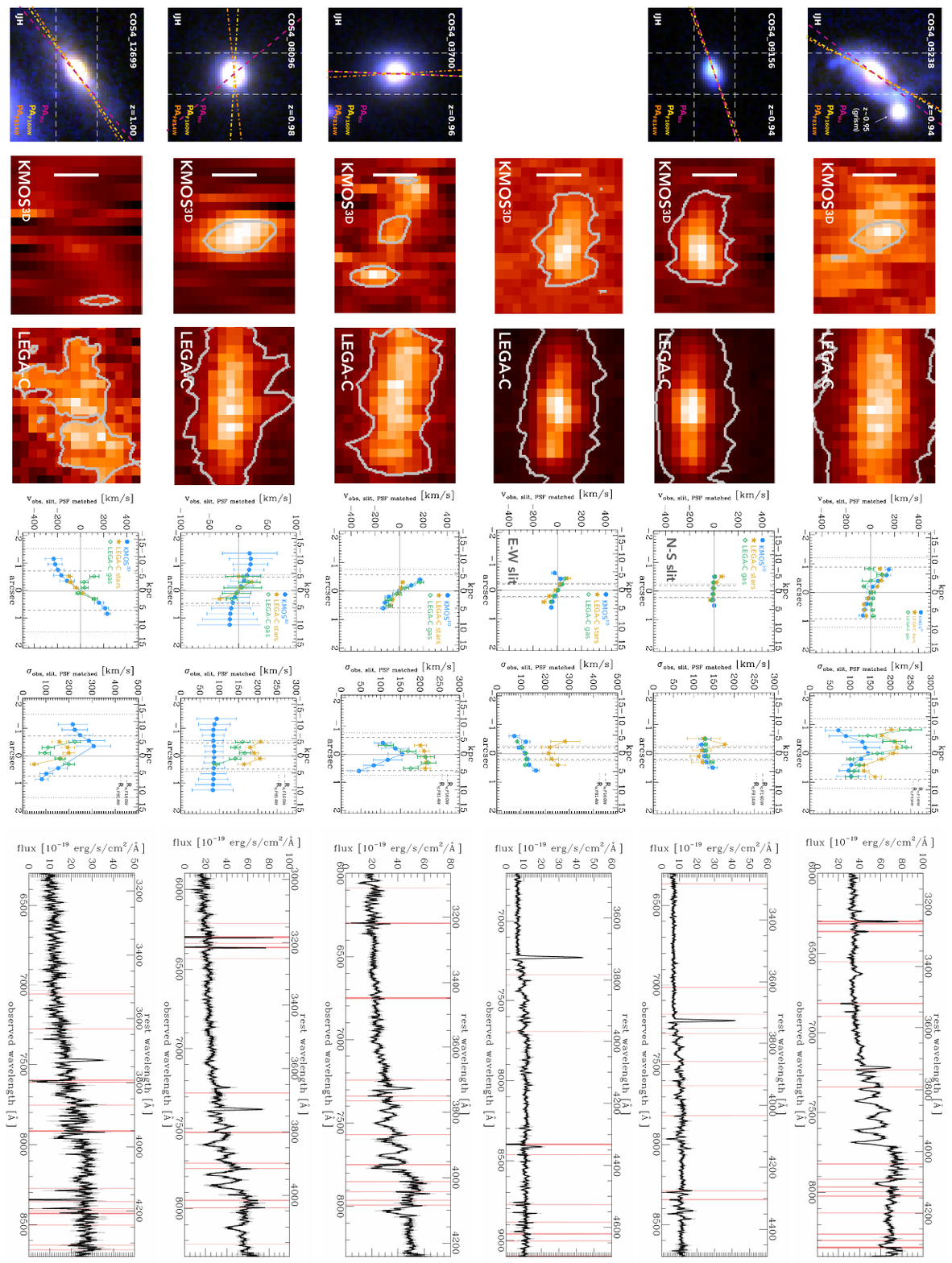}
	\caption{Continuation of Figure~\ref{f:gallery1}. IJH {\it HST} color-composite image, 2D PV diagrams from fixed (pseudo-)slit extractions after PSF-matching with $S/N=3$ contours, corresponding 1D LOS velocities and velocity dispersion profiles, and integrated 1D spectra for the LEGA-C data. See Appendix~\ref{a:gallery} and Fig.~\ref{f:obsprof} for details. The 1D LOS kinematic profiles from the fixed (pseudo-)slit extractions (see Sections~\ref{s:kmoskin} and \ref{s:lgckin}) form the basis for the measurements described in Section~\ref{s:measure}.}
	\label{f:gallery3}
\end{figure*}

\section{Impact of high-ionisation Neon transitions on pPXF fits}\label{a:nev}

\begin{figure*}
	\centering
	\includegraphics[width=0.95\textwidth]{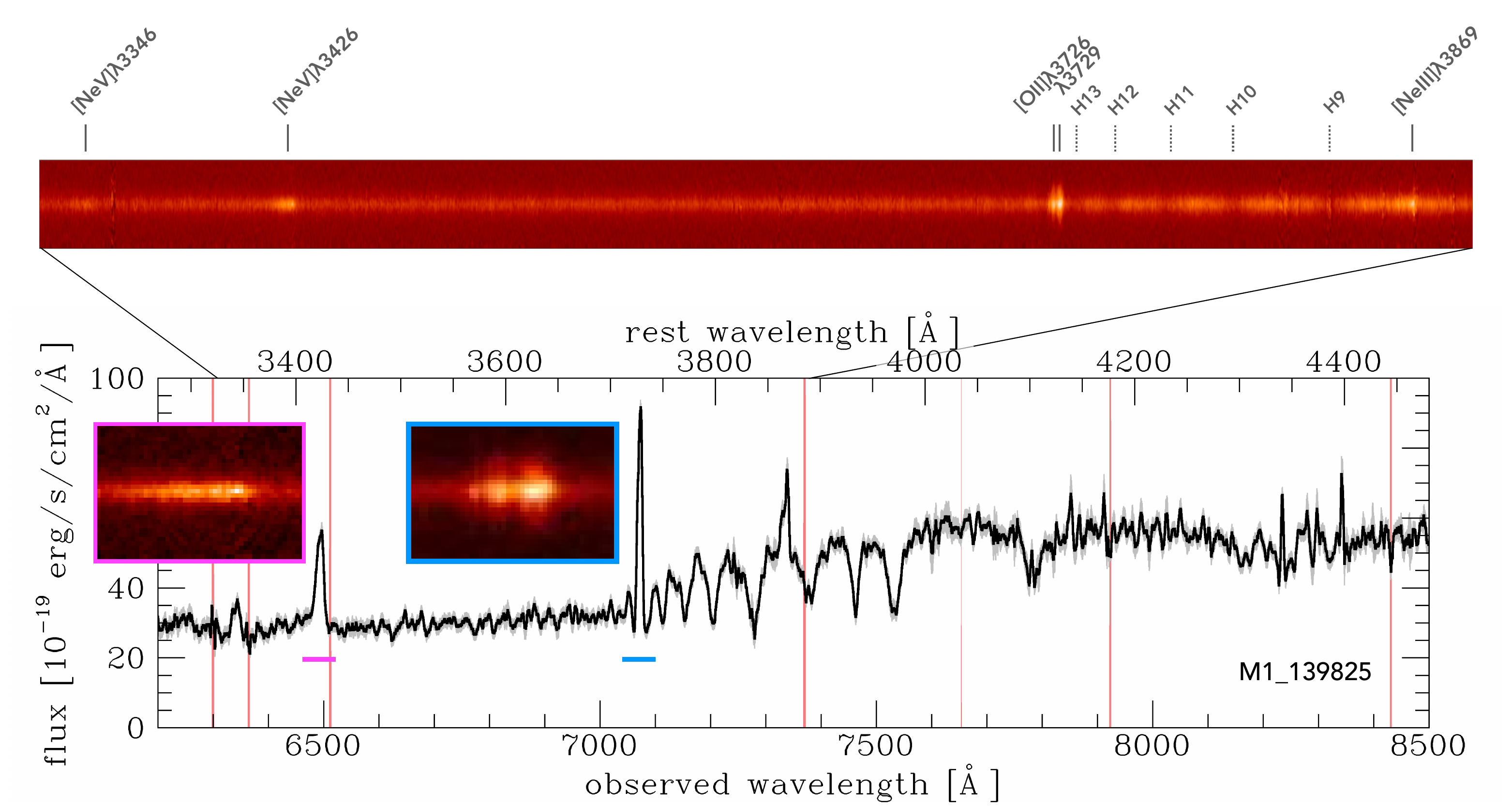}
	\caption{Cut-out of the 2D LEGA-C spectrum for galaxy M1\_139825, including the high-ionisation [NeV]$\lambda3347$, [NeV]$\lambda3427$ and [NeIII]$\lambda3870$ emission lines (top), and the full collapsed 1D spectrum with pink vertical lines indicating low quality regions (bottom). The insets in the bottom panel are zoom-ins on the [NeV]$\lambda3427$ and [OII]$\lambda\lambda3727,3730$ emission lines, showcasing the different kinematics and spatial distributions traced by these lines. The [NeV]$\lambda3427$ emission in particular is centrally concentrated but has a broad velocity distribution, whereas the [OII] emission is spatially extended with shallow velocity gradient.}
	\label{f:nev_pv}
\end{figure*}

We show a LEGA-C PV diagram and collapsed 1D spectrum for one of the two galaxies with strong [NeV]$\lambda3347$, [NeV]$\lambda3427$ and [NeIII]$\lambda3870$ emission in our sample (Figure~\ref{f:nev_pv}). The emission of these high-ionisation lines is centrally concentrated and shows a broad velocity distribution, indicative of a nuclear outflow powered by an AGN. It is clearly decoupled from the disc kinematics, traced here primarily by the [OII]$\lambda\lambda3727,3730$ doublet. 

\begin{figure*}
	\centering
	\includegraphics[width=0.98\textwidth]{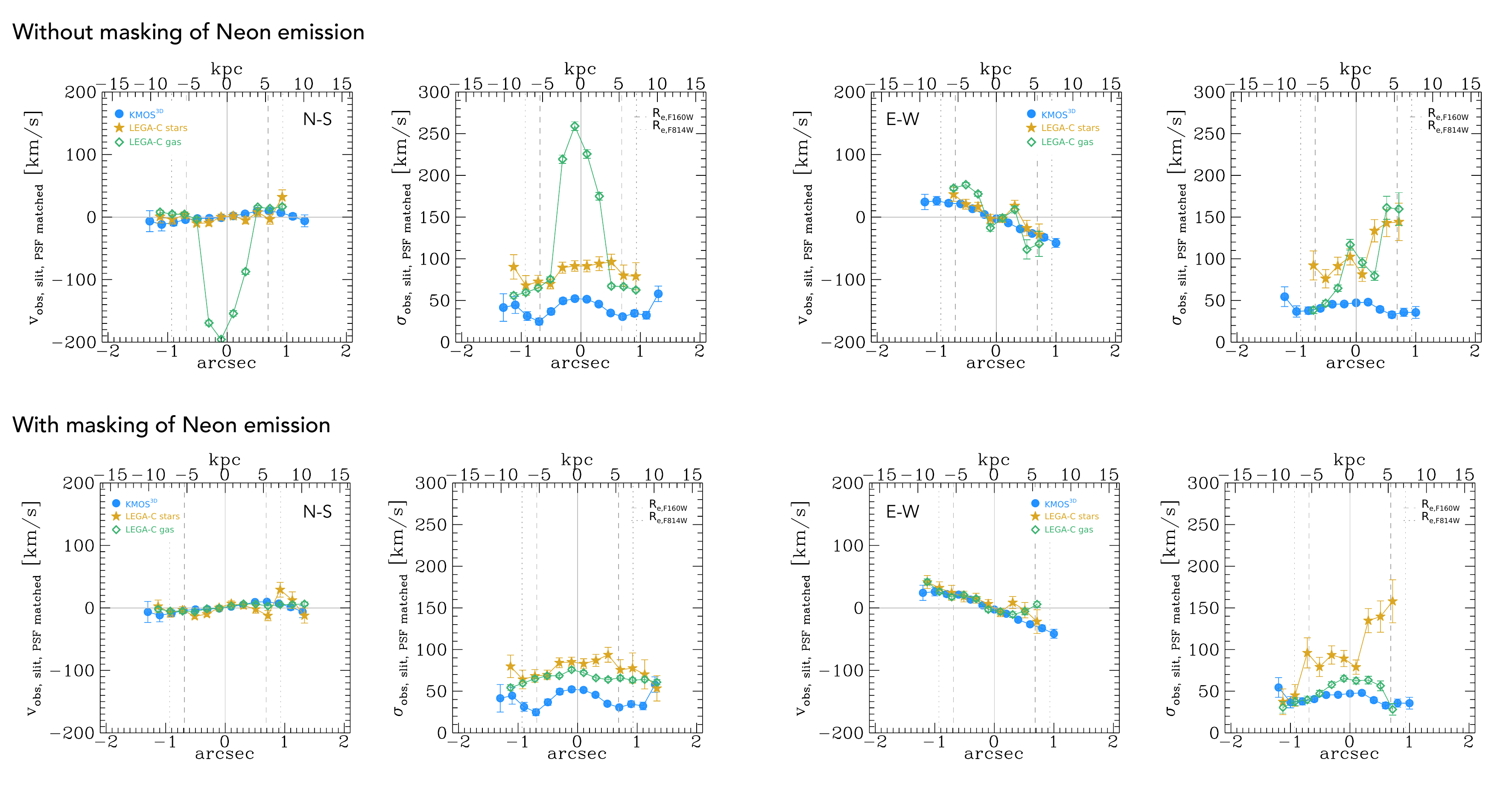}
	\caption{Comparison of 1D kinematic profiles without (top row) and with (bottom row) masking of the [NeV]$\lambda3347$, [NeV]$\lambda3427$ and [NeIII]$\lambda3870$ emission lines before extracting LEGA-C gas and stellar kinematics (open green diamonds and filled golden stars, respectively). The \kd extractions (filled blue circles) are the same in both rows. We show the observed velocity and velocity dispersion measurements along the N-S (pseudo-)slit in the left panels, corresponding to the 2D and 1D spectra shown in Figure~\ref{f:nev_pv}. The right panels show extractions in E-W direction. The \kd observations have been convolved to match the LEGA-C PSFs before extracting kinematics, as described in Section~\ref{s:kinex}. The impact of the [NeV]$\lambda3347$, [NeV]$\lambda3427$ and [NeIII]$\lambda3870$ emission on the extracted LEGA-C gas kinematics is particularly evident for the N-S slit observations in the left panels, where the difference in observed velocities and velocity dispersions is up to 200~km/s in the central regions of the galaxy.}
	\label{f:nev_prof}
\end{figure*}

In Figure~\ref{f:nev_prof} we contrast the effect of masking the Neon lines before fitting the spectra with pPXF (bottom row) to the standard extraction (top row). The impact on the extracted ionised gas kinematics is evident. For our comparison of kinematics `as observed' in Section~\ref{s:compasobs}, we use kinematics extracted from masked spectra. Since the stellar kinematic fits are only minimally affected by the masking of the Neon emission ($2-5$ per cent difference in $\sigma_*$), we perform our comparison of dynamical mass estimates in Section~\ref{s:compmdyn} with the original pPXF fits.

\section{Kinematic maps of the modelled KMOS$^{\rm 3D}$ galaxies}\label{a:maps}

\begin{figure*}
	\centering
    \begin{subfigure}{0.4\textwidth} 
       \includegraphics[width=\textwidth]{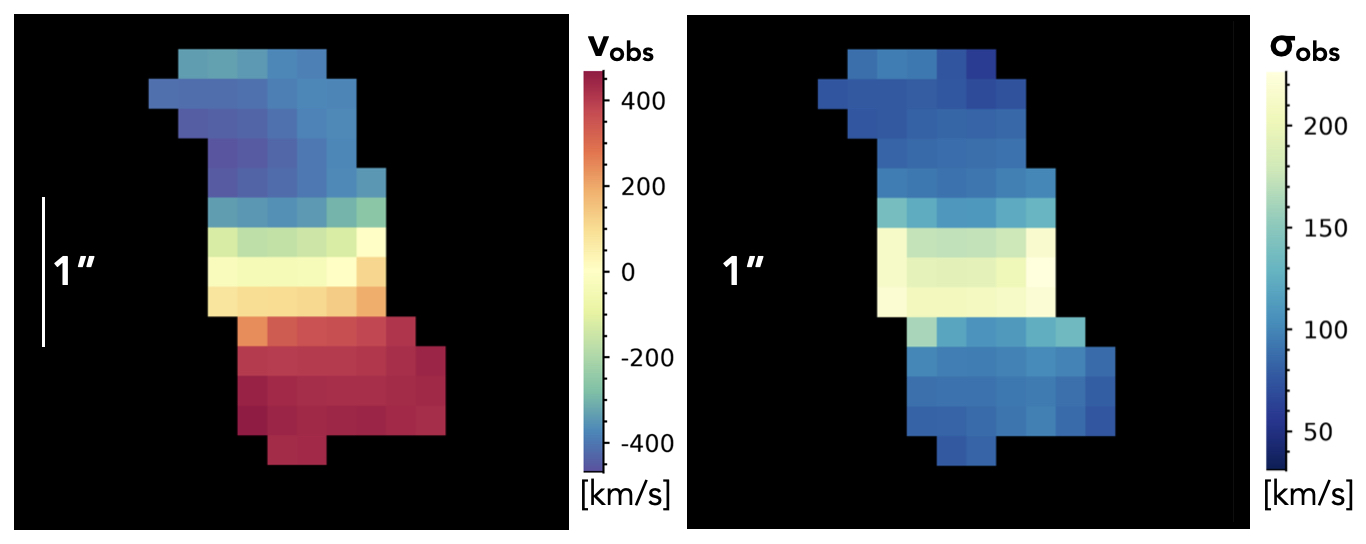}
      \caption{COS4\_03493}
    \end{subfigure}
    \begin{subfigure}{0.4\textwidth} 
       \includegraphics[width=\textwidth]{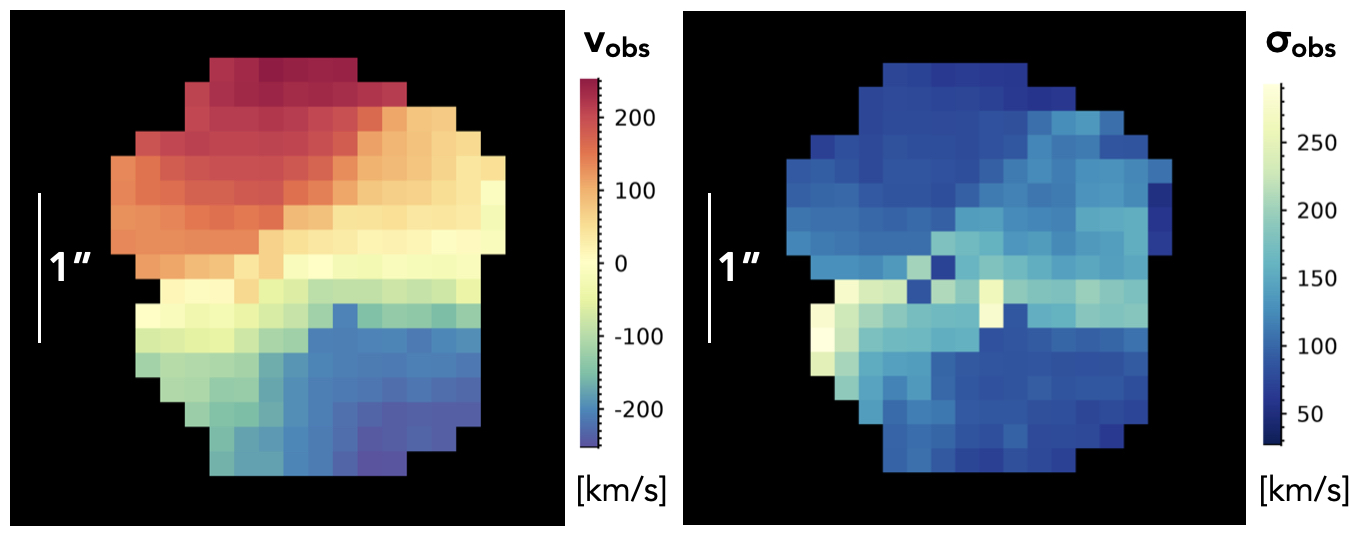}
      \caption{COS4\_13901}
    \end{subfigure}
    \begin{subfigure}{0.4\textwidth} 
       \includegraphics[width=\textwidth]{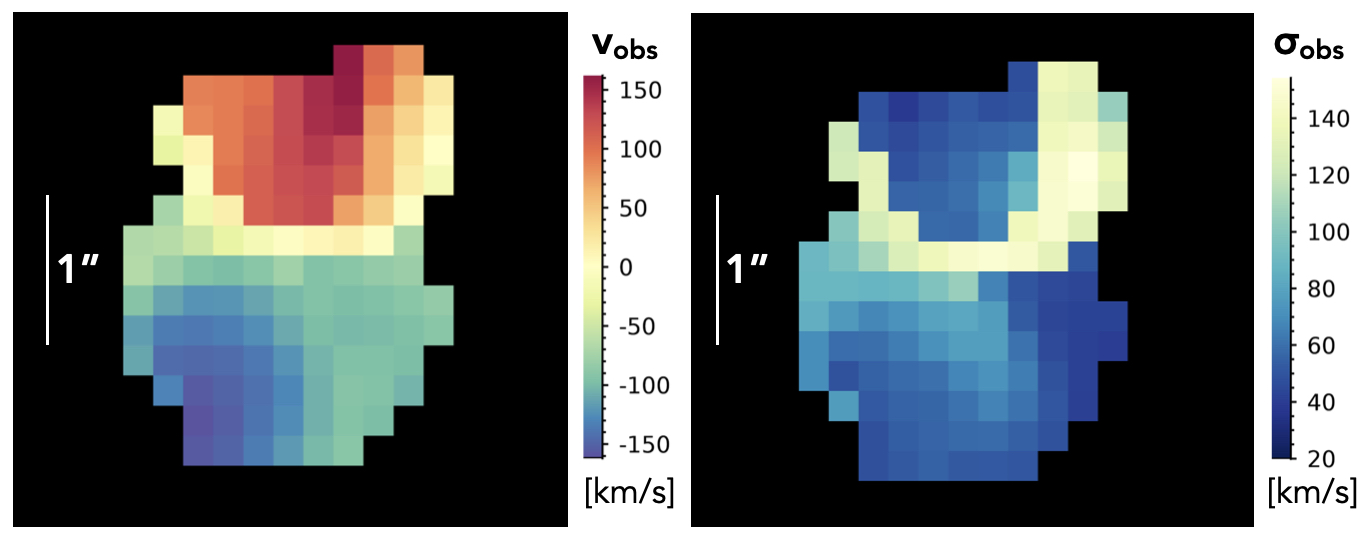}
      \caption{COS4\_04943}
    \end{subfigure}
    \begin{subfigure}{0.4\textwidth} 
       \includegraphics[width=\textwidth]{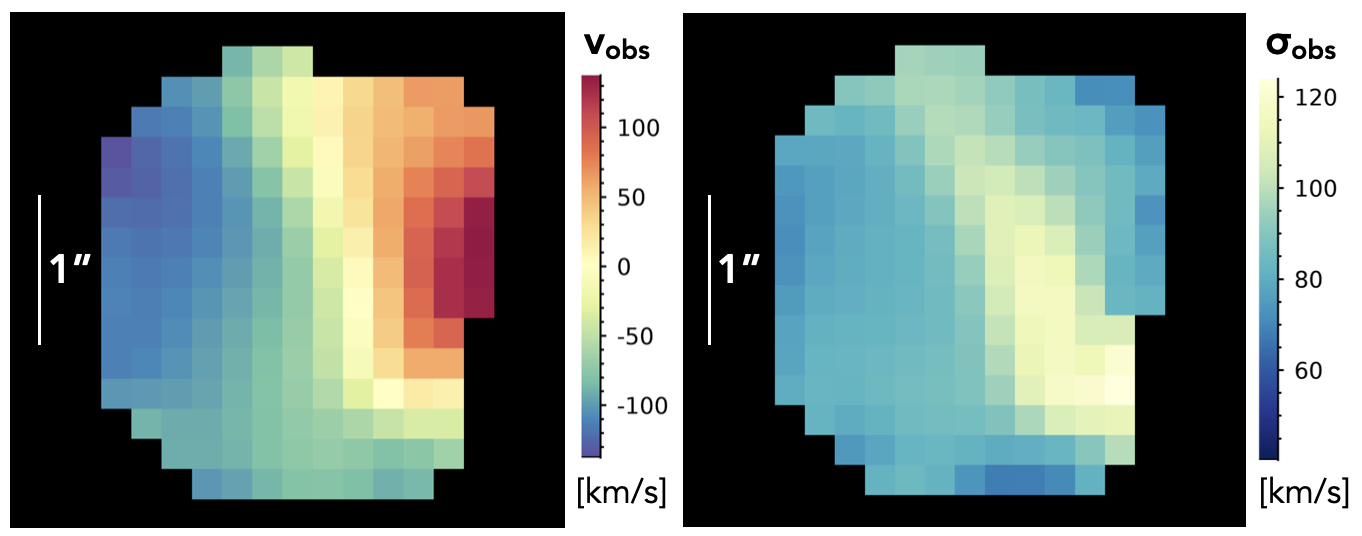}
      \caption{COS4\_19648}
    \end{subfigure}
    \begin{subfigure}{0.4\textwidth} 
       \includegraphics[width=\textwidth]{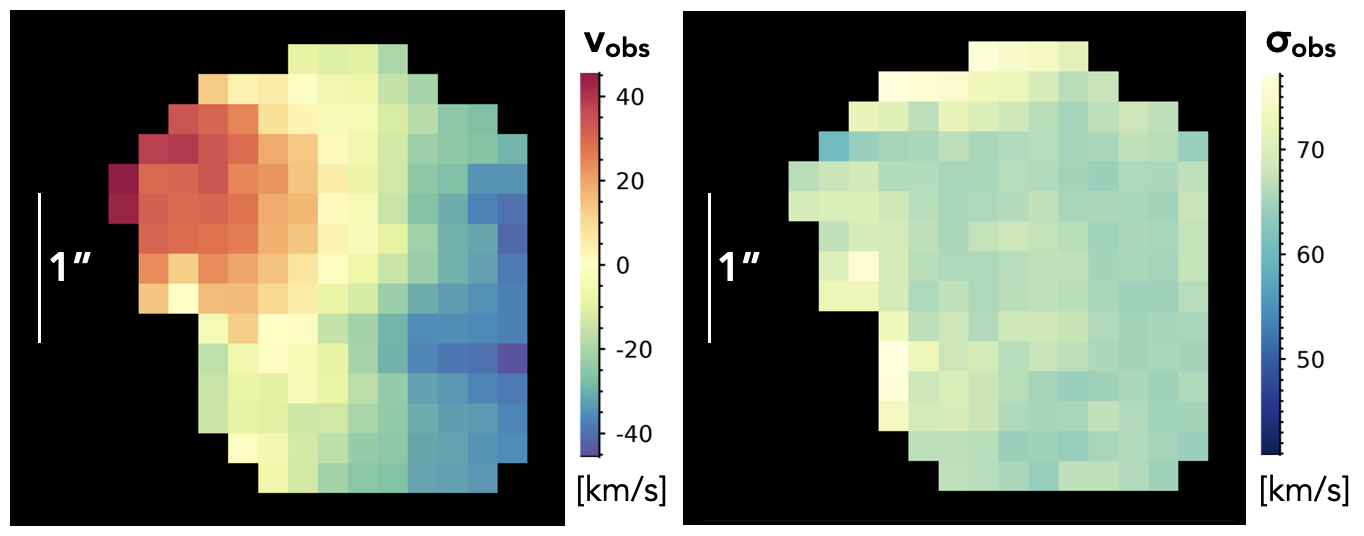}
      \caption{COS4\_25353}
    \end{subfigure}
    \begin{subfigure}{0.4\textwidth} 
       \includegraphics[width=\textwidth]{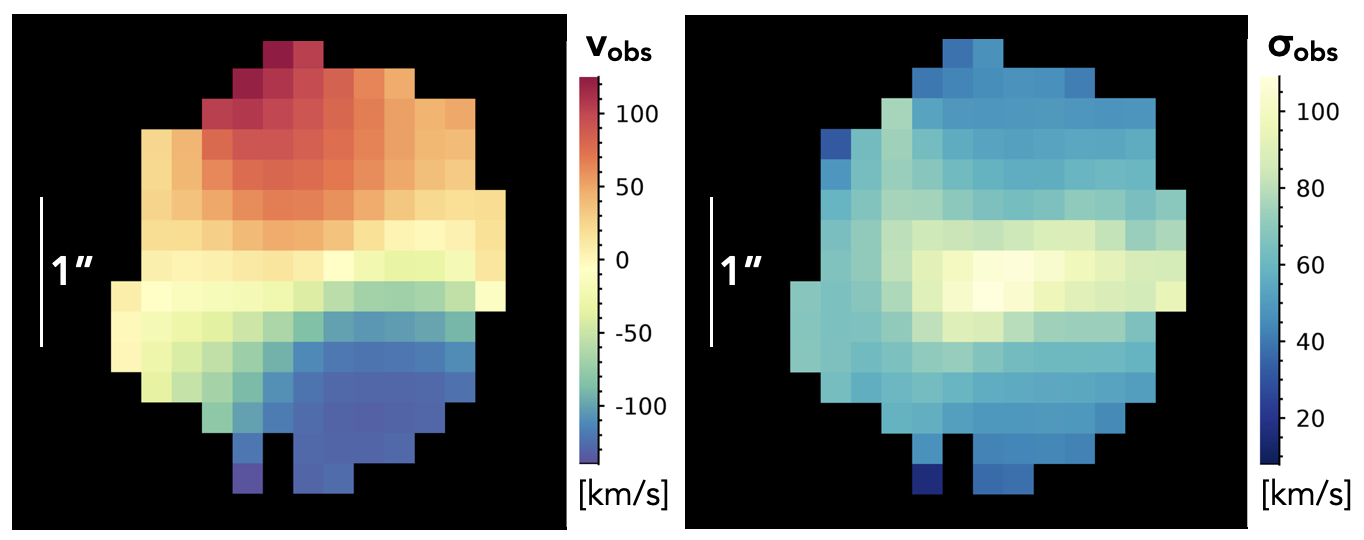}
      \caption{COS4\_17628}
    \end{subfigure}
    \begin{subfigure}{0.4\textwidth} 
       \includegraphics[width=\textwidth]{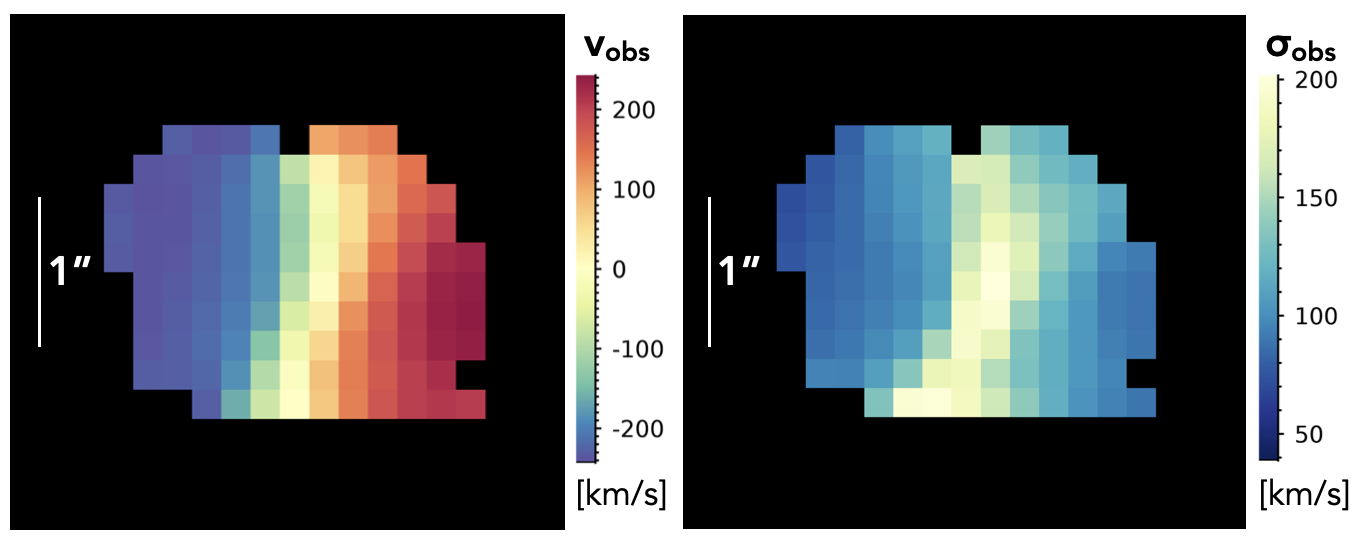}
      \caption{COS4\_06487}
    \end{subfigure}
    \begin{subfigure}{0.4\textwidth} 
       \includegraphics[width=\textwidth]{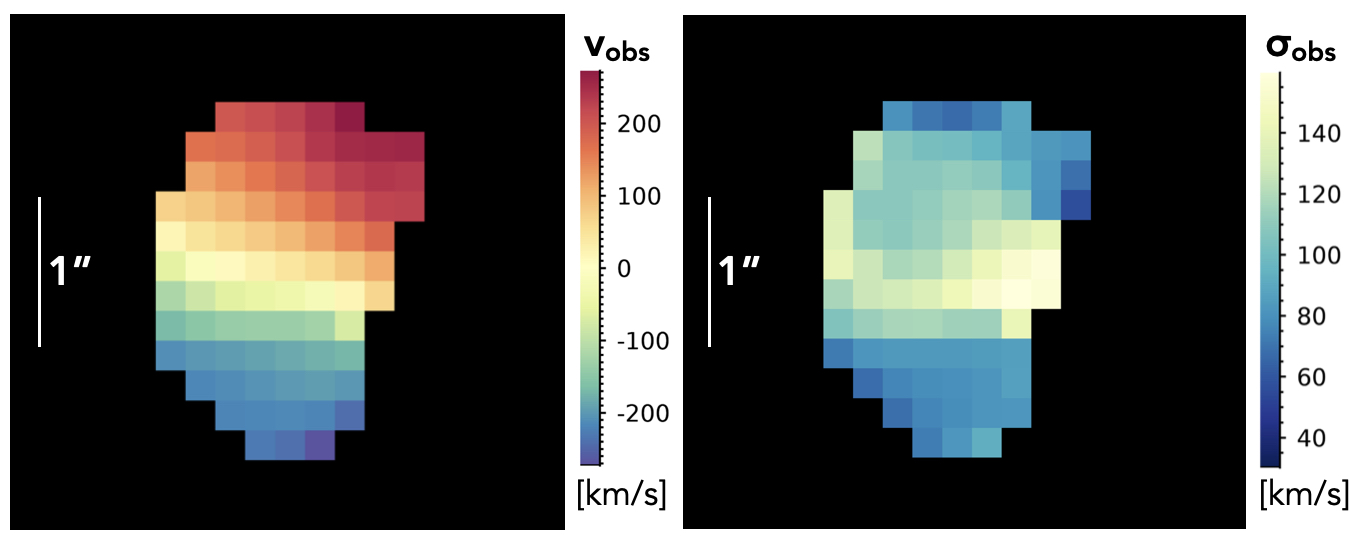}
      \caption{COS4\_05296}
    \end{subfigure}
    \begin{subfigure}{0.4\textwidth} 
       \includegraphics[width=\textwidth]{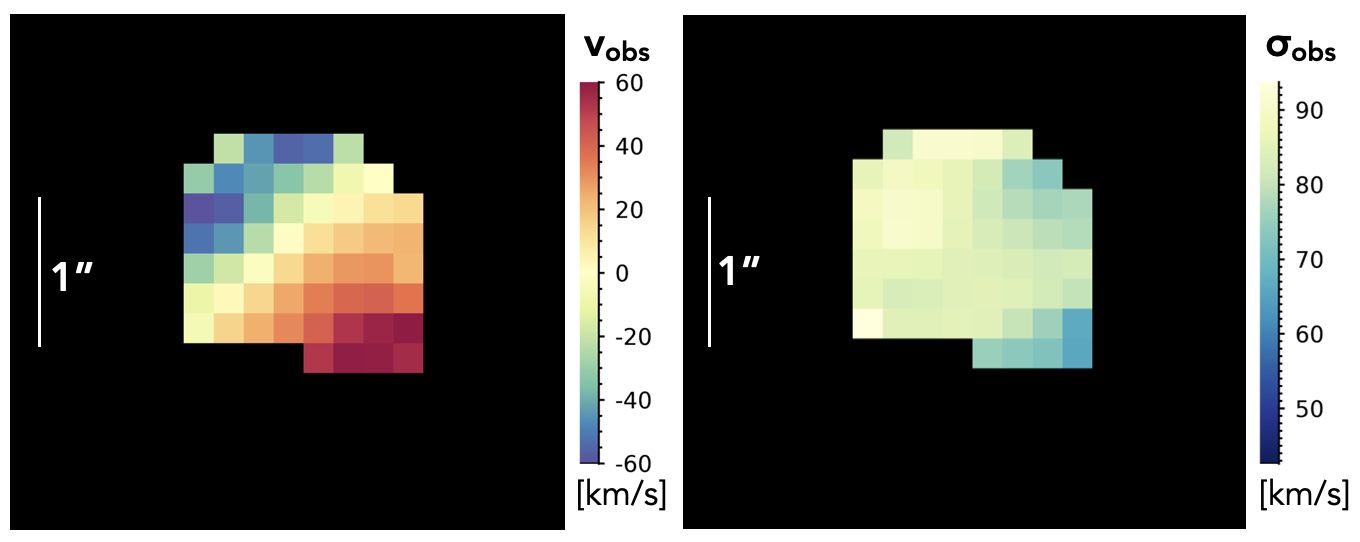}
      \caption{COS4\_08096}
    \end{subfigure}
    \begin{subfigure}{0.4\textwidth} 
       \includegraphics[width=\textwidth]{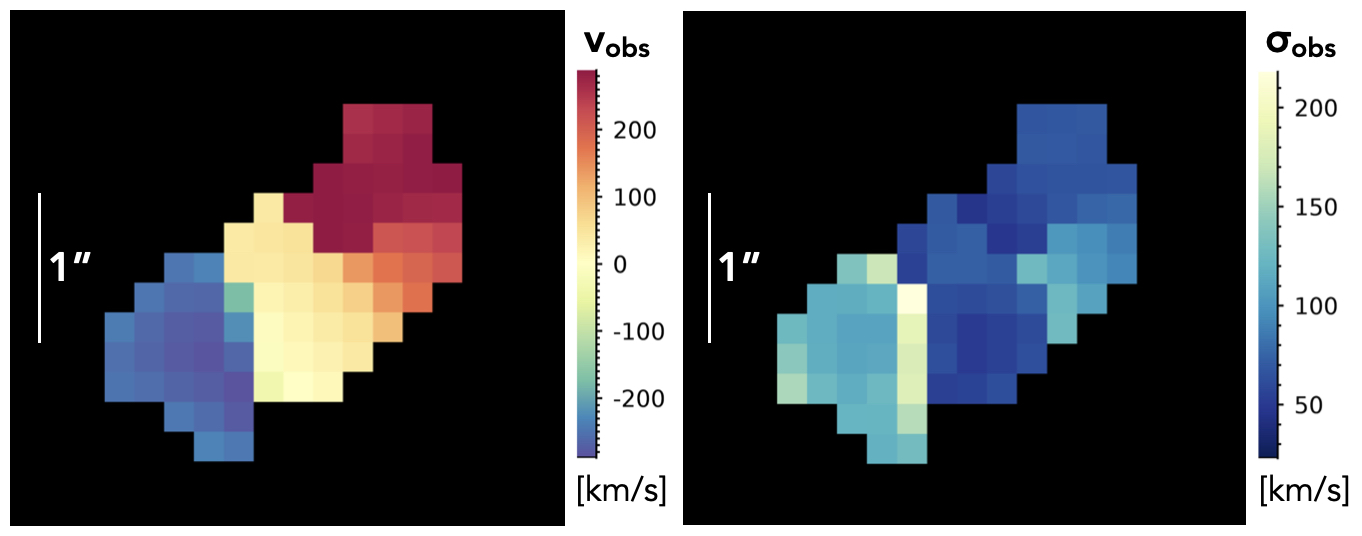}
      \caption{COS4\_12699}
    \end{subfigure}
   	\caption{For illustrative purposes, we show smoothed 2D maps of velocity and velocity dispersion for the ten \kd galaxies for which we construct dynamical models.}
	\label{f:maps}
\end{figure*}

In Figure~\ref{f:maps} we show the velocity and velocity dispersion maps of the ten KMOS$^{\rm 3D}$ galaxies for which we construct dynamical models. In this work, we model the H$\alpha$ major axis kinematics (see Section~\ref{s:mdynk3d}). The 2D maps are shown only for illustrative purposes.

\section{MCMC posterior distributions of the KMOS$^{\rm 3D}$ dynamical models}\label{a:corner}

In Figures~\ref{f:c1} and \ref{f:c2} we show the MCMC posterior distributions for the \kd dynamical modelling performed for ten galaxies. Due to the heterogeneous data quality in our sample, we have fixed the structural parameters for all objects based on available F160W imaging data (see Section~\ref{s:mdynk3d} for details). The free parameters of our models are the total baryonic mass $M_{\rm bar}$, the intrinsic velocity dispersion $\sigma_0$, and the central dark matter fraction $f_{\rm DM}(<R_{e,\rm F160W})$. For galaxies COS4\_04943 and COS5\_05296 we constrain upper limits on $\sigma_0$ through the upper $2\sigma$ boundary of the marginalised posterior distributions (black outlines in Figure~\ref{f:dmdyn3}). In addition, the marginalised posterior distribution of $\sigma_0$ for galaxies COS4\_03493, COS4\_13901 and COS4\_12699 is non-Gaussian and dominated by low values. These galaxies are indicated by grey outlines in Figure~\ref{f:dmdyn3}. The true structural distribution of mass might be different from what is constrained through the F160W imaging, and this can impact the recovery of our free model parameters. In particular, a model with different or free structural parameters might result in different best-fit values for $M_{\rm bar}$, $f_{\rm DM}$, and even $\sigma_0$. This is evident in particular through the anti-correlations seen in all models between the posterior distributions of $M_{\rm bar}$ and $f_{\rm DM}$. However, the total enclosed mass $M_{\rm dyn}(<r)$, the main focus of our dynamical mass comparison in this paper, has been shown to be relatively insensitive to the detailed mass decomposition \citep[e.g.][]{WuytsS16, Price21}. See Appendix~\ref{a:f814w_modelling} for further discussion.

\begin{figure*}
	\centering
    \begin{subfigure}{0.37\textwidth} 
	   \includegraphics[width=\textwidth]{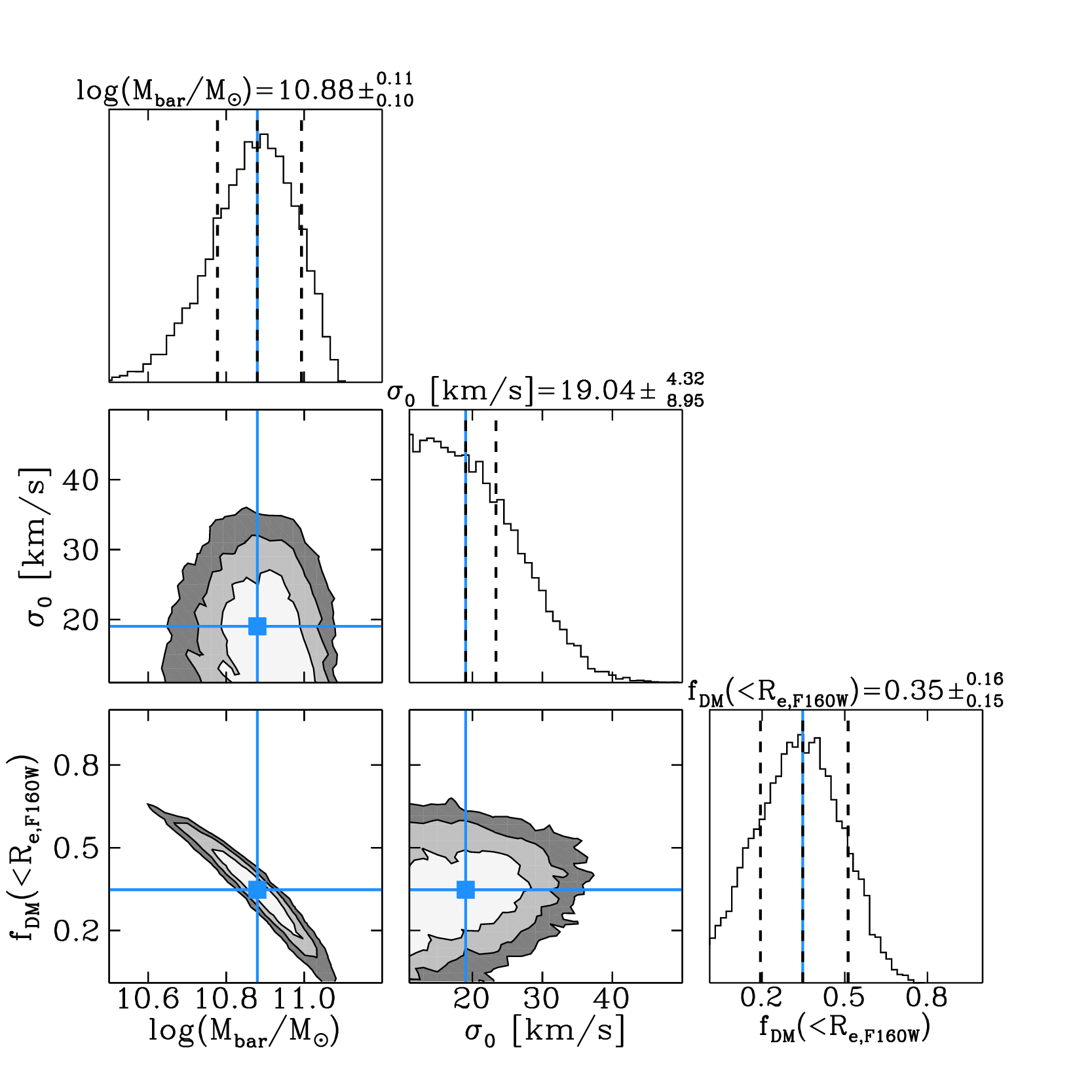}
      \caption{COS4\_03493}
    \end{subfigure}
    \begin{subfigure}{0.37\textwidth} 
    \includegraphics[width=\textwidth]{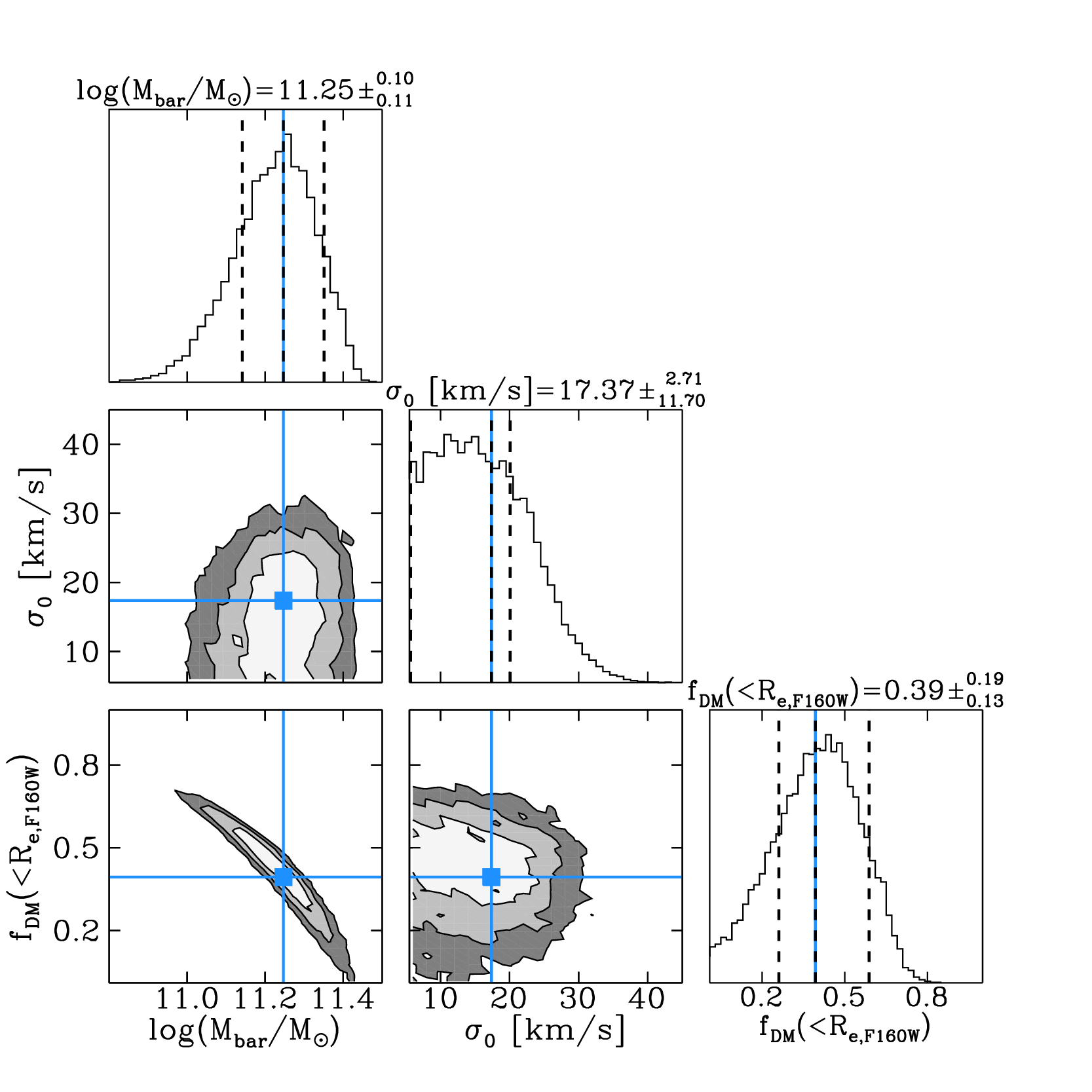}
      \caption{COS4\_13901}
    \end{subfigure}
    \begin{subfigure}{0.37\textwidth} 
    \includegraphics[width=\textwidth]{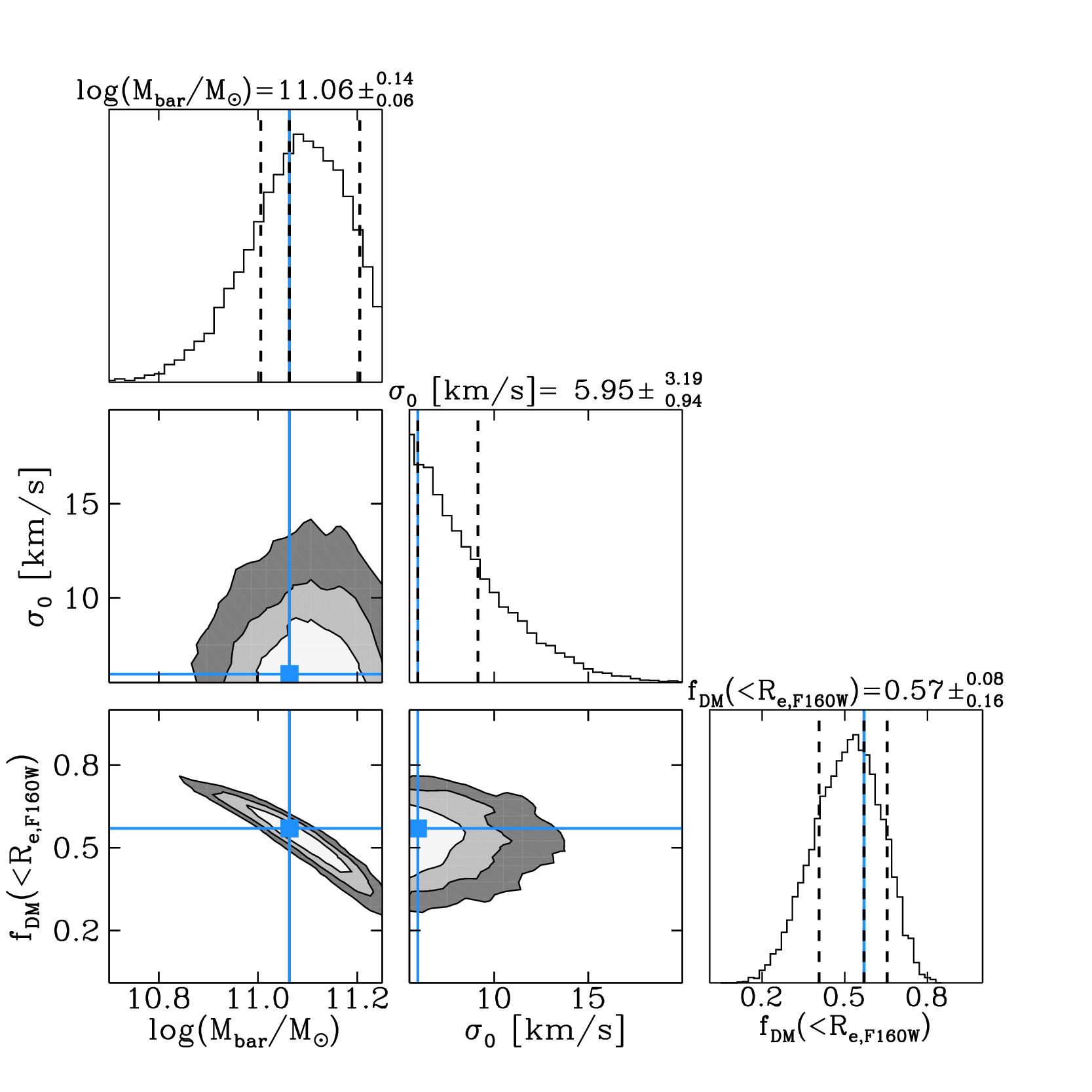}
      \caption{COS4\_04943}
    \end{subfigure}
    \begin{subfigure}{0.37\textwidth} 
    \includegraphics[width=\textwidth]{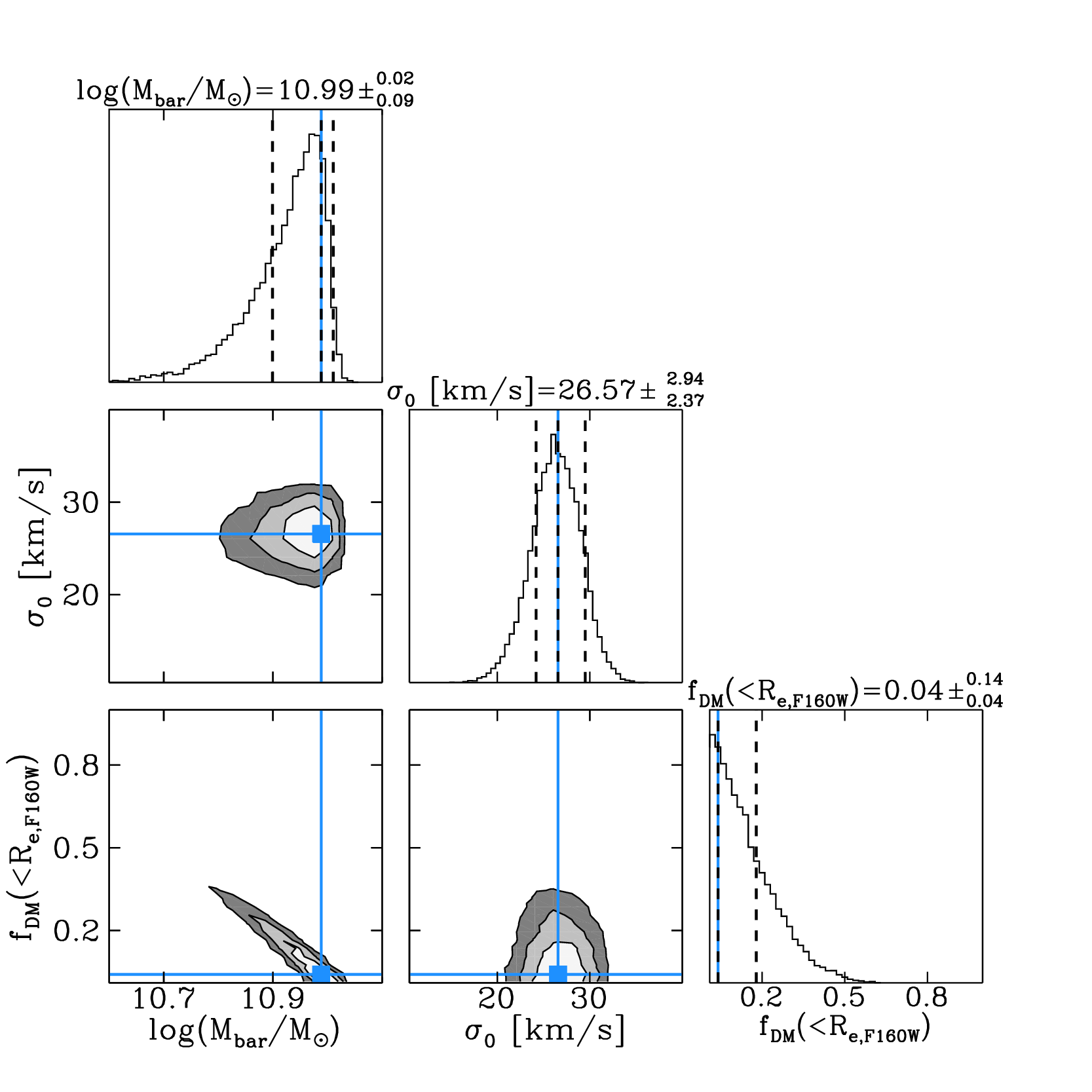}
      \caption{COS4\_19648}
    \end{subfigure}
    \begin{subfigure}{0.37\textwidth} 
    \includegraphics[width=\textwidth]{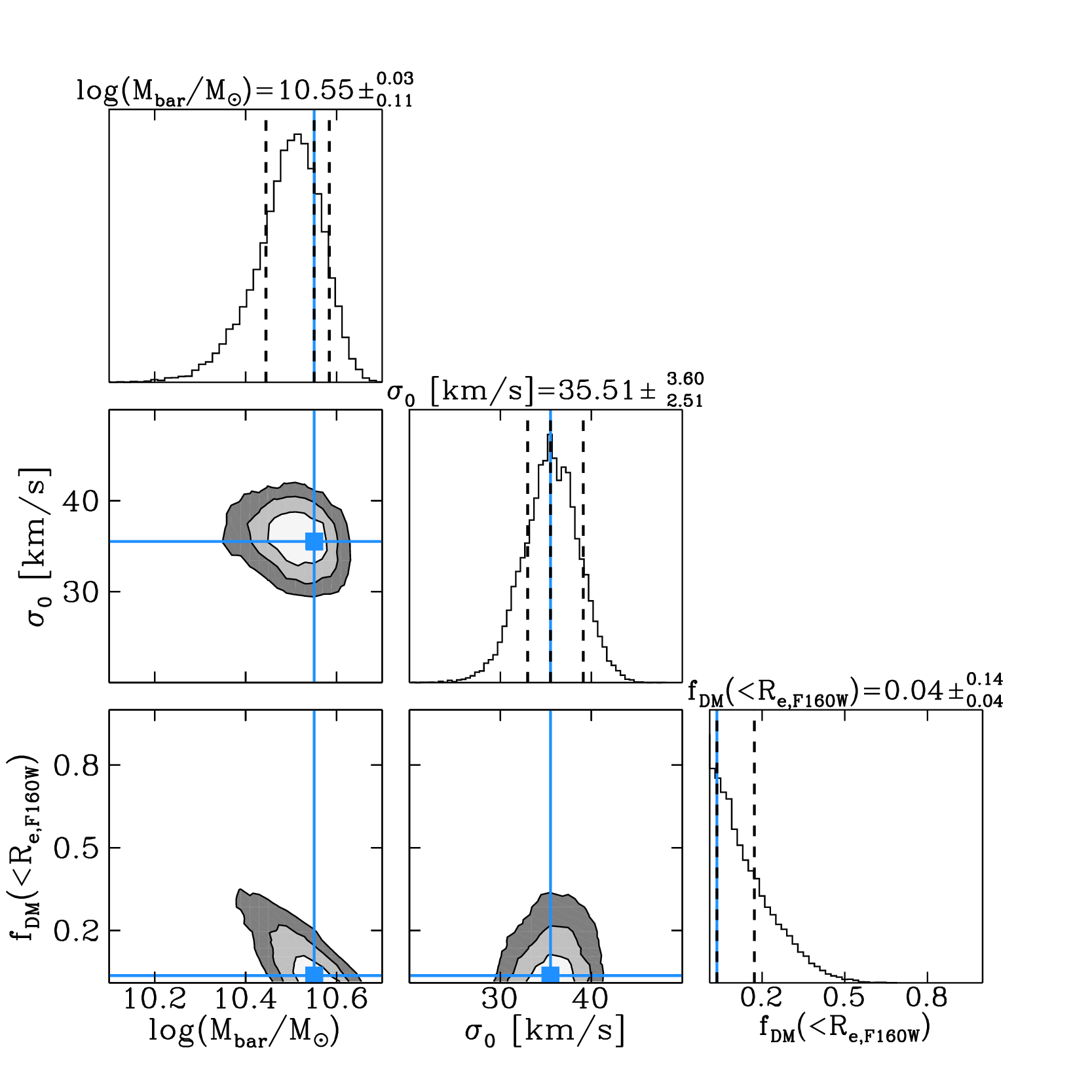}
      \caption{COS4\_25353}
    \end{subfigure}
    \begin{subfigure}{0.37\textwidth} 
    \includegraphics[width=\textwidth]{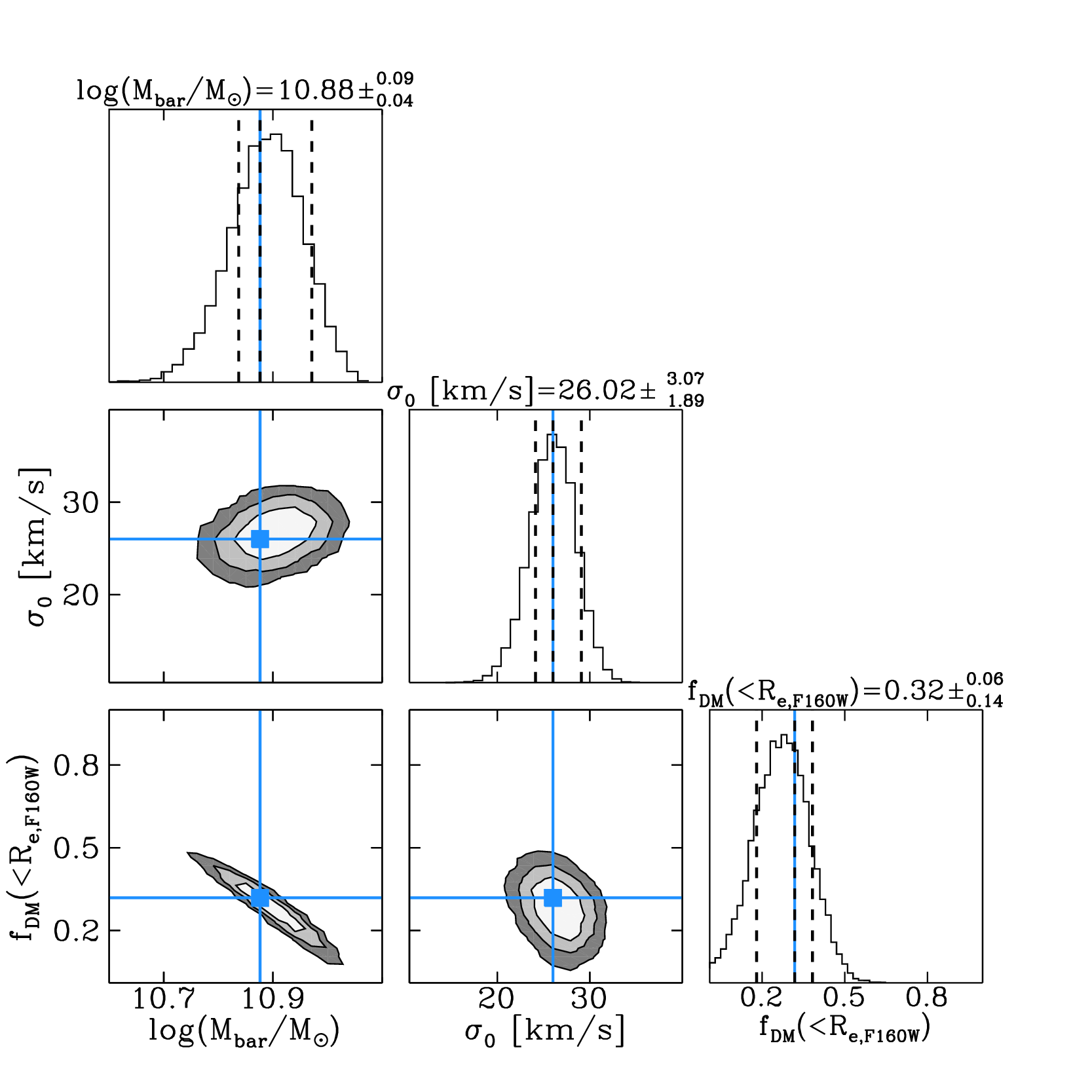}
      \caption{COS4\_17628}
    \end{subfigure}
   	\caption{MCMC sampling of the joint posterior probability distributions of the fiducial model parameters $M_{\rm bar}$, $\sigma_0$, and $f_{\rm DM}(<R_{e,\rm F160W})$ for the \kd dynamical modelling. We indicate the maximum a posteriori value, found by joint posterior analysis \citep[see][]{Price21}, as blue vertical lines in the 1D histograms, and as blue squares in the 2D histograms. Uncertainties on the best-fit parameters (the $68^{\rm th}$ percentiles) are indicated by dashed black lines in the 1D histograms. The 2D distributions show as contours 1, 1.5, and 2 standard deviations.}
	\label{f:c1}
\end{figure*} 

\begin{figure*}
	\centering
    \begin{subfigure}{0.37\textwidth} 
       \includegraphics[width=\textwidth]{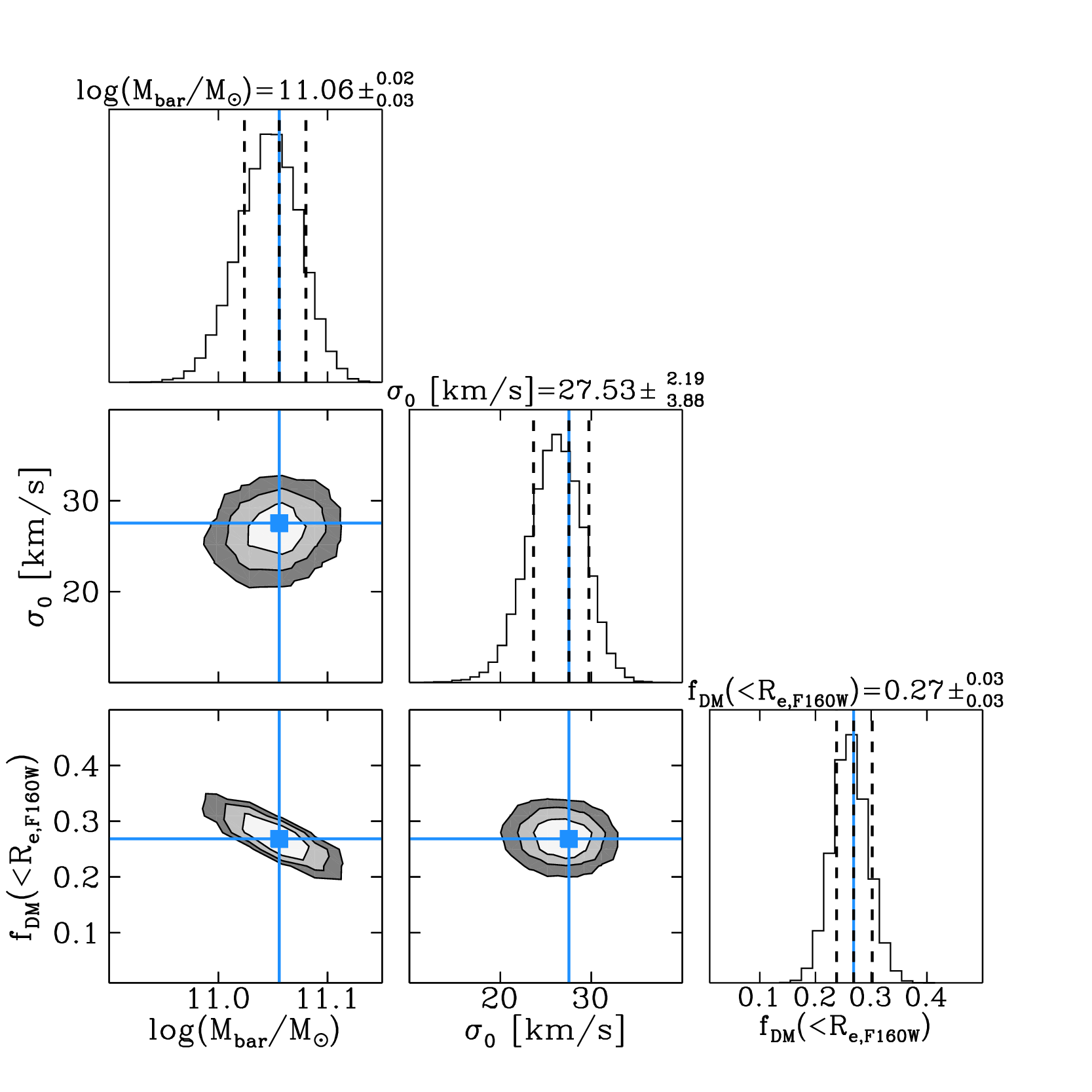}
      \caption{COS4\_06487}
    \end{subfigure}
    \begin{subfigure}{0.37\textwidth} 
    \includegraphics[width=\textwidth]{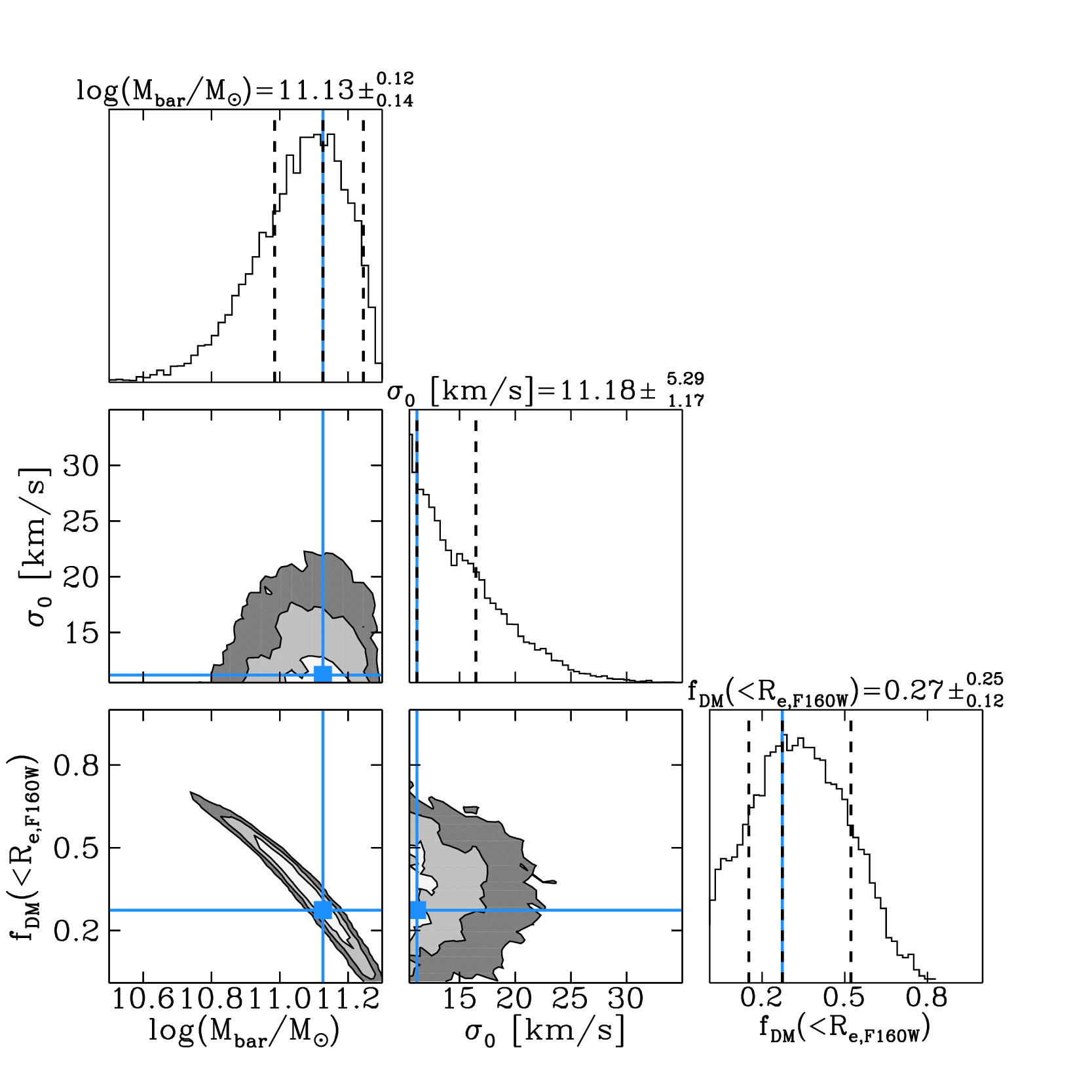}
      \caption{COS4\_05296}
    \end{subfigure}
    \begin{subfigure}{0.37\textwidth} 
    \includegraphics[width=\textwidth]{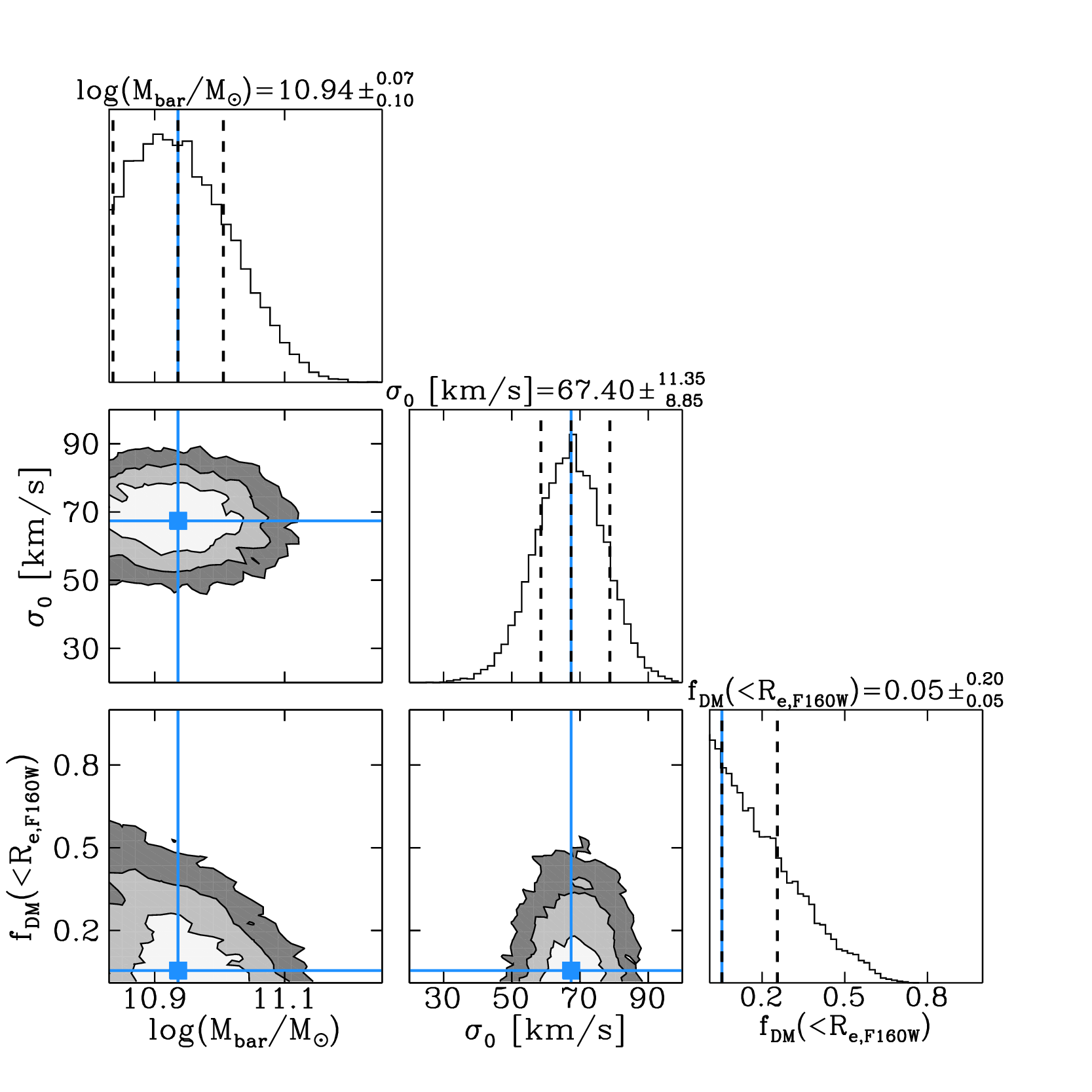}
      \caption{COS4\_08096}
    \end{subfigure}
    \begin{subfigure}{0.37\textwidth} 
    \includegraphics[width=\textwidth]{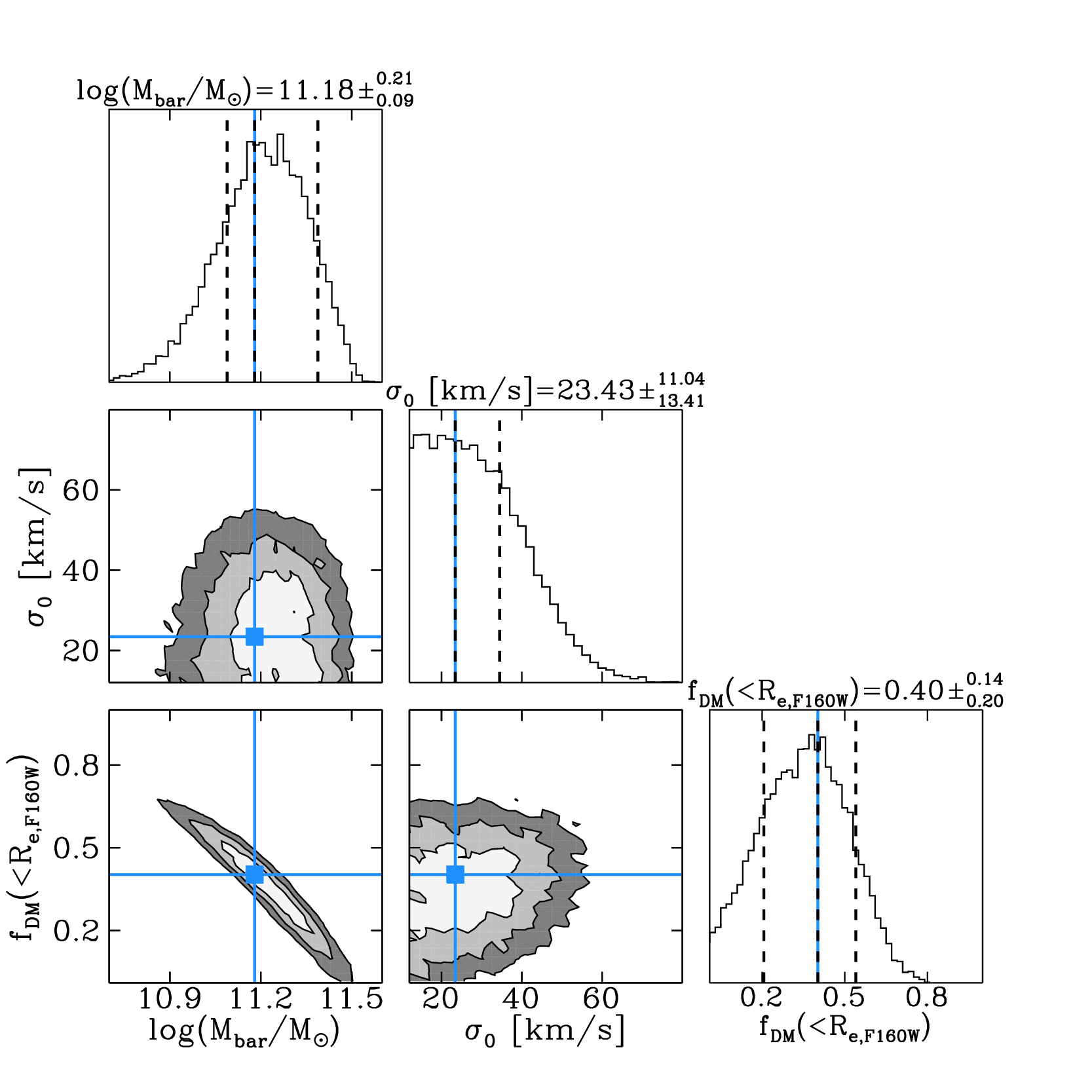}
      \caption{COS4\_12699}
    \end{subfigure}
   	\caption{Continuation of Figure~\ref{f:c1}. MCMC sampling of the joint posterior probability distributions of the fiducial model parameters $M_{\rm bar}$, $\sigma_0$, and $f_{\rm DM}(<R_{e,\rm F160W})$ for the \kd dynamical modelling. We indicate the maximum a posteriori value, found by joint posterior analysis \citep[see][]{Price21}, as blue vertical lines in the 1D histograms, and as blue squares in the 2D histograms. Uncertainties on the best-fit parameters (the $68^{\rm th}$ percentiles) are indicated by dashed black lines in the 1D histograms. The 2D distributions show as contours 1, 1.5, and 2 standard deviations.}
	\label{f:c2}
\end{figure*}

\section{Structural parameters from F160W and F814W}\label{a:structure}

\begin{figure*}
	\centering
	\includegraphics[width=0.3\textwidth]{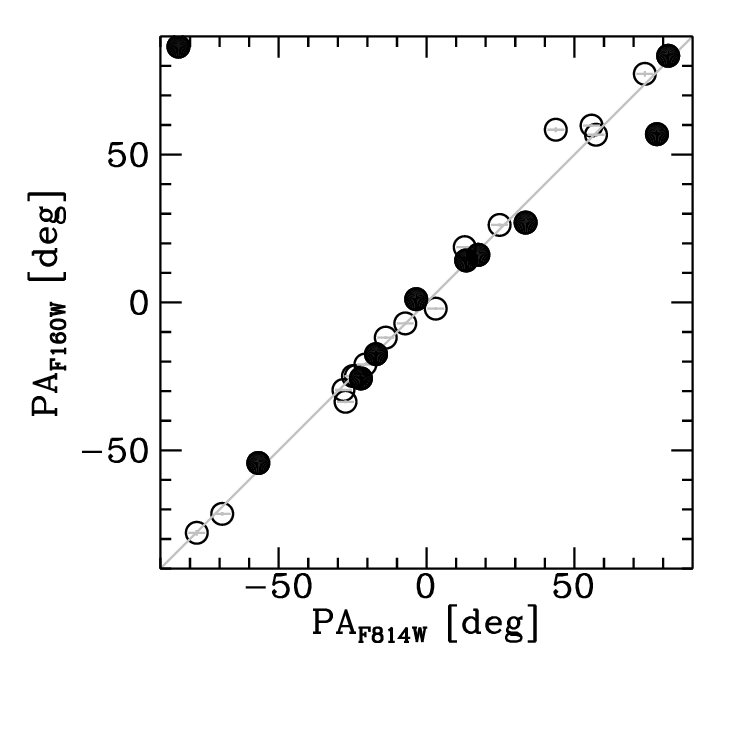}
	\includegraphics[width=0.3\textwidth]{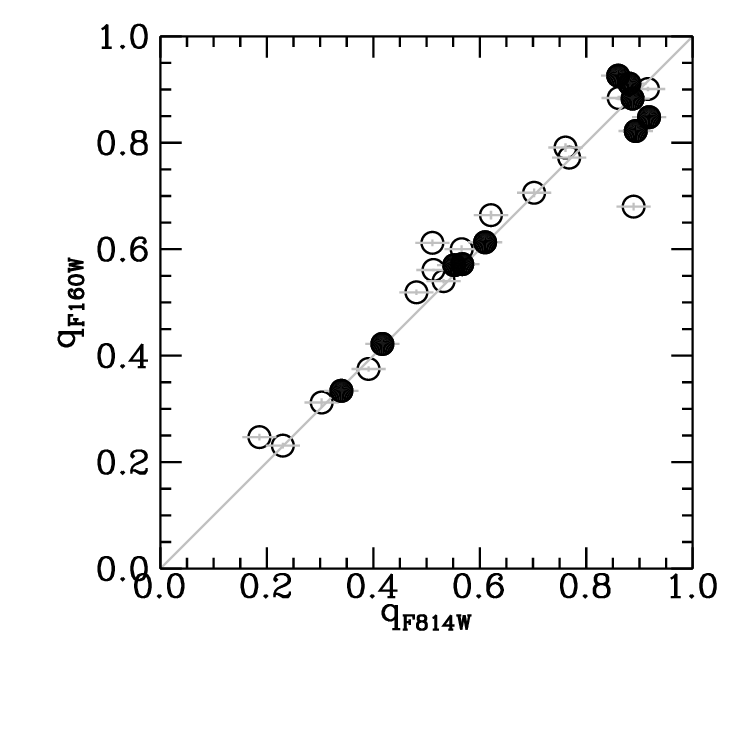}
	\includegraphics[width=0.3\textwidth]{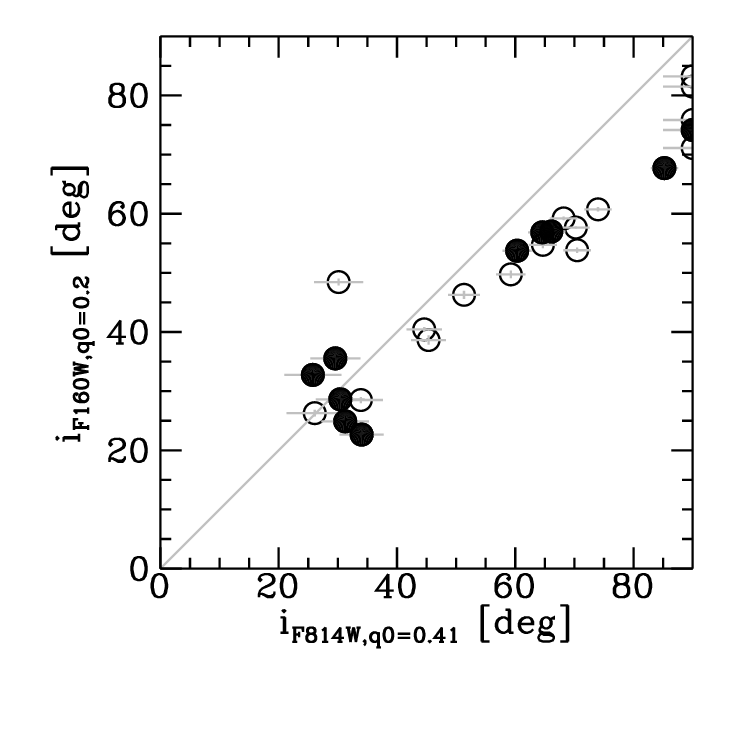}
	\includegraphics[width=0.3\textwidth]{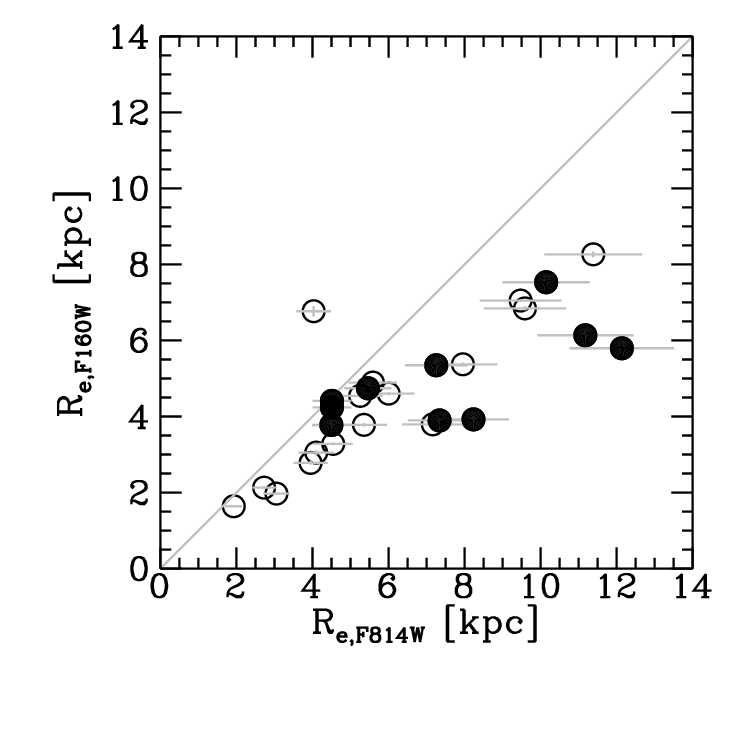}
	\includegraphics[width=0.3\textwidth]{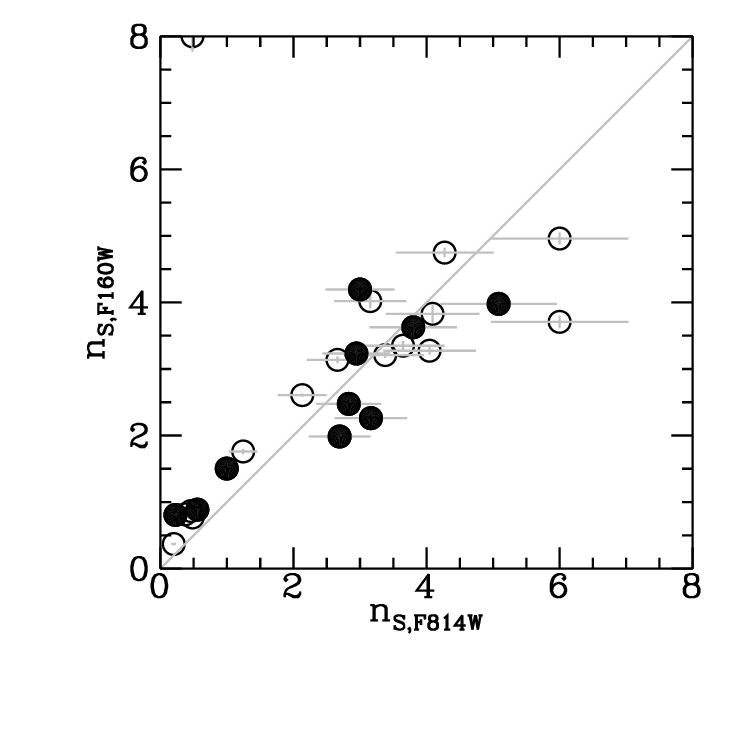}
	\caption{Comparison pf structural parameters constrained from F160W imaging {\it vs.\,} F814W imaging. From top left to bottom right: position angle PA, projected axis ratio $q=b/a$, inclination inferred from $b/a$ and assuming an intrinsic thickness $q_0$, effective radius $R_e$, and Sérsic index $n_S$, for all galaxies (open circles) and the dynamical mass sample (filled circles). Position angles, axis ratios, and Sérsic indices agree reasonably well between the two filter bands, but sizes are systematically larger from F814W, and increasingly so towards larger sizes. This could be due to more recent star-formation in the outer disc regions and/or enhanced central dust obscuration. The LEGA-C JAM models assume an intrinsically thicker distribution for the stellar component which is used as a prior for the inclination (not fixed). The comparison shown here illustrates the effect of the different assumptions on intrinsic thickness if translated directly into inclination.}
	\label{f:comp_structure}
\end{figure*}

We compare structural parameters constrained from F160W and F814W imaging for galaxies in our sample (Figure~\ref{f:comp_structure}). Morphological position angles (top left), axis ratios $q=b/a$ (top middle), and Sérsic indices (bottom right) agree reasonably well. In the top right panel we compare inferred inclinations based on different assumptions about the intrinsic thickness of galaxies. For KMOS$^{\rm 3D}$, this corresponds to the inclinations that are used in the fiducial dynamical modelling. For LEGA-C, we show inclinations inferred by assuming $q_0=0.41$. This value corresponds to the centre of the normally distributed prior on $q_0$ for the JAM modelling described by \cite{vHoudt21}. \cite{vHoudt21} note that the inclination is typically unconstrained by their data, and the prior mainly enters the modelling to account for this uncertainty in the error budget.
This figure demonstrates the general effect of assuming an intrinsically thicker distribution, which is increasingly prominent for more inclined systems. Since higher inclination $i$ results in lower total mass estimates for the same observed velocity ($v_{\rm obs}=v_{\rm rot}\cdot\sin(i)$), we might expect a general effect such that higher$-i$ systems have relatively lower dynamical masses when comparing to estimates assuming intrinsically thinner distributions. 

Considering effective radii measured from F160W and F814W, we see a clear trend of larger estimates from the F814W imaging (bottom left). For the KMOS data we used F160W imaging following work by \cite{Wisnioski15, Wisnioski18, Wisnioski19, WuytsS16, Genzel17, Genzel20, Lang17, Uebler17, Uebler19,Price21} to approximate the baryonic mass distribution. 
\cite{WuytsS12} have shown that the stellar mass distribution of SFGs at this redshift is more concentrated compared to $H-$band light. On the other hand, \cite{NelsonE16a} and \cite{Wilman20} find that effective radii measured from H$\alpha$ emission, tracing recent star formation and thereby the cold gas distribution, are larger by a factor $\sim1.1-1.2$, even after dust correction \citep{Tacchella18, NelsonE21}. 
The shallower F814W imaging used to constrain the dynamical modelling of the LEGA-C data is more sensitive to younger star-forming regions and to substructure \citep[see e.g. examples by][]{WuytsS16}. Therefore, it is not surprising that almost all size measurements for our sample from $i-$band data are larger than those from $H-$band. However, these differences in effective radii should be of little concern as in general the total dynamical mass is insensitive to the relative distributions of baryons and dark matter \citep[see e.g.][]{WuytsS16, Uebler19, vHoudt21}. Since we compare dynamical mass estimates for \kd and LEGA-C models at the same physical radii (see Section~\ref{s:compmdyn}), we expect negligible effects from the different effective radii utilized in the modelling procedures (see also Appendix~\ref{a:f814w_modelling}).

\section{Impact of slit mis-centering and underestimated velocity gradients}\label{a:misalignment}

\begin{figure}
	\centering
	\includegraphics[width=\columnwidth]{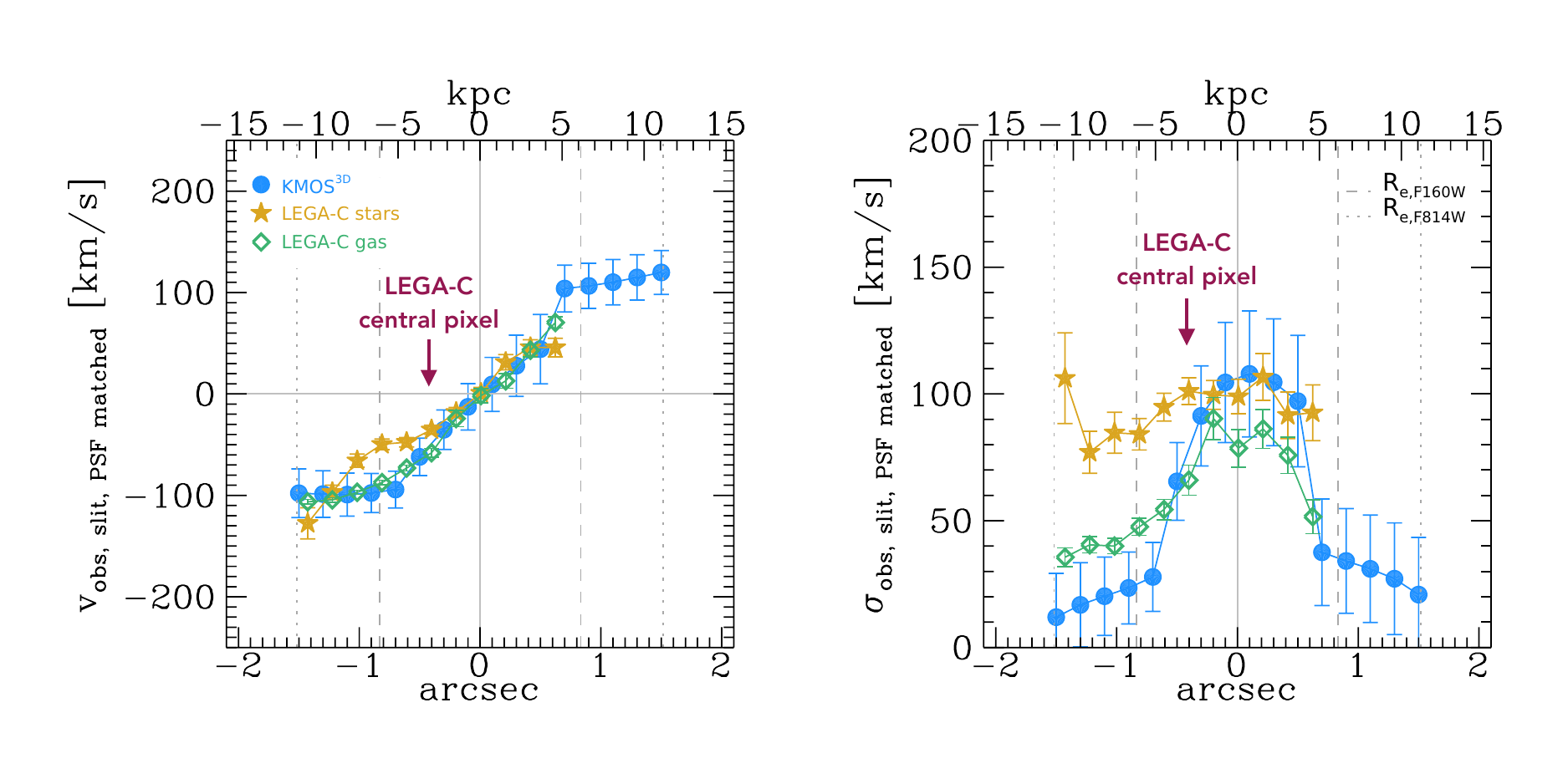}
	\caption{1D kinematic profiles extracted along a N-S (pseudo-)slit for galaxy COS4\_04943-M3\_122667. The LEGA-C stellar and gas kinematics (filled golden stars and open green diamonds, respectively), have been aligned with the \kd data (filled blue circles) based on the profile shapes, the minimum and maximum velocities, and the dispersion peak. The extraction corresponding to the central pixel row of the LEGA-C 2D spectrum is offset from the kinematic centre of the \kd data by $\sim0.4\arcsec$.}
	\label{f:COS3_05062}
\end{figure}

In Figure~\ref{f:COS3_05062} we show kinematic extractions for galaxy COS4\_04943-M3\_122667 along the N-S (pseudo-)slit. For this object, the primary emission feature in the LEGA-C spectrum is the H$\beta$ line, and we use the gas emission line profiles to align the LEGA-C data to the \kd extractions. We indicate in the figure the central pixel of the LEGA-C 2D spectrum corresponding to the 1D extractions shown, relative to the kinematic centre from the \kd extractions. Based on this alignment, the LEGA-C data do not sample the full extent of the rotation curve.

This galaxy has a dynamical mass estimate from \kd that is higher by 0.44~dex compared to the LEGA-C estimate (see discussion in Section~\ref{s:compmdyn}). The comparison of the \kd and LEGA-C extracted kinematics suggests that the integrated stellar velocity dispersion might be underestimated, which could explain this large difference. Indeed, from the observed \kd kinematics along the kinematic major axis we get $v_{\rm rms}\sim151$~km/s, while the integrated stellar velocity dispersion amounts to $\sigma_{\star,\rm int}\sim104$~km/s (both estimates not accounting for inclination). 

This galaxy has also a dynamical mass estimate from LEGA-C based on the modelling of the resolved ionised gas kinematics by \cite{Straatman22}, which is also lower than the \kd estimate by 0.41~dex. Since in their analysis the kinematic profiles are symmetrized around the central pixel before fitting, the partial coverage of the full velocity gradient as seen in Figure~\ref{f:COS3_05062} would explain the lower dynamical mass estimate.

We note that the JAM measurement, however, corresponds well within the uncertainties to the \kd measurement (see Section~\ref{s:mcum} and the top left panel of Figure~\ref{f:mcum}). This is likely due to the flexibility of the JAM model to find a centre position based on the photometry other than the central pixel.

%%%%%%%%%%%%%%%%%%%%%%%%%%%%%%%%%%%%%%%%%%%%%%%%%%

% Don't change these lines
\bsp	% typesetting comment
\label{lastpage}
\end{document}